\documentclass[aps,prb,twocolumn,nofootinbib,superscriptaddress,10pt]{revtex4-2}
\pdfoutput=1 
\usepackage[utf8]{inputenc}
\usepackage[english]{babel}
\usepackage[T1]{fontenc}
\usepackage[pdftex]{graphicx}
\usepackage{amsmath}
\usepackage{cancel}
\usepackage{amsthm}
\usepackage{amssymb}
\usepackage{bbm}
\usepackage{braket}

\newcommand{\maxstate}{\phi_{\text{max}}}
\newcommand{\minstate}{\phi_{\text{min}}}
\newcommand{\ctc}{\Lambda(\hat M)}
\newcommand{\doubleE}{\mathbb{E}}

\newtheorem{lemma}{Lemma}

\usepackage{xcolor}
\usepackage{dcolumn}
\usepackage{bbm}
\usepackage{bbold}
\usepackage[normalem]{ulem}
\usepackage{footmisc}
\usepackage{xcolor}
\usepackage{booktabs}
\usepackage{comment}
\usepackage{multirow}
\usepackage[colorlinks=true,linkcolor=blue,citecolor=blue,urlcolor=blue,unicode]{hyperref}

\newcommand{\Tr}{\mathrm{Tr}}

\def\lp{\left (}
\def\rp{\right )}

\begin{document}
\title{Charge Transport Capacity as a Probe of Resonances in Models of Many-Body Localization
}

\author{Jessica K. Jiang}
\affiliation{Department of Physics and Institute for Quantum Information and Matter,
California Institute of Technology, Pasadena, California 91125, USA}
\author{Federica M. Surace}
\affiliation{Department of Physics and Institute for Quantum Information and Matter,
California Institute of Technology, Pasadena, California 91125, USA}
\affiliation{School of Physics, Trinity College Dublin, Dublin 2, Ireland}
\affiliation{Universität Innsbruck, Institut für Theoretische Physik, Technikerstraße 21a, 6020 Innsbruck, Austria}
\author{Olexei I. Motrunich}
\affiliation{Department of Physics and Institute for Quantum Information and Matter,
California Institute of Technology, Pasadena, California 91125, USA}
\date{\today}

\begin{abstract}
The fate of Many-Body Localization (MBL) in the thermodynamic limit remains elusive, partly because numerical studies suffer from unexplained finite-size effects.
We introduce and numerically study the \emph{charge transport capacity} (CTC)---a quantity that upper bounds the number of particles that can ever be transported across a central cut of a one-dimensional lattice.
For ergodic systems of spinless fermions, the CTC is linear in the system size $L$, while we expect it to be $O(1)$ for localized models.
Surprisingly, in the interacting Anderson model for numerically accessible $L$, the disorder-averaged CTC is small, but grows with $L$ at an increasing rate. 
Moreover, this growth rate appears to be independent of the disorder strength $W$ at very large $W$.
We find that, for these system sizes, this growth occurs because, as $L$ increases, many-body resonances that transport more charge across the cut become more likely.
Using a diagonally-improved perturbative model for the weakly interacting regime, we provide an understanding of the microscopic origins of the growth of these \emph{charge transport resonances} (CTRs).
We find that these resonances are sensitive to charge configurations over a spatial region whose size is set by the range of the resonance, not by $W$, and that numerics cannot access system sizes where their behavior will converge.
However, this effective model is consistent with a regime of strong disorder where, for large $L$, resonances are exponentially suppressed in their size.
Finally, we study measures of average charge transport and suggest that for strong enough disorder, average product states can only transfer $O(1)$ charge.
Our work suggests that the unsettled growth of short-ranged many-body resonances with $L$ contributes to the finite-sized drift towards thermalization at numerically accessible system sizes, and provides an understanding of how they can remain controlled or eventually destabilize the MBL phase.
\end{abstract}

\maketitle

\section{Introduction}
Interacting quantum systems subject to strong quenched disorder are conjectured to exhibit Many-Body Localization (MBL). 
If it exists, MBL would be the only \emph{stable} phase of matter known to violate the Eigenstate Thermalization Hypothesis (ETH), giving rise to useful, unique properties.
The first arguments for the stability of Anderson localization~\cite{anderson_absence_1958, lagendijk_fifty_2009} under interactions were put forth in~\cite{gornyi_interacting_2005,basko_metalinsulator_2006}, eventually followed by a proposed proof of its existence in a one-dimensional disordered Ising spin chain, but under the assumption of limited level attraction~\cite{imbrie_many-body_2016,imbrie_diagonalization_2016}. 
Extensive numerical studies also showed signatures of localization at large disorder strengths~\cite{oganesyan_localization_2007,znidaric_many-body_2008,pal_many-body_2010,bardarson_unbounded_2012,bauer_area_2013,de_luca_ergodicity_2013,kjall_many-body_2014,khemani_nonlocal_2015,bera_many-body_2015,bar_lev_absence_2015,serbyn_criterion_2015,khemani_critical_2017,doggen_many-body_2018,mace_multifractal_2019, sierant_polynomially_2020,morningstar_avalanches_2022, colbois_interaction-driven_2024}.
Finally, experiments have explored various aspects of the MBL phenomenology~\cite{schreiber_observation_2015,smith_many-body_2016,choi_exploring_2016,luschen_observation_2017,lukin_probing_2019,rispoli_quantum_2019,guo_observation_2021,chiaro_direct_2022,leonard_probing_2023}.
For more comprehensive reviews of MBL, see Refs.~\cite{nandkishore_many-body_2015,alet_many-body_2018,abanin_colloquium_2019,sierant_many-body_2025}.

In recent years, the status of MBL as a phase in the thermodynamic limit has become controversial.
The root of this controversy is the presence of strangely persistent finite-size effects observed in numerical studies, where the apparent critical disorder strength $W^*$ moves to larger values as the system size $L$ increases.
Although this drift was observed in early numerical studies of MBL~\cite{oganesyan_localization_2007,bar_lev_absence_2015}, its importance was brought back to light when a series of works used this observation to argue that MBL does not exist in the thermodynamic limit~\cite{suntajs_quantum_2020,suntajs_ergodicity_2020,sels_dynamical_2021}, triggering a debate over MBL as a phase.
The counter arguments put forth in favor of MBL argued that extrapolating from small-size numerics was unreliable~\cite{panda_can_2020,sierant_thouless_2020,abanin_distinguishing_2021,sierant_challenges_2022}, but also pointed out that these finite-size drifts can be compatible with MBL in the thermodynamic limit~\cite{crowley_constructive_2022}.
It was then found, based on the understanding and simulations of avalanches~\cite{de_roeck_stability_2017, thiery_many-body_2018, morningstar_avalanches_2022, sels_bath-induced_2022}, that the \emph{lower bound} of $W^*$ is at much higher disorder strengths than previously believed~\cite{de_roeck_stability_2017,thiery_many-body_2018,morningstar_avalanches_2022, sels_bath-induced_2022}. 
The resolution of this debate~\cite{kiefer-emmanouilidis_evidence_2020, aceituno_chavez_ultraslow_2024, luitz_absence_2020, ghosh_resonance-induced_2022, weisse_operator_2025} and thus the status of MBL remains unclear.

Amidst this controversy over the existence of MBL, an important question remains unanswered: what are the physical mechanisms that push finite-size MBL regimes towards thermalization in the first place? 
The answer to this question is not only important for resolving the debate over MBL, but also could be useful for understanding why quantum many-body systems thermalize in a broader sense.
On the theoretical side, the current accepted theory of how MBL is destabilized in the thermodynamic limit is the avalanche mechanism~\cite{de_roeck_stability_2017, thiery_many-body_2018}.
However, numerically accessible system sizes are too small to host the rare thermal regions and their signatures that form a central part of the avalanche theory.
Thus, it cannot explain the observed thermal drift at numerically accessible $L$.
On the numerical side, it is clear that these destabilizing mechanisms cannot be captured using spectral properties averaged across eigenstates, which observe the system to be localized at finite sizes even at disorder strengths believed to be thermal in the thermodynamic limit.
Instead, one is interested in the statistics of \emph{rare} events, neglected by aggregate measures, which could proliferate and thus destabilize the MBL phase at larger system sizes~\cite{morningstar_avalanches_2022,colbois_statistics_2024, biroli_large-deviation_2024}.
To date, the exact mechanisms for this finite-size drift are still unclear.

One class of such rare events is \emph{many-body resonances} (MBRs).
In the simplest case, these resonances are rare pairs of nearly degenerate many-body eigenstates that are orthogonal superpositions of two different configurations of charges or Anderson LIOMs.
An MBR is system-wide if the two configurations differ on a region of size $O(L)$.
We say interchangeably that the two eigenstates, or the corresponding charge configurations, are in resonance.
Many-body resonances were recognized early on for their role in the unusual conductivity properties in the MBL phase~\cite{gopalakrishnan_low-frequency_2015, agarwal_rareregion_2017}. 

The influence of these MBRs on various aspects of the MBL phase diagram has been explored in many previous works~\cite{agarwal_rareregion_2017,khemani_critical_2017,kjall_many-body_2018, villalonga_eigenstates_2020, garratt_local_2021,garratt_resonant_2022,crowley_constructive_2022,long_phenomenology_2023,ha_many-body_2023, biroli_large-deviation_2024}.
In particular, the statistics and properties of \emph{system-wide} resonances at numerically accessible system sizes have been well-characterized numerically, as they were recognized as a possible key player in destabilizing the MBL phase at larger system sizes~\cite{morningstar_avalanches_2022,ha_many-body_2023, biroli_large-deviation_2024,laflorencie_cat_2025, padhan_long-range_2025}.
On the theoretical side, the presence of resonances of all ranges in the large $L$ regime and their influence on local observables were studied in Refs.~\cite{garratt_local_2021, garratt_resonant_2022}. 
The presence of these MBRs was, furthermore, proposed as an explanation for the properties of the numerical crossover regime in Ref.~\cite{crowley_constructive_2022}.
These works furthermore explored numerically and analytically the cumulative effects of these resonances on observable quantities.

Many properties of these MBRs remain elusive.
An open question is that it is unclear how to reconcile the observations of many-body resonances at finite system sizes and the asymptotic instabilities of MBL (some connections between avalanches and MBRs were made in~\cite{ha_many-body_2023}).
Importantly, the direct microscopic resolution of short- and intermediate-range MBRs at numerically accessible system sizes has been lacking, largely because these local resonances are much more numerous in the spectrum and difficult to resolve individually.
Thus, past work has mostly focused on the aggregate consequences of these local resonances, but it is unclear how their microscopic properties evolve with the system size.
In this work, we advance our understanding in these directions, where we find that, surprisingly, charge transport resonances (CTRs) that transport larger amounts of charge systematically become more probable with $L$ for numerically accessible $L$, even at extremely strong disorder strengths. 
We explain this behavior through microscopic mechanisms and extend this understanding to the asymptotically large $L$ limit. 
We argue that these finite-sized drifts of CTRs are an inherent feature of the MBL phase.

Specifically, we take a new perspective on the finite-size destabilization of MBL by studying the \emph{charge transport capacity} (CTC).
The CTC is an upper bound on the net charge transfer across a cut, maximized over all possible initial states and times.
The CTC was first studied in our previous work~\cite{jiang_quasiconservation_2025}, and originally drew inspiration from the integrated heat current operator in~\cite{de_roeck_absence_2024}.
The CTC is a useful probe of ergodicity breaking---specifically, the (disorder-averaged) CTC scales as $L/2$ in ergodic systems but remains $O(1)$ in non-interacting localized systems.
The intuition for the scaling of the CTC is simple:
Ergodic systems are capable of moving an extensive amount of charge, whereas the absence of transport in localized systems means that only $O(1)$ charge can be moved. 
Compared to other measures studied in ED, the CTC is unique in that it captures the worst-case scenario of transport, making it highly sensitive to rare transport events.\footnote{We note that the seemingly related problem of closed-system charge transport and conductivity in MBL systems has been studied in numerous previous works. 
Many of these studies focused on the time decay of autocorrelation functions under quench dynamics starting from a product state~\cite{hauschild_domain-wall_2016, gopalakrishnan_griffiths_2016, luitz_extended_2016, agarwal_rareregion_2017, lezama_apparent_2019, lezama_equilibration_2021, sierant_challenges_2022}.
Another recent point of interest is the apparent growth of the \emph{number entropy}, associated with particle-number fluctuations of a subsystem with time~\cite{kiefer-emmanouilidis_evidence_2020}, the interpretation of which remains the subject of debate~\cite{luitz_absence_2020, ghosh_resonance-induced_2022, aceituno_chavez_ultraslow_2024}. 
On the other hand, recent rigorous results have ruled out the possibility of diffusive transport in strongly disordered interacting chains, while not excluding, in principle, anomalous subdiffusive transport~\cite{de_roeck_absence_2024}.
Our work explores a distinct regime, where we consider closed-system charge transport processes over all possible time scales, including processes that can happen on times that are exponentially long in the system size, which are typically not considered important in usual transport characterizations.}
In particular, the eigenstates of the CTC operator uniquely correspond to \emph{charge transport resonances} (CTRs): MBRs which transport net charge across the cut.
To our knowledge, this provides the first numerical characterization of how the microscopic details of finite-range MBRs evolve with system size.

Our main results are as follows: 
As the system size increases, CTRs that move increasing amounts of charge across the central cut in the lattice become more probable, at a rate appearing to be constant in the disorder strength $W$.
This manifests in the dependence of the CTC on $L$, where, in stark contrast to the case of Anderson localization, the disorder-averaged CTC remains small but grows apparently without saturation with the system size once interactions are present.
This behavior persists at very high disorder strengths, deep into the regime where MBL is presumed.

However, the story of this unbounded growth appears to be more nuanced. 
We also observe a disorder-independent sensitivity of the largest detected resonances to the configurations of fixed ``background'' charges that are frozen between the two resonating product states, even at the largest system sizes accessible.
The proliferation of the CTRs can be explained by a simple model based on a \emph{diagonally-improved} first-order perturbation theory~\cite{lin_slow_2020}.
The CTC in this toy model, despite its apparent linear growth at accessible system sizes, is strictly bounded in the thermodynamic limit by an $O(1)$ number, showing that the asymptotic fate of the CTC remains unknown.
Furthermore, we argue that this model is consistent with a strong disorder regime where, in the $L\rightarrow\infty$ limit, the probability of resonance occurrence is exponentially suppressed in the spatial range of the resonance, which is compatible with MBL.
The implications of this are that system sizes accessible to ED cannot observe settled behavior for the resonances, which we suggest is a reason for the observed finite-size numerical drifts.

We then perform a complementary study of the charge transport of \emph{average states}, showing evidence that average states cannot transport more than $O(1)$ charge above a critical disorder strength. 
We focus on two related measures. 
First, we study the Frobenius norm of the CTC operator, which we find appears to be saturated $O(1)$ with the system size for large enough disorder strengths.
We then study an object we call the \emph{state charge transfer bound} (SCTB), measuring the maximum amount of charge a given initial state can transfer.
From this, our results suggest that for large enough disorder strengths, average product states as initial states cannot transfer extensive amounts of charge at any later time. 

Our work provides a new picture that the growing size of CTRs pushes finite-size \emph{strongly disordered} spin chains towards apparent thermalization at larger system sizes.
Whether these resonances will eventually proliferate in a controlled or uncontrolled way at large $L$ is not directly known from our data, but we argue that this is consistent with a picture where resonances are controlled as $L \to \infty$ at large enough disorder.
Our work highlights the importance of characterizing the growth of \emph{finite-range} resonances in destabilizing the MBL phase, highlighting difficulties in studying system-wide resonances at these finite sizes.
This suggests a physical mechanism for the observed finite-size drifts that have plagued MBL studies, and invites future work on how to interpret numerical data about MBL.

This paper is structured as follows.
In Sec.~\ref{sec:ctc_introduction} we introduce the charge transport capacity (CTC). 
In Sec.~\ref{sec:ctc_probe}, we establish the scaling of the CTC with $L$ in delocalized and localized systems. 
In Sec.~\ref{sec:ctc_interacting_models}, we focus on the CTC in the interacting Anderson model, discussing our numerical results in Sec.~\ref{sec:ctc_interacting_anderson_model}.
We then identify and characterize charge-transport resonances in Sec.~\ref{sec:ctr_properties}, and describe a toy model for them in Sec.~\ref{sec:ctc_model}.
In Sec.~\ref{sec:charge_transport_typical_states}, we characterize the \emph{average} charge transport.
First, we study the Frobenius norm of the CTC operator in Sec.~\ref{sec:ctcb_frobenius_norm}, then the state charge transfer bound (SCTB) in Sec.~\ref{sec:sctb}. We conclude in Sec.~\ref{sec:conclusion} with a discussion of how our work expands on our current understanding of MBL, and possible near-term directions.

\section{The Charge Transport Capacity}\label{sec:ctc_introduction}
We begin by introducing the main quantity we study in this work: the charge transport capacity (CTC), which first appeared in our previous work~\cite{jiang_quasiconservation_2025}.
In this work, we will use the words ``charge'' and ``particle'' interchangeably.
Consider a one-dimensional (1D) chain of length $L$ with open boundary conditions (OBC), with a Hamiltonian $\hat{H}$ that has $U(1)$ symmetry (the specific form of the Hamiltonian is not important at this point), and fix a site $i_0$ where $1\leq i_0\leq L-1$.
The charge transport capacity operator across the link $i_0$ (i.e., between sites $i_0$ and $i_0 + 1$) is defined as:
\begin{equation}
    \hat M\equiv \hat N_R-[\hat N_R]_{\text{diag}}.
\end{equation}
Here, $\hat N_R \equiv \sum_{i=i_0+1}^{L}\hat{c}_i^{\dagger}\hat{c}_{i}$ is the total number operator for all the sites to the right of $i_0$. Similarly, $\hat N_L \equiv \sum_{i=1}^{i_0}\hat{c}_i^{\dagger}\hat{c}_{i}$ is the total number operator for all the sites to the left of $i_0$.
We will assume the $i_0$ dependence implicitly whenever we write $\hat N_R$ or $\hat N_L$.
The notation $[\hat N_R]_{\text{diag}}$ refers to parts of $\hat N_R$ which are diagonal with respect to the Hamiltonian, i.e., $[\hat O]_{\text{diag}}=\sum_n \langle n|\hat O|n\rangle |n\rangle\langle n|$ with $\{|n\rangle \}$ being the eigenstates of $\hat H$.

We are primarily interested in the \emph{charge transport capacity}---the ``spectral diameter'' of $\hat M$, 
\begin{equation}
    \ctc\equiv \lambda_{\text{max}}-\lambda_{\text{min}}~.
\end{equation} 
Here, $\lambda_{\text{max}}$ and $\lambda_{\text{min}}$ are the maximum and minimum eigenvalues of $\hat M$, respectively.
Operationally, the charge transport capacity acts as an upper bound on how much charge can change on one side of $i_0$, over all possible initial conditions and all time. 

To see this, let us consider the maximal magnitude of the expectation value of the change in the amount of charge on the right of $i_0$, over all possible initial conditions $\ket{\psi(0)} \equiv \ket{\psi_0}$ in the Hilbert space $\mathcal{H}$, and times $t\in\mathbb{R}$:
\begin{equation}\label{eq:true_capacity}
\mathcal{Q} = \sup_{t\in\mathbb{R}, \psi_0\in \mathcal{H}} |\langle\psi_0 | \Delta \hat N_R (t) |\psi_0 \rangle| 
= \sup_{t\in\mathbb{R}} \|\Delta\hat N_R(t)\|_{\text{op}} ~.
\end{equation}
Here, $\Delta\hat N_R(t) := N_R(t) - N_R(0)$ and
$\|\cdot\|_{\text{op}}$ denotes the operator norm.
We will later also allow restriction to particular initial states, defining $\mathcal{Q}(\psi_0)\equiv\sup_{t\in\mathbb{R}} [|\langle\psi_0|\Delta \hat N_R (t)|\psi_0\rangle|]$. 
In practice, for a generic system, $\mathcal{Q}$ is difficult and impractical to compute, both analytically and numerically.
Instead, we study a convenient upper bound to $\mathcal{Q}$.
To do this, consider an integral of motion $\hat D$, where $[\hat D, \hat H]=0$.
Now, define $\hat M_D =\hat N_R+\hat D$.
Then,
\begin{align}\label{eq:inequality_ctc}
    \left\|\Delta \hat N_R(t)\right\|_{\text{op}}&=\left\|\hat M_{\hat D}(t)-\hat M_{\hat D}(0)\right\|_{\text{op}} \\ &\leq \lambda_{\text{max}}(\hat M_{\hat D})-\lambda_{\text{min}}(\hat M_{\hat D})~.
\end{align} 
Here, $\lambda_{\text{max}}({\hat M_{\hat D}})$ is the largest eigenvalue of $\hat M_{\hat D}=\hat N_R + \hat D$ and $\lambda_{\text{min}}(\hat M_{\hat D})$ is its minimum eigenvalue.
We denote $\Lambda(\hat M_{\hat D})\equiv \lambda_{\text{max}}(\hat M_{\hat D})-\lambda_{\text{min}}(\hat M_{\hat D})$ as the ``spectral diameter'' of $\hat M_{\hat D}$.\footnote{We note that using $\Lambda(\hat M)$ is a slight improvement over the straightforward triangular inequality bound $\left\|\hat M(t)-\hat M(0)\right\|_{\text{op}}\leq2\left\|\hat M\right\|_{\text{op}}$, used in Ref.~\cite{jiang_quasiconservation_2025}, by subtracting a multiple of $I$ from $\hat M$ which minimizes its operator norm, i.e., $\lambda_{\text{max}}(\hat M)-\lambda_{\text{min}}(\hat M)=\min_{s\in \mathbb{R}} 2\|\hat M-sI\|$.}

To get a good estimate for $\mathcal{Q}$, we choose $\hat D$ to tighten the upper bound on $\|\Delta\hat N_R(t)\|_{\text{op}}$.
One simple (possibly nonoptimal) choice is to simply remove the diagonal part of $\hat N_R$:
\begin{equation}
\hat M \equiv \hat M_{-[\hat N_R]_{\text{diag}}}=\hat N_R-[\hat N_R]_{\text{diag}}.
\end{equation}
We refer to $\hat M$ as the \emph{charge transport capacity operator}, and the spectral diameter of $\hat M$
\begin{equation}\label{eq:ctc_definition}
    \ctc\equiv\lambda_{\text{max}}(\hat M)-\lambda_{\text{min}}(\hat M)
\end{equation}
as the \emph{charge transport capacity} (CTC). As mentioned earlier, this quantity upper bounds the charge transfer across $i_0$.
It is important to note that $\Lambda(\hat M)$ is an upper bound on the true maximal charge transfer [as in Eq.~(\ref{eq:true_capacity})], and that this upper bound is not necessarily tight. It is thus not the true ``capacity'' of the system in the sense that $\mathcal{Q}$ is, but for the sake of clarity and conciseness, we will call this quantity the charge transport capacity. However, the reader should always keep in mind that the CTC is an \emph{upper bound} on the true transport capacity.

We briefly comment on three important properties of the CTC and $\hat M$. First, $\ctc$ is strictly (possibly not tightly) upper-bounded by $L$ for a system of spinless fermions with $U(1)$ symmetry.
Second, we note that the choice of $\hat D=-[\hat N_R]_{\text{diag}}$ minimizes the Frobenius norm of $\hat M_{\hat D}$ over $\hat D$, $\sqrt{\mathrm{Tr}[\hat M_{\hat D}^\dagger \hat M_{\hat D}]}$.
Finally, note that because the CTC is a quantity that takes into account all energy eigenstates, all of the discussion in this work effectively corresponds to the infinite temperature regime.

The charge transport capacity is a useful quantity to study in localization problems. 
In some sense, it is an adversarial quantity with respect to localization: it represents the worst-case scenario, an upper bound on \emph{all} possible charge transport processes across site $i_0$. 
If the disorder-averaged CTC is finite in the $L\rightarrow \infty$, that implies that regardless of initial conditions or observation time, only a finite amount of charge can ever traverse $i_0$, which is highly nonergodic behavior and therefore direct evidence of localization.

\section{Charge Transport Capacity as a Probe for (Non)ergodicity}\label{sec:ctc_probe}
\begin{figure*}[t]
    \centering
    \includegraphics[width=\linewidth]{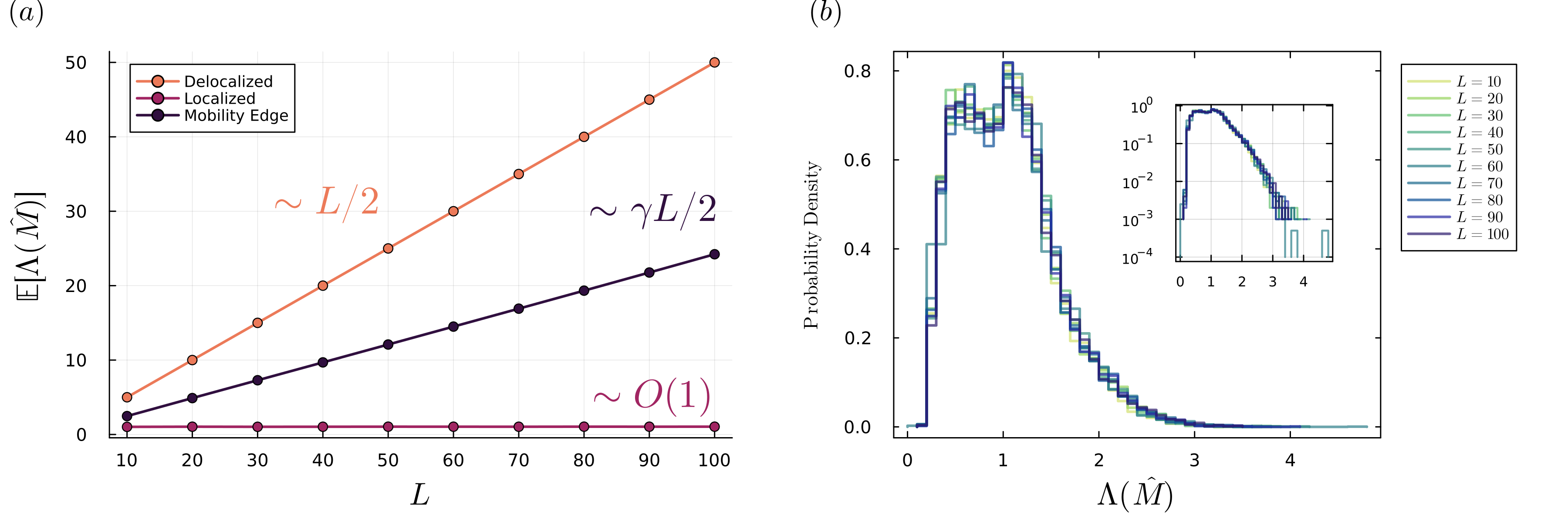}
    \caption{A summary of the behavior of the charge transport capacity (CTC) $\ctc$ for different free-fermion systems.
    Panel (a): The expected scaling of the averaged CTC with system size $L$ for different systems (in this case, in order of top to bottom of the legend labels: the quasiperiodic Aubry-André model in its delocalized phase, the Anderson model at disorder strength $W=6$, and the generalized Aubry-André model in its phase with a mobility edge, cf.~App.~\ref{app:free_fermion_ctcb}). 
    What is specifically averaged over depends on the specific model.
    One sees that delocalized systems, including those that obey ETH, have $L/2$ scaling of the CTC (orange line).
    On the other hand, full or partial localization changes the scaling: when the system has a mobility edge, the CTC scales extensively with $L$, but with a reduced slope (black line) compared to the fully delocalized case.
    For the fully localized model, such as the Anderson insulator, the averaged CTC is $O(1)$ with system size (magenta line) once the system size is large enough compared to the localization length.
    Panel (b): The probability distributions of the CTC in the Anderson insulator, with disorder strength $W=6$.
    In the inset, we plot the same distribution but with the $y$-axis on a logarithmic scale, where it is clear that $P[\ctc] \sim \exp(-c \ctc)$.
    The key is that the distributions of $\ctc$ converge to a fixed distribution at large enough $L$, where large $\ctc$ is exponentially suppressed in $\ctc$.
    }
\label{fig:summary_plot_ctcb}
\end{figure*}

In this section, we establish that the charge transport capacity is a useful probe of ergodic and nonergodic dynamics. 
We summarize its expected behavior in ergodic and (partially) localized systems. 
Specifically, in ergodic systems, the CTC empirically scales as $\sim L/2$, whereas in localized systems (such as the Anderson insulator), the disorder-averaged CTC is a constant $O(1)$ with $L$.
For systems with a mobility edge, the CTC scales as $\sim\gamma L/2$, where $\gamma$ is the fraction of non-interacting orbitals that are delocalized. 
For detailed numerics of the CTC in specific non-interacting models, we refer the reader to Appendix~\ref{app:free_fermion_ctcb}.
In Fig.~\ref{fig:summary_plot_ctcb}, we illustrate a summary of the CTC's behavior in systems with varying degrees of localization.

\subsection{Delocalized Systems}\label{sec:ctc_ergodic_systems}
As illustrated by the labeled curve in Fig.~\ref{fig:summary_plot_ctcb}(a), delocalized systems exhibit a CTC that grows linearly in system size.
We observe that these systems have a CTC which empirically scales as $L/2$. 
This scaling holds exactly for any Hamiltonian with inversion symmetry\footnote{There are subtleties here when degenerate systems are involved. See Appendix~\ref{app:degenerate_hamiltonians} for a discussion.} (see Appendix~\ref{app:derivation_inversion_symmetric} for the proof). 
We also observe empirically that systems obeying the ETH exhibit $\ctc \sim L/2$, though a formal proof of this is beyond the scope of this work.
In particular, we verified numerically that when the eigenvectors closely resemble those drawn from the Gaussian Orthogonal Ensemble, the CTC scales as $L/2$ at sufficiently large $L$.

The intuition for the $\sim L/2$ scaling is that an ergodic system explores all possible configurations and thus can transfer an extensive number of particles.
One can imagine preparing a \emph{domain wall} state with $L/2$ particles to the left of $i_0=\lfloor L/2\rfloor$ and time-evolving it under an ergodic Hamiltonian with $U(1)$ symmetry. 
In the long-time limit, one would typically observe $\sim L/2$ particles evenly spread out across the lattice, with $L/4$ particles on each side of $i_0$. 
Thus, at least $L/4$ particles are capable of crossing the link $i_0$, implying a linear $L$ dependence of the CTC on $L$ greater than or equal to $L/4$. 
Thus, a sub-$L/4$ scaling of $\ctc$ indicates some level of nonergodic behavior.

It must be noted that the converse does not hold: $L/2$ scaling of the CTC does not imply the ETH.
For example, integrable models, such as the free fermion model with nearest-neighbor density--density interactions ($\sum_i \hat{n}_i \hat{n}_{i+1}$), do not thermalize in the usual chaotic sense due to the presence of extensively many conserved quantities, yet still have a CTC that scales as $L/2$ due to the delocalization of the eigenstates.

\subsection{Fully and Partially non-interacting Localized Systems}
Just as the scaling of the CTC detects delocalization, it also serves 
as a powerful measure of full and partial localization.
By full localization, we mean that all of the eigenstates of the system are exponentially localized, whereas partial localization means that there is a mobility edge present.
The discussion in this section pertains to non-interacting models which are provably (partially) localized with a Hamiltonian of the form
\begin{equation}\label{eq:free_fermion_model}
    \hat H = -t\sum_{i=1}^{L-1} \left(\hat c^\dagger_{i+1} \hat c_i + c^\dagger_i c_{i+1} \right) + \sum_{i=1}^{L-1} h_i\hat c^\dagger_i \hat c_i~.
\end{equation}
Here, we fix open boundary conditions. For this Hamiltonian, the CTC operator $\hat M$ admits an exact analytical form in terms of the single-particle orbitals, and many of its properties can be inferred from those expressions. 
We refer the reader to Appendix \ref{app:free_fermion_ctcb} and~\cite{jiang_quasiconservation_2025} for additional details of these analytical expressions.
Some examples of provably localized models with Hamiltonians of the above form are the Anderson insulator~\cite{anderson_absence_1958} with the choice of $h_i\sim\mathrm{Unif}[-W, W]$, or the quasiperiodic Aubry-André model with $h_i=h\cos[2\pi k(i+\phi)]$ with an irrational $k$ and $h>2$~\cite{aubry_analyticity_1980}.
For detailed numerical calculations of the CTC on these models, we refer the reader to Appendices \ref{app:anderson_model_ctc} and \ref{app:quasiperiodic_noninteracting_ctcb}.

As illustrated by the corresponding curve in Fig.~\ref{fig:summary_plot_ctcb}(a), non-interacting localized systems show a constant $O(1)$ disorder/phase-averaged CTC at system sizes $L$ much larger than the localization length $\xi$.
This is the average worst-case charge transfer across disorder realizations.
Its value represents the worst-case charge transfer in a typical disorder realization, and as the distributions show, this is also representative of typical samples.
At the operator level, this result can be understood as a consequence of $\hat M$ being quasilocal around $i_0$ in localized systems.
This is because subtracting out the diagonal part of $\hat N_R$, which is effectively a sum of LIOMs to the right of the cut, is equivalent to removing support on LIOMs that are far away from $i_0$.
One can find a more detailed argument in Appendix~\ref{app:free_fermion_ctcb} and our previous work~\cite{jiang_quasiconservation_2025}.

An $O(1)$ CTC averaged over disorder realizations is a strong statement about transport in localized systems: it says that in an average disorder realization, one cannot move extensive amounts of charge, no matter how long one waits.
The probability distributions of the CTC depend on the localization model in question. 
For the Anderson insulator, as in Fig.~\ref{fig:summary_plot_ctcb}(b), $P[\ctc]$ shows exponentially decaying tails and a peak in the distribution at 1 for the parameters studied.
This peak represents situations where there is a hybridization of orbitals around the cut $i_0$, allowing for a short-ranged transport process where 1 particle can hop across the cut back and forth.\footnote{A scenario for this is a two-site resonance across the cut in the single-particle problem.
Note that a CTC of exactly $1$ occurs when the resulting two orbitals each have equal weight on the two sites (e.g., symmetric/antisymmetric orbitals at perfect resonance), while generically one has different weights on the two sites.
However, the perfect resonance is an extremum point as a function of possible parameters, and as such translates to enhanced probability for CTC close to one, which explains the appearance of the peak at 1.
}

Secondly, the CTC is also capable of detecting the presence of mobility edges.
Specifically, in this situation, the (averaged) CTC empirically scales with $L$ as $\sim\gamma L/2$, where $\gamma$ is the fraction of delocalized eigenstates in the spectrum, exemplified by the labeled curve in Fig.~\ref{fig:summary_plot_ctcb}(a).
The intuition here is that in the presence of a mobility edge, the system has $(1-\gamma) L$ localized degrees of freedom, and so one is effectively working with a delocalized model on a lattice with $\gamma L$ sites, leading to a $\ctc\sim \gamma L/2$ scaling.
We investigate the specific example of the generalized Aubry-André model~\cite{ganeshan_nearest_2015} in Appendix~\ref{sec:ctc_gaa_model} for a system with a mobility edge.

\section{Charge Transport Capacity in Interacting Models}\label{sec:ctc_interacting_models}
In this section, we numerically investigate the behavior of the charge transport capacity in the interacting Anderson model. Having established that the CTC is a powerful probe of ergodicity and localization, we now examine how systems expected to exhibit MBL fit within this framework.
We will begin by showing and analyzing the numerical results, then end with an argument that $\doubleE[\ctc]$ should be $O(1)$ as $L\rightarrow\infty$ deep in the MBL phase. 

\subsection{Methods}\label{sec:interacting_model_methods}
For all of the quantities calculated in this work, we employ Exact Diagonalization (ED) on system sizes $L=4$ to 16 to compute $\hat M$ and related quantities.
We fix open boundary conditions and $i_0=\lfloor L/2\rfloor$. We employ the Julia programming language~\cite{bezanson_julia_2017}.

The Hamiltonian we consider has $U(1)$ symmetry. All of our calculations are performed in the half-filling sector with $\hat N_{\text{tot}}=\lfloor L/2\rfloor$, where $\ctc$ is maximized.
We perform calculations only at even system sizes with $4\leq L\leq 16$ to avoid fluctuations in transport arising due to the cut $i_0$'s placement in the odd or even lattices; for such even $L$, the cut then divides the system into two halves.

\subsection{Charge Transport Capacity in the Interacting Anderson Model}\label{sec:ctc_interacting_anderson_model}
Now, we turn to one of the main results of this paper and show how the CTC behaves in the standard model for many-body localization: a 1D chain of spinless fermions with nearest-neighbor hopping and nearest-neighbor density-density interactions:
\begin{equation}\label{eq:interacting_hamiltonian}
\begin{split}
\hat H ={}& -t \sum_{i=1}^{L-1} \big[ \hat c_i^\dagger \hat c_{i+1} + \hat c_{i+1}^\dagger \hat c_i \big] 
+ \sum_{i=1}^{L} h_i \, \hat c_i^\dagger \hat c_i \\
&+ J \sum_{i=1}^{L-1} \hat c_i^\dagger \hat c_{i+1}^\dagger \hat c_{i+1} \hat c_i ~.
\end{split}
\end{equation}
Here, we use open boundary conditions and fix $t=1$ uniformly across all sites. The interaction $J$ is also a positive number fixed across all sites.
The onsite potentials $h_i$ are independently sampled from the uniform distribution between $[-W, W]$ where $W\in \mathbb{R}$. 

It is conjectured that in the $L\to\infty$ limit, for a fixed interaction strength $J$, this model has a critical disorder strength $W^*$ where the system experiences a transition from an ergodic phase whose eigenstates obey ETH (and hosts a prethermal regime at intermediate disorder strengths~\cite{long_phenomenology_2023}), to an MBL \emph{phase}, where all of the eigenstates are localized, and the ETH breaks down~\cite{sierant_many-body_2025}. 
Many details of this phase diagram remain poorly understood --- including the critical disorder strength $W^*$ itself, for which only lower bounds are currently known, and it is still controversial whether or not $W^*$ even exists.

We wish to study the charge transport capacity in this model.
For the main text, we consider the regime of fairly weak interactions, $J=0.1$, where the currently known lower bound for the transition to the MBL phase is approximately $W\sim 4-6$~\cite{colbois_interaction-driven_2024}. 
The behavior of the CTC in this regime is qualitatively similar for the regime with a larger interaction strength. 
We will focus on $J=0.1$, which is particularly convenient as we will also use perturbation theory in the interaction strength later on in this work to understand some behaviors. 

The weakly interacting regime is less studied in the literature, whose focus has mostly been on the Heisenberg (or XXX) chain, corresponding, in our notation, to the interacting Anderson model with parameters $t=0.5, J=1$
\cite{sierant_many-body_2025}.
For a more direct connection to the literature, we present data for the CTC in this parameter regime in Appendix~\ref{app:large_interaction_strength}.
Many of our results do not qualitatively change in either regime of strong or weak disorder (relative to the other parameters of the model).

\begin{figure*}[t]
    \centering
    \includegraphics[width=0.85\linewidth]{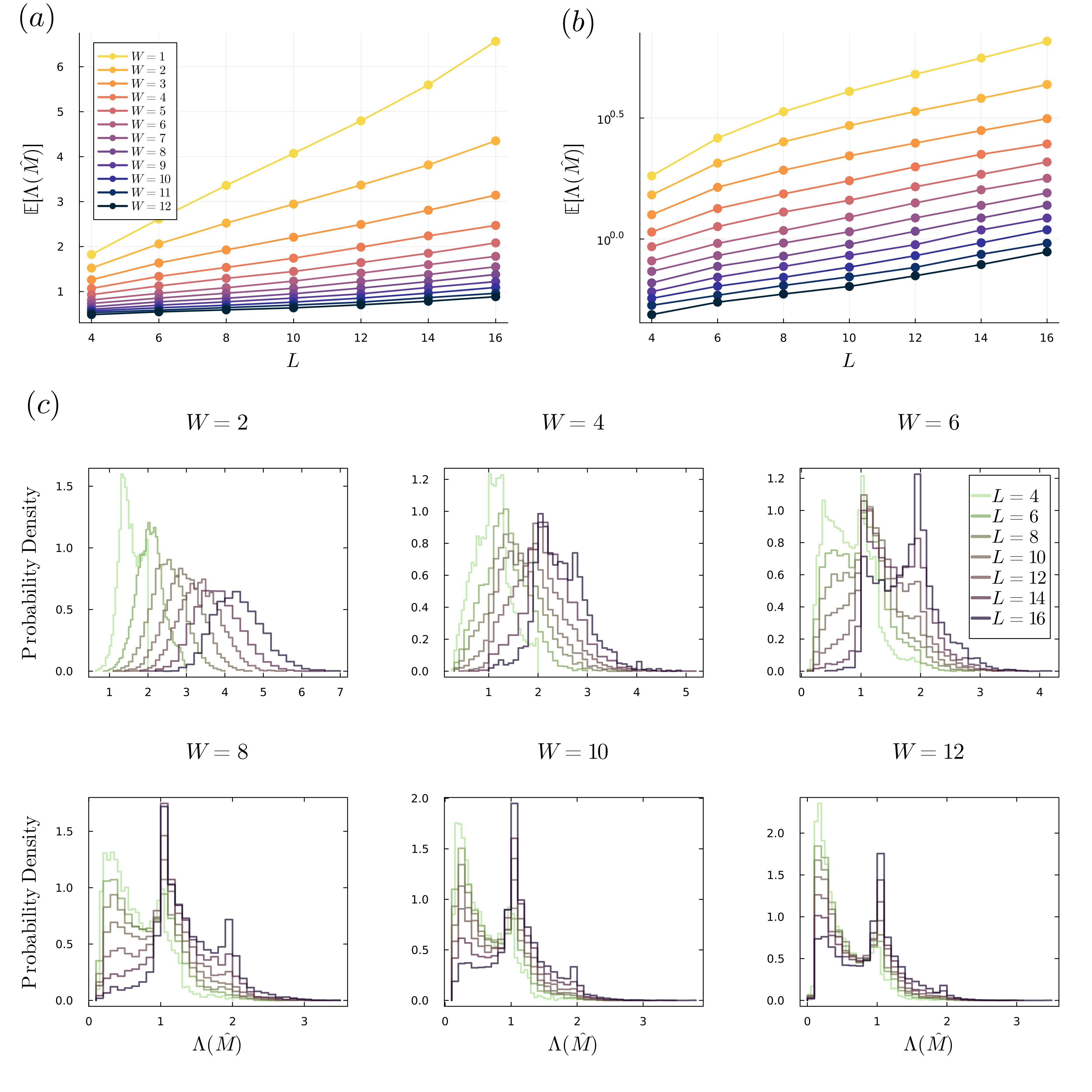}
    \caption{
    Analysis of the charge transport capacity (CTC) for the interacting Anderson model with interaction strength $J=0.1$. 
    Panel (a): The dependence of the disorder-averaged CTC $\doubleE[\ctc]$ on $L$. 
    Each curve represents a different disorder strength from $W=1$ to $W=12$ in steps of 1, and each point is the average of the CTC over 8000 disorder realizations for $L\leq 14$. For $L=16$, there are 2000 disorder realizations for $W\leq 4$, >10000 disorder realizations for $W\leq 6$, and 12000 disorder realizations otherwise. 
    Panel (b): The same data as in (a), but with the y-axis on a log-scale.
    Panel (c): Calculated probability density functions of the charge transport capacity $\ctc$ over disorder realizations. Each panel represents a different disorder strength $W$, and the distribution is estimated using the same disorder realizations as in (a) and (b).
    Within a panel, each colored curve represents a different system size, ranging from $L=4$ to $L=16$ in steps of 2.
    }
\label{fig:ctcb_main_results_J_01}
\end{figure*}
In Fig.~\ref{fig:ctcb_main_results_J_01}(a) and (b), we show data on the $L$-dependence of the disorder-averaged charge transport capacity. 
The calculated PDFs of the CTC are shown in Fig.~\ref{fig:ctcb_main_results_J_01}(c).
The disorder strengths we sample for $J=0.1$ lie well before and well beyond the known lower bounds on the MBL transition.
We will split our analysis into two parts: 
First, we will discuss the results deep in the ergodic regime with low $W$. 
Subsequently, we will then focus on the regime deep into what is presumed to be the MBL phase for very high $W$.

\subsubsection{Charge transport capacity in the ergodic regime}\label{sec:ergodic_interacting_ctc_results}
The regime $W\leq 4$ can be considered to be safely on the ergodic side of the phase diagram. 
For these disorder strengths, surprisingly, it appears that $\mathbb{E}[\ctc]$ is underneath the characteristic ETH scaling of $L/2$, even at very small disorder strength $W=1$.
For $W\geq 3$ on the ergodic side, the values of the CTC are smaller than what would be consistent with diffusion (i.e., $\sim L/4$ scaling).
On the other hand, one can see that the slope of $\mathbb{E}[\ctc]$ in this regime may be increasing with $L$, a behavior that cannot continue asymptotically (remember that the CTC is upper bounded by $L$). 
Consistent with previous numerical studies~\cite{panda_can_2020}, the smallness of the CTC highlights that there is non-ETH-like behavior in the weak disorder regime at these finite sizes, but the increasing slope shows that there are significant finite-size effects pushing the system towards more thermal behavior.

We speculate that the unusual scaling of the CTC may offer a different perspective on mobility edges in this model, which have been seen at finite-sizes numerically~\cite{naldesi_detecting_2016,chanda_many-body_2020}, 
even though theoretical arguments suggest that this many-body mobility edge cannot survive in the thermodynamic limit~\cite{de_roeck_absence_2016}. 
In Sec.~\ref{sec:ctc_probe}, we established that the sub-$L/2$ linear scaling of the CTC in the ergodic regime is a signature of the presence of a mobility edge.
Assuming this interpretation carries over in a direct manner (a claim that we do not attempt to investigate further in this work), one can interpret the increasing slope for larger $L$ as a trend towards the erasure of this finite-size ``mobility edge'' in the thermodynamic limit. 
It would be interesting to understand the mechanisms of this mobility-edge erasure as $L\rightarrow\infty$, something we leave for future work.

To summarize, the results of the CTC on the ergodic side highlight the existence of the known ``numerical crossover regime''--- a range of disorder strengths where the system appears localized at small system sizes (i.e., sub $L/4$ scaling of the CTC), but will eventually thermalize at large $L$ (i.e. the increasing rate of growth with $L$). The results also point towards an avenue for understanding the mechanisms of the asymptotic fate of the finite-size mobility edge.

\subsubsection{Charge transport capacity in the MBL regime}\label{sec:mbl_ctc_results}
Now, for the rest of the paper, we turn our attention to what is presumed to be deep in the MBL phase: here, for $J = 0.1$, we refer to disorder strengths $W>8$, with $W\geq 4-6$ being the literature estimated lower-bound for the transition.

\begin{figure}[t]
    \centering
    \includegraphics[width=\linewidth]{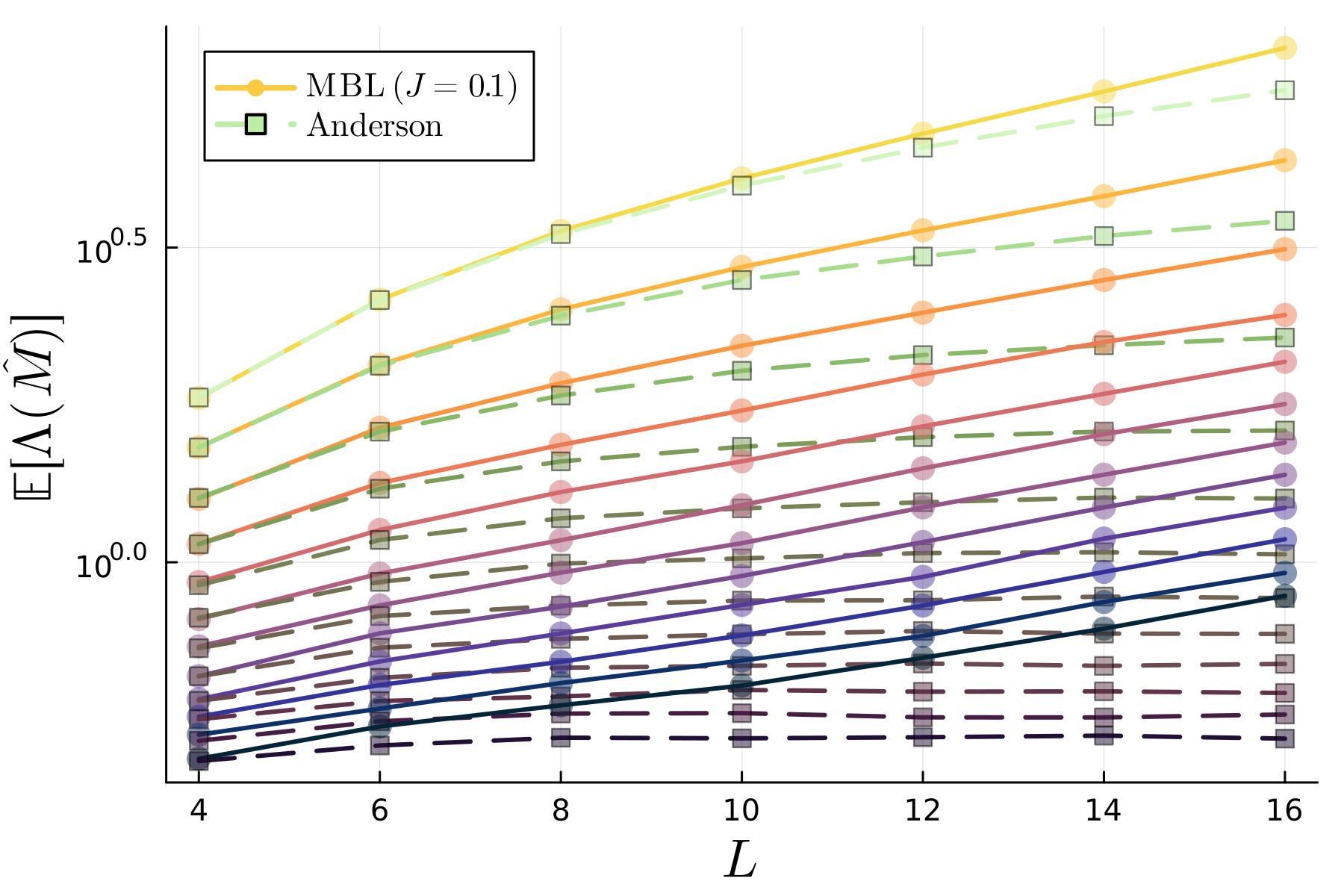}
    \caption{Comparison of the disorder-averaged charge transport capacity $\ctc$ as a function of $L$ between the non-interacting Anderson model (dashed lines, square markers) and the interacting problem (solid lines, circle markers).
    The data for the interacting problem are identical to those in Fig.~\ref{fig:ctcb_main_results_J_01}.
    Each curve represents a different disorder strength from $W=1$ to 12 in steps of 1, with lighter colors representing weaker disorder.
    }
    \label{fig:comparison_mbl_anderson}
\end{figure}

Observing Fig.~\ref{fig:ctcb_main_results_J_01}(a), surprisingly, even for the largest disorder strengths tested (up to $W=12$, which is presumed to be very deep into the MBL phase), $\mathbb{E}[\ctc]$ continues to grow at an increasing rate with $L$. 
This increasing growth cannot continue forever, as the CTC is upper-bounded by $L$, highlighting significant finite-size effects at play.
In Fig.~\ref{fig:ctcb_main_results_J_01}(b), we show the same data as in (a), but with a logarithmic scale on the $y$-axis. 
From this plot, two very counterintuitive features emerge at strong disorder: 
First, the CTC increases apparently exponentially with $L$.
This observation is starkly different from the CTC of the Anderson insulator, where the disorder-averaged CTC shows signs of saturation in the Anderson insulator at these same small system sizes $L\leq 16$ for the same disorder strengths. 
A comparison of the data between the two cases can be found in Fig.~\ref{fig:comparison_mbl_anderson}.
Second, the rate of the exponential growth shows little apparent dependence on $W$, evidenced by the slopes appearing identical across $W$ for the strong disorder regime. 
This is particularly counterintuitive, since one would expect some suppression of this growth at larger disorder strengths.\footnote{Note, however, that the dependence of the localization length on $W$ is rather weak for large $W$, cf.\ the ansatz 
\begin{equation}\label{eq:anderson_ll}
    \xi \approx 1/[\ln\{1+(W/W_0)^2\}]~,
\end{equation} 
where $W_0\approx 2.26$. See the supplementary material of~\cite{colbois_interaction-driven_2024}.}

To understand the origin of this growth, we examine the probability distributions of the CTC in Fig.~\ref{fig:ctcb_main_results_J_01}(c).
For example, observing the panel corresponding to $W=10$ in Fig.~\ref{fig:ctcb_main_results_J_01}(c), 
one sees that for small system sizes $L=4, 6$, the distribution of the CTC bears resemblance to those of the non-interacting Anderson model [Fig.~\ref{fig:summary_plot_ctcb}(b)]:
The distributions have exponentially decaying tails and are sharply peaked at two values: a very small nonzero value, and at the integer value $1$ representing cases where a LIOM near the cut is hybridized, causing the cut at $i_0$ to slice right through the LIOM and allowing one particle to hop across the link.
As the system size is increased, the peak at the small value lowers, while the peak at the integer value $1$ increases, and the tail visibly starts to broaden towards the right.
At $L=14$, as the peak at the small nonzero value of CTC is greatly suppressed, a peak at the next integer value, $2$, starts to rise. 
We also observe a very small peak at $4$ particles for $L=16, W=4$, as well as a small bump at $3$ particles crossing the link beginning to form for $W=6$ and $L=16$, meaning that these integer peaks can indeed extend to beyond $2$ particles--- unfortunately, the ED system size limitations prevent us from seeing larger integer-valued peaks for strong disorders.
This feature of shifting peaks persists for all disorder strengths tested outside the deeply ergodic regime and drives the growth of the disorder-averaged CTC.
We note that aside from the integer valued peaks, there also appears to be a systematic rightward shift of the bulk of the distribution. 
However, it is unclear if the underlying mechanism for this systematic rightward shift is related to those of the shifting integer peaks. 

These numerical results are surprising for many reasons. 
Firstly, the CTC growing with $L$ in the interacting problem starkly contrasts with that of the non-interacting problem, which shows signs of saturation of the CTC at these sizes.
Secondly, as discussed previously, the growth apparently being exponential and at the same rate across $W$, even at very strong disorder strengths, is an unintuitive feature.
Thirdly, as we will discuss in detail in Sec.~\ref{sec:l-bit-model-results}, one would expect that if the MBL Hamiltonian can be rewritten in terms of quasilocal conserved quantities, then one would obtain that $\hat M$ is quasilocal around $i_0$ and thus has a finite disorder-averaged CTC~\cite{jiang_quasiconservation_2025}.
The growth of the CTC is contrary to this simple LIOM picture of the MBL phase.
As we will discuss later, there are various possible scalings of the CTC with $L$ which could be compatible with the MBL phase.
Finally, the peaks at integer values are peculiar, and the focus in the next few sections will be on establishing the mechanism behind them.

We caution that, from our current data, even though the curves appear to grow without saturation, the story in reality is much more complex and does not fit with any simple extrapolation of the data.
We will discuss a diagonally-improved perturbative model which partially reproduces the CTC data in Sec.~\ref{sec:ctc_model}.

\subsubsection{The tightness of $\ctc$ as a bound}\label{sec:ctc_is_a_tight_bound}
\begin{figure*}[t]
    \centering
    \includegraphics[width=\linewidth]{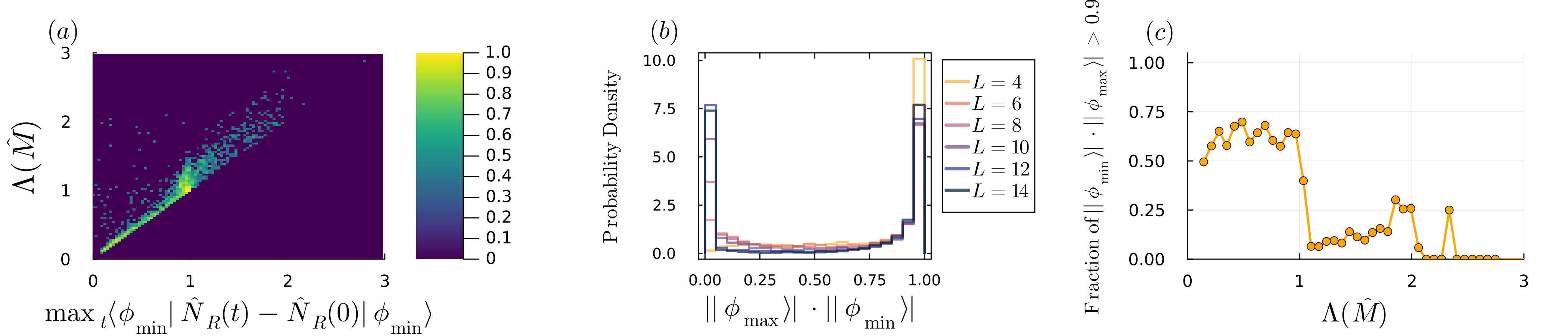} 
    \caption{A summary of results showing that the CTC's growth captures true charge transfer dynamics in the system for large disorder strengths $W$. All of the results in this figure are for $W=10$, $J=0.1$, $L=14$. Panel (a) was plotted with 2000 disorder realizations, while panel (b) has 4000 disorder realizations; data for other large disorder strengths ($W\geq 8$) look qualitatively similar to those presented here, and can be found in Appendix~\ref{app:overlap}. 
    Panel (a): a heat map comparing the CTC ($y$-axis) with the maximum of $\max_t{|\langle\hat{N}_R(t)-\hat{N}_R(0)\rangle|}$ over a quench dynamics simulation ($x$-axis) in the strong disorder regime.
    The quench evolves $|\phi_{\min}\rangle$ as an initial state, and $t\in [0,10^5]$ in steps of $\Delta t=10$. The color scale, shown on a logarithmic scale, ranges from black (low density of disorder realizations) to yellow (high density).  
    Panel (b): the probability distribution of the absolute-value-state overlap $|\ket{\maxstate}|\cdot|\ket{\minstate}|$. 
    Each colored curve represents a different system size as indicated by the legend.
    Panel (c): the fraction of disorder realizations with $|\ket{\maxstate}|\cdot|\ket{\minstate}|\geq 0.9$ as a function of $\ctc$ of the disorder realization.
    }
\label{fig:ctcb_is_tight_bound}
\end{figure*}
It is a priori unclear whether the growth of the CTC genuinely reflects physical processes that transfer more charges. 
A natural concern is that, because $\ctc$ is an upper bound on the maximal net charge transferred $\mathcal{Q}$, its growth may reflect a loosening of the bound rather than genuine charge transport. 
However, here, we provide evidence that at sufficiently strong disorder and at the available system sizes, the CTC faithfully captures the system's capacity to transfer charge.
For the purposes of this study, then, the CTC should be viewed as a tight bound in the strong disorder regime.

A useful fact is that there is only one physical scenario that can saturate the bound. Let $\ket{\maxstate}$ and $\ket{\minstate}$ be the eigenstates of $\hat M$ with the largest and minimum eigenvalues, respectively.
It is straightforward to see that the system transfers charge equal to the CTC if and only if these two states are connected by time evolution, i.e., there exists a time $t$ such that $\ket{\maxstate} = e^{-i\hat H t}\ket{\minstate}$ [considering for concreteness processes with $\Delta N_R(t) = \lambda_{\text{max}} - \lambda_{\text{min}} > 0$ and assuming that these eigenvalues have degeneracy of $1$].

Motivated by this fact, we can test if the system is capable of transporting an amount of charge close to the CTC by simulating quench dynamics under $\hat H$ from initial state $\ket{\minstate}$ and obtain a numerical estimate of the maximal $\langle\hat N_R(t)\rangle$ over $t$. 
During the quench experiment, we keep track of $\langle\Delta\hat  N_R(t)\rangle$ for times $t\in[0,t_{\text{max}}]$ in steps of $\Delta t$. Now, denote $t^*=\mathrm{argmax}_{t\in[0,t_{\text{max}}]}(\langle\Delta\hat N_R(t)\rangle)$. 
The quantity $\langle\Delta\hat N_R(t^*)\rangle$ is a lower bound on the maximal charge transfer of the system.
Thus, we can measure the tightness of the CTC by measuring how close it is to $\langle\Delta\hat N_R(t^*)\rangle$.

We show an analysis of $\left|\langle\Delta\hat N_R(t^*)\rangle\right|$ compared to $\ctc$ in Fig.~\ref{fig:ctcb_is_tight_bound}(a) for $W=10$ and $L=14$.
Data for other disorder strengths can be found in Appendix~\ref{app:overlap}.
We note that  $\left|\langle\Delta\hat N_R(t^*)\rangle\right|$ and $\ctc$ almost correspond one-to-one for $\ctc\leq 1$.
Even when $\ctc$ is greater than 1, and the CTC potentially worsens as an upper bound, in most instances the system still transfers a value that is comparable to $\ctc$.
On the flip side, there are instances where there is a discrepancy:
We note that there are some cases where the $\ctc$ is large, but $\left|\langle\Delta\hat N_R(t^*)\rangle\right|$ is rather small.
Furthermore, we empirically observe that the CTC becomes a worse bound for larger system sizes $L$.
Of course, we also cannot rule out the possibility that each of these instances will reach a charge transfer closer to $\ctc$ if time evolved for long enough, or at smaller time steps, or from a modified initial state.
Regardless, even though the CTC is not a perfectly tight bound, this data strongly suggests that the growth of the CTC with $L$ in the strong disorder regime should be interpreted as a true growth of the system's capability to transfer charges with $L$.
In this sense, it is a tight bound for our purposes.

We now present some numerical evidence that there are significantly many cases where $\ket{\maxstate}$ and $\ket{\minstate}$ are approximately connected by time evolution in the strongly disordered regime.
Rewriting  $\ket{\maxstate}=\sum_na_n|n\rangle$ and $\ket{\minstate}=\sum_nb_n|n\rangle$ in terms of energy eigenstates $\{|n\rangle\}$, we require that the ``absolute-value-state overlap'' between $\ket{\maxstate}$ and $\ket{\minstate}$
\begin{equation}\label{eq:absolute_value_overlap}
|\ket{\maxstate}|\cdot|\ket{\minstate}| \equiv \sum_{n}|a_n||b_n|
\end{equation}
must be equal to $1$.
In other words, the energy distribution of the two states must match, $|a_n|^2 = |b_n|^2$.
We note that $|\ket{\maxstate}|\cdot|\ket{\minstate}|=1$ does not necessarily imply that $\ket{\minstate}$ and $\ket{\maxstate}$ are connected by time evolution.
Nonetheless, the measure provides a first proxy for it.
It is also important to bear in mind that a low overlap between $\ket{\maxstate}$ and $\ket{\minstate}$ does not necessarily imply that the CTC is a poor bound---it only shows that the bound is not perfectly tight.

In Fig.~\ref{fig:ctcb_is_tight_bound}(b), we show the probability distributions of $|\ket{\maxstate}|\cdot|\ket{\minstate}|$ for different system sizes at $W=10$.
Interestingly, the distributions of $|\ket{\maxstate}|\cdot|\ket{\minstate}|$ show that this quantity is either very close to $1$ or to $0$, with a lot fewer instances in between.
Generally, as the system size increases, the number of instances with 0 overlap increases.

One might worry that $|\ket{\maxstate}|\cdot|\ket{\minstate}| \approx 1$ occurs only when $\ctc$ is small, so that large $\ctc$ would always coincide with cases where $\ket{\maxstate}$ and $\ket{\minstate}$ are not good descriptions of the worst-case charge transfer process. However, this does not appear to be the case. Figure~\ref{fig:ctcb_is_tight_bound}(c) shows that among disorder realizations with $\ctc \approx 2$, a significant fraction---about $1/4$ at this disorder strength---satisfy $|\ket{\maxstate}|\cdot|\ket{\minstate}| \geq 0.9$, confirming that larger $\ctc$ does not necessarily correspond to the eigenstates of the CTC becoming a worse description for the true worst-case charge transfer.
For the same calculations for other disorder realizations, see Appendix~\ref{app:overlap}.

In summary, the sets of data in Fig.~\ref{fig:ctcb_is_tight_bound} provide strong evidence that the growth of the charge transport capacity deep in the MBL regime captures real transport physics, captured by the time evolution of $\ket{\minstate}$ to $\ket{\maxstate}$. 
This strongly hints that the states $\ket{\minstate}$ and $\ket{\maxstate}$ are special and important players in the growth of the CTC.
Indeed, in the next few sections, we show that $\ket{\minstate}$ and $\ket{\maxstate}$, for a significant fraction of disorder realizations, are participating in an interaction-mediated \emph{charge transport resonance} that allows the transfer of anomalously large integer numbers (mostly 2 for our sizes) of charges.

As an aside, we emphasize that a complementary helpful perspective is to view $\hat M\equiv \hat N_R-[\hat N_R]_{\text{diag}}$ as simply an object of interest in the MBL phase, irrespective of its interpretation as an upper bound on the maximal charge transfer. 
We know that certain properties of the MBL phase, such as the presence of LIOMs and resonances, will restrict what behaviors the CTC object should exhibit.
Thus, its scaling with $L$ is useful as a probe of these properties regardless of whether it remains a loose or tight upper bound as $L$ increases.
In particular, the matrix elements of $\hat M$ and its extremal eigenvalues provide information about resonances, whose presence must be controlled for the stability of the MBL phase.

\subsection{Properties of Charge Transport Resonances}\label{sec:ctr_properties}
\begin{figure*}[t]
    \centering
    \includegraphics[width=0.85\linewidth]{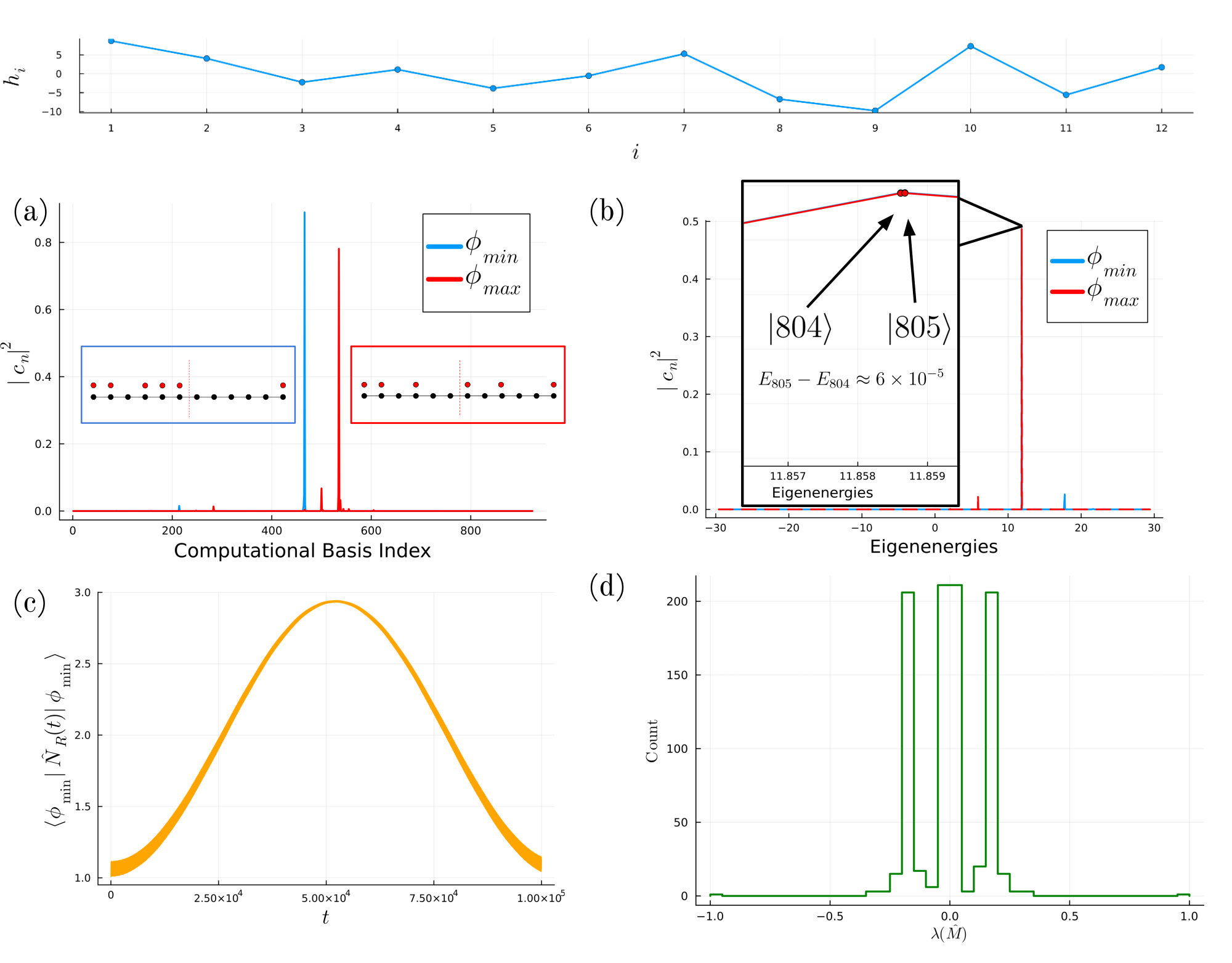}
\caption{Figure summarizing characteristics of a specific disorder realization of the interacting Anderson model hosting a CTR of 2 particles. This realization has $\ctc=1.98$, $L=12, W=10$, $|\ket{\maxstate}| \cdot |\ket{\minstate}| \approx 1$.
The disorder landscape, the onsite potentials $h_i$ as a function of position $i$, is plotted at the top of this figure.
Panel (a): The weights of $\ket{\maxstate}$ and $\ket{\minstate}$ expanded in the computational basis. The insets, with the blue outline corresponding to $\ket{\minstate}$ and the red corresponding to $\ket{\maxstate}$, illustrate the specific computational basis states which have significant weight in these two states. 
Panel (b): The expansion of $\ket{\maxstate}$ and $\ket{\minstate}$ in the Hamiltonian eigenstate basis, plotted against the energy.
The inset zooms into the shared peak in (b), showing that $\ket{\maxstate}$ and $\ket{\minstate}$ are almost equal superpositions of two consecutive energy eigenstates labelled $|804\rangle, |805\rangle$ with energy splitting $\Delta E_{{\text{res}}} \approx 6\times 10^{-5}$, markedly smaller than the calculated average level spacing $E_n-E_{n-1}\approx 0.064$. 
Panel (c): Quench experiment, starting from $\ket{\minstate}$ at $t=0$, measuring the number of particles to the right of $i_0$, $\langle \minstate|\hat N_R(t)|\minstate\rangle$ on the $y$-axis as a function of time $t$ on the $x$-axis.
Starting from $\ket{\minstate}$, which has initially 1 particle on the right, after $O(10^{4})$ ($\approx 2\pi/\Delta E_{\text{res}}$) units of time, 3 particles are now to the right of $i_0$.
In this process, 2 particles were moved across $i_0$, showing that the calculated $\ctc$ indeed can be realized by quench dynamics starting from $\ket{\minstate}$.
We note that in the plot, small oscillations are superimposed on the dominant one, though they are difficult to resolve at this scale, giving the line its apparent thickness.
Panel (d): the density of states of $\hat M$ for this particular disorder realization. 
Only one pair of eigenvalues lies on the extreme ends $\lambda(\hat M)\approx\pm 1$, corresponding to the calculated $\ctc\approx 2$. 
}
\label{fig:example_ctr}
\end{figure*}
As the distributions of the CTC show, the growth of the disorder-averaged CTC deep in the MBL phase is driven by growing integer numbers of particles hopping across the link, essentially during the time evolution from $\ket{\minstate}$ to $\ket{\maxstate}$. 
In this section, we show that these peaks at integer values $m$ are caused by the presence of charge transport resonances (CTRs) between $\ket{\minstate}$ and $\ket{\maxstate}$ which transfer $m$ particles between each other.

To illustrate some characteristic properties of these CTRs, we will focus on one example of them
lying inside the 2-particle peak at $W=10$ in Fig.~\ref{fig:ctcb_main_results_J_01}(c) and at $L=12$.
This example is summarized in Fig.~\ref{fig:example_ctr}.
In this instance, $\ket{\minstate}$ and $\ket{\maxstate}$ are connected by time evolution and transfer 2 particles across $i_0$.
This example is one of a ``perfect'' resonance [$\theta=\pi/4$ as in Eq.~(\ref{eq:resonance_structure})].
We also note that resonances of the form of this example are indeed representative of realizations with $\ctc\approx 2$.
A key diagnostic of these resonances is to check whether $|\ket{\maxstate}| \cdot |\ket{\minstate}| \approx 1$, and we can see that in Fig.~\ref{fig:ctcb_is_tight_bound}(c), nearly half of the disorder realizations with $\ctc\approx 2$ at $W=10, L=14$ satisfy this condition. 
Thus, instances like the one in Fig.~\ref{fig:example_ctr} account for a significant fraction of the peak at $\ctc\approx 2$ in the probability density function, and so this CTR serves as a representative example of other CTRs. 
This justifies our choice of the example in Fig.~\ref{fig:example_ctr} to study.
Our goal is to use this CTR, which represents the simplest case for a resonance, as a starting point for the discussion.
For two additional examples of CTRs, see Appendix~\ref{app:more_ctr_examples}.

Before we analyze the main example in Fig.~\ref{fig:example_ctr}, we begin by summarizing the concept of a CTR and highlighting several important considerations.
A CTR of $m$ particles is an \emph{interaction-mediated} resonance between LIOM configurations $|a\rangle, |b\rangle$ which, between them, transfer charge $m$ across $i_0$.
In many cases, $m$ is approximately an integer.
Here, we assume that the spectrum is approximately diagonal in a properly defined LIOM basis, except for such rare resonances.
In these cases, the spectrum contains a nearly degenerate pair of eigenstates $\ket{\mu}, \ket{\nu}$ approximately described by a two-level hybridization,
\begin{align}\label{eq:resonance_structure}
    \ket{\mu}&= \cos{\theta}\,|a\rangle+\sin{\theta}\,|b\rangle ~, \\ \nonumber
    \ket{\nu} &= -\sin{\theta}\,|a\rangle +\cos{\theta}\,|b\rangle ~, \nonumber
\end{align}
so that unitary time evolution approximately tunnels between $|a\rangle$ and $|b\rangle$ and thereby transports approximately $m$ particles across $i_0$. 
More precisely, $m$ particles are transported exactly when $\theta = \pi/4 + n\pi$ for any $n\in\mathbb{Z}$.
In fact, if the discussed measure $|\ket{\maxstate}|\cdot|\ket{\minstate}|$ is close to $1$ (which represents a significant fraction of disorder realizations transporting integer charges), that implies that $\theta\approx\pi/4$ and thus, a near-perfect resonance.
Operationally, if a CTR is the mechanism that transfers the most charge in the system (which may not always be the case), such a CTR can be identified by diagonalizing the charge-transport operator $\hat{M}$ and extracting its extremal eigenstates $\ket{\maxstate}$ and $\ket{\minstate}$.

Indeed, consider the idealized perfect resonance situation with $\theta = \pi/4$ and where $\ket{a},\ket{b}$ are essentially computational basis states (which is appropriate at strong disorder where the LIOMs are essentially site occupation operators).
Suppose $j > i_0$ is one of the $m$ sites where $\ket{a}$  and $\ket{b}$ have different occupations:
$\hat n_j \ket{a} = \ket{a}, \hat n_j \ket{b} = 0$.
Consider now the space $\text{span}\{\ket{\mu},\ket{\nu}\} = \text{span}\{\ket{a},\ket{b}\}$.
Clearly, $\hat n_j$ acts within this space, given simply by $\ket{a}\!\bra{a}$.
Furthermore, $[\hat n_j]_{\text{diag}}$ also acts within this space, with its action given by
\begin{equation*}
\bra{\mu} \hat n_j \ket{\mu} \ket{\mu}\!\bra{\mu} + \bra{\nu} \hat n_j \ket{\nu} \ket{\nu}\!\bra{\nu} = \frac{1}{2} (\ket{a}\!\bra{a} + \ket{b}\!\bra{b}) ~. 
\end{equation*}
Hence $\hat n_j - [\hat n_j]_{\text{diag}}$ acts in this space, given by
\begin{equation}
\hat n_j - [\hat n_j]_{\text{diag}}~: \quad \frac{1}{2} (\ket{a}\!\bra{a} - \ket{b}\!\bra{b}) ~.
\end{equation}
Assuming that we have $m$ such sites, the resulting contribution to $\hat{M}$ is
\begin{equation}
\hat{M}~: \quad \frac{m}{2} (\ket{a}\!\bra{a} - \ket{b}\!\bra{b}) ~,
\end{equation}
from which we can read off its highest and lowest eigenvalues and the corresponding eigenvectors:
$\ket{\maxstate} = \ket{a}, \lambda_{\text{max}} = m/2$ and $\ket{\minstate} = \ket{b}, \lambda_{\text{min}} = -m/2$.
Note that we do not need to know the action of $\hat{M}$ on the rest of the states as long as the above-identified part gives the extremal eigenstates of $\hat M$.

The notion of a charge transport resonance is challenging to define precisely, but a few things are important to note. 
First, a CTR is interaction-mediated and is not present in the non-interacting problem.
In particular, in the non-interacting Anderson model, resonances can occur between single-particle orbitals which are nearly degenerate~\cite{mott_conduction_1968, ivanov_hybridization_2012}.
This near degeneracy can arise, for example, when the onsite potential energies between two or more sites are approximately equal.
These resonances are responsible for the exponentially decaying tail of the distributions of the CTC in the Anderson insulator and the peak at 1, as seen in Fig.~\ref{fig:summary_plot_ctcb}(b).
On the other hand, the addition of interactions allows for the mixing of \emph{configurations} of orbitals, giving rise to exponentially more possibilities for resonances.
Our CTRs fall into the latter category--- they are resonances between \emph{configurations} of charges.

Because of this, it is important to distinguish when CTRs are responsible for an abnormally large CTC and when they are not.
In the non-interacting problem, the primary mechanism for these large CTCs, as mentioned previously, is regions of flat potential energy near the link $i_0$.
These regions occur with exponentially small probability, and will still show up in the interacting problem and allow some disorder realizations with large $\ctc$.
However, we would not classify these large CTC instances as being caused by a CTR, since this is a phenomenon independent of interactions.
The CTRs primarily contribute to the observed significant peaks at integer values in Fig.~\ref{fig:ctcb_main_results_J_01}, which appear uniquely when interactions are present.

Finally, another important point is that a given disorder realization may have many CTRs in its spectrum, not just the one that is directly responsible for the CTC. 
These other CTRs can be identified as other pairs of eigenstates of $\hat M$.
We will quantify the importance of these additional CTRs when we characterize the full spectrum of the CTC operator via the Frobenius norm.

\subsubsection{Structure of $\ket{\minstate}$ and $\ket{\maxstate}$}\label{sec:localization_of_ctrs}
We now discuss the characteristics of the representative example of a CTR provided in Fig.~\ref{fig:example_ctr}, at system size $L=12$, involving a specific disorder realization with $W=10$ and $J=0.1$. 
This disorder realization has a CTC of $1.98$ with $\ket{\maxstate}$ and $\ket{\minstate}$ having an absolute-value-state overlap $|\ket{\maxstate}|\cdot|\ket{\minstate}|\approx 0.98$.

In Fig.~\ref{fig:example_ctr}, we show $\ket{\maxstate}$ and $\ket{\minstate}$ expanded in both the computational and Hamiltonian eigenstate bases and plot the absolute value squared of the weights against the indexing of the basis.
One can immediately see that these eigenstates are very localized in both bases.
For the computational basis as in Fig.~\ref{fig:example_ctr}(a), one can see that $\ket{\minstate}$ primarily has weight on the computational basis state $\ket{110111000001}$, and meanwhile $\ket{\maxstate}$ only has significant weight on the state $\ket{110100101001}$.
We note that between these two computational basis states, $2$ charges are transferred from sites $5$ and $6$ to sites $7$ and $9$ (with index starting at $i=1$ from the left).

Meanwhile, in Fig.~\ref{fig:example_ctr}(b), we expand $\ket{\maxstate}$ and $\ket{\minstate}$ in the energy eigenstate basis, with the corresponding energy marked on the $x$-axis.
One observes that the states $\ket{\minstate}$ and $\ket{\maxstate}$ are nearly both equal superpositions of the consecutive eigenstates indexed $|804\rangle$ and $|805\rangle$, with energies in the middle--upper end of the spectrum.
Importantly, we note that $|804\rangle$ and $|805\rangle$ are \emph{nearly degenerate}, with energy level spacing of $\Delta E_{\text{res}}\approx 6\times 10^{-5}$, much smaller than the average level spacing in this spectrum, which is around $|E_n-E_{n-1}|\sim 0.064$.

Furthermore, the overlap of each of these energy eigenstates with $\ket{\maxstate}$ and $\ket{\minstate}$ gives a value very close to $\approx±1/\sqrt{2}$.
One can verify that these energy eigenstates are hybridized almost perfectly, as follows:
\begin{align}
    |804\rangle &\approx \sqrt{\frac{1}{2}}|\maxstate\rangle + \sqrt{\frac{1}{2}}|\minstate\rangle ~, \nonumber \\
    |805\rangle &\approx \sqrt{\frac{1}{2}}|\maxstate\rangle - \sqrt{\frac{1}{2}}|\minstate\rangle \nonumber~.
\end{align}
This perfect hybridization not only affirms that the two states $\ket{\minstate}$ and $\ket{\maxstate}$ are connected by time evolution, but also supports a perfect transfer of particles between $\ket{\minstate}$ and $\ket{\maxstate}$, which aligns with the CTC being close to integer value $2$.
In Fig.~\ref{fig:example_ctr}(c),  one can see that starting from $\ket{\minstate}$ as an initial condition and time evolving under unitary dynamics of the Hamiltonian, two particles are transferred across $i_0$.
Thus, one can infer that the system evolves periodically between $\ket{\minstate}$ and $\ket{\maxstate}$, with a period of $\sim 2\pi/\Delta E_{\text{res}}$.

One may wonder whether most of the CTRs indeed only involve charge transfer processes localized near $i_0$ or if there is a possibility for these resonances to be long-ranged.
For example, looking at this particular CTR, it is clear that 2 particles are transferred very close to the cut.
We provide some evidence in Appendix~\ref{app:dipole} that many of these resonances are indeed long-ranged, and become longer-ranged as the system size increases.
However, due to the small system sizes accessible to ED, we cannot draw any definitive conclusions.

\subsubsection{Resonance collars of CTRs}\label{sec:ctc_resonance_collar}
The notion of a \emph{resonance collar} plays a role in pictures of resonances in the MBL regime~\cite{imbrie_many-body_2016,de_roeck_many-body_2017,garratt_local_2021,de_roeck_absence_2024}\footnote{The resonance collar is known as the ``buffer zone'' in Refs.~\cite{de_roeck_many-body_2017,garratt_local_2021}.}.
It measures the spatial range around resonating LIOMs in which the existence of the resonance remains sensitive to the occupations of the background charges that do not participate in the resonance.
Here we define the notion of a resonance collar more rigorously and examine it in the context of charge-transport resonances (CTRs).

Consider two LIOM configurations $\ket{I}$ and $\ket{J}$ on a lattice $L$, and let $X_i$ denote the LIOM flip operator centered at position $i$. 
Suppose that $\ket{I}$ and $\ket{J}$ are involved in a CTR. 
We define the active LIOMs as the subregion $R\subseteq L$ that contains all sites on which the LIOM occupations of $\ket{I}$ and $\ket{J}$ differ (for instance, the example in Fig.~\ref{fig:example_ctr} has range $r=5$).
The resonance range $r$ is then the size of this subregion, $r=|R|$.
A resonance collar $C$, where $R\subseteq C\subseteq L$, is a subregion containing the active region such that for any $i\notin C$, flipping a background LIOM does not destroy the resonance, i.e., $X_i\ket{I}$ and $X_i\ket{J}$ remain in resonance.
(Note that for simplicity, we include the resonating region $R$ in the collar $C$.)
Equivalently, the resonance should be insensitive to the occupations of LIOMs outside $C$ (the ``background'' LIOMs).
In a localized system, one expects $C$ to be finite; otherwise, one can tune arbitrarily many background LIOMs to bring any set of active LIOMs into resonance, which would cause resonances to proliferate and destroy MBL.

At the accessible system sizes, however, our observed CTRs appear to be surprisingly sensitive to the background configuration, suggesting that the resonance collar (if it exists) is large compared to the localization length. 
One can observe this in the same CTR example provided in Fig.~\ref{fig:example_ctr}.
Observing the density of states of $\hat M$ in Fig.~\ref{fig:example_ctr}(d), one sees that $\ket{\maxstate}$ and $\ket{\minstate}$ are the only eigenstates with eigenvalues near $1$ or $-1$, respectively, while the next-largest and next-minimum eigenvalues already lie in the bulk of the DOS far away from the extremal eigenvalues. 
Thus, this is the only CTR pair in the spectrum that transports two charges.
If the resonance were insensitive to the background, one would expect the same CTR to reappear multiple times in the spectrum, corresponding to different occupations of LIOMs distant from the active resonance. 
Moreover, we verified that in this disorder realization, the CTR is destroyed when the end sites are removed, but it is insensitive to adding additional sites with random fields and to small variations of the fields at the ends of the chain.
A similar story occurs when one looks at other CTRs at $L=16$, the largest available system size. 

Taken together, these observations suggest that the existence of the CTR of 2 particles depends strongly on the particular background configuration of charges not directly involved in the transport event. 
In particular, the fact that truncating a single site from an $L=16$ disorder realization eliminates a two-particle CTR suggests that the resonance collar for those CTRs spans, at minimum, the full accessible system around $i_0$ for $L=16$.
This is a very counterintuitive result, given that, at $W=10$, the non-interacting localization length is $\sim 0.4$, less than a lattice spacing.
This seems contradictory to the fact that flipping LIOMs that are at least 4 lattice spacings away from the active LIOMs in the resonance can affect whether the given two configurations are in resonance.
In Sec.~\ref{sec:ctc_model}, we introduce a simple toy model to count these charge transport resonances and use the model to provide a crude argument for why the resonance collar, once the resonance is observed to exist, should be roughly independent of the localization length and only dependent on the range of the resonance.

\subsubsection{Other examples of charge transport resonances}
\label{subsubsec:otherCTRs}
The example of a CTR presented in this section is a very simple example of a resonance.
In some sense, it is an idealized resonance which is an equal superposition of two product states.
The reality is that many resonances do not appear this ``clean,'' as the original Anderson orbitals may not be perfectly localized.
We refer the reader to Appendix~\ref{app:more_ctr_examples} for illustrations and discussions of two additional examples of CTRs that transfer 2 particles.
The conclusions from this section largely transfer over to those examples as well.

\subsubsection{The probabilities of charge transport resonances}
We now revisit the results of Fig.~\ref{fig:ctcb_main_results_J_01} in light of the connection between the observed integer-valued peaks in $P[\ctc]$ and CTRs.
The observations of $P[\ctc]$ reveal some rather surprising properties of these charge transport resonances.

First, the observation that these distributions exhibit larger weight on larger integers as $L$ increases shows that at system sizes accessible to ED, the observed CTRs appear to be transporting more and more charges, seemingly with no signs of stopping.
Second, even stranger is the fact that the CTRs grow larger seemingly at an exponential rate, seemingly independent of the disorder strength.
This is particularly bizarre since it implies that adding sites far away from $i_0$ influences probabilities of resonances that appear to primarily reside closer to $i_0$, even at extremely large disorder strengths where localization lengths can be smaller than a lattice spacing.

Together, taken at face value, these observations appear to be rather concerning for the asymptotic fate of MBL.
It would therefore be useful to understand the underlying mechanisms that give rise to these CTRs, which will be the focus of the next section.

\subsection{Diagonally-Improved Perturbation Theory for Counting Charge Transport Resonances}
\label{sec:ctc_model}
Now that we understand the physical properties of a CTR, we wish to understand the physical mechanisms underlying their growth with $L$, which at first glance appears to be unintuitive and worrisome for the fate of MBL in the thermodynamic limit.

To this end, we present a toy model, incorporating a diagonally-improved first-order perturbation theory~\cite{lin_slow_2020} of the weakly interacting Anderson model, that partially explains the growth of the CTC in the MBL regime on available sizes (we will explain this model momentarily).
This model uses fictitious level dynamics starting from the Anderson insulator and tuning the interaction strength, approximating forbidden level crossings between Anderson LIOM configurations using first-order estimates of the energy levels.
The idea of studying levels as some parameter is varied has been explored in various contexts in MBL, and the avoided crossings can usually be linked to some property of the physical system~\cite{maksymov_energy_2019, monthus_many-body-localization_2017, monthus_level_2016, filippone_drude_2016, serbyn_spectral_2016}.
Our procedure draws inspiration from past works~\cite{kjall_many-body_2018, garratt_local_2021, moudgalya_perturbative_2020}. 
The conceptual differences here compared with these previous works are that we treat interactions as perturbations on top of the Anderson insulator, and perform an explicit numerical counting of resonances in a context directly relevant for accessible $L$ of the widely studied disordered XXZ chain.

We will show that, by construction, 
our toy model
has a finite CTC of $\max{[\text{CTC}_{\text{Anderson}},2]}$ in each sample, and hence its average is bounded in the thermodynamic limit.
Despite this, at system sizes accessible to ED, the disorder-averaged CTC in this model grows roughly linearly with $L$, with no signs of saturation (while it is still significantly below the above maximal value).
We also provide an approximate understanding of the counting of such resonances, where the key ingredient is that the ``resonance collar'' of CTRs of 2 particles is larger than the system sizes accessible to ED.
This result shows that properties of resonances at these system sizes cannot be extrapolated to the thermodynamic limit.

We note that in this section, we will refer to Anderson orbital configurations as LIOM configurations, where LIOM refers to the LIOMs in the non-interacting Anderson problem.

\subsubsection{Forbidden level crossings and charge transport resonances}
\begin{figure*}[t]
    \centering
    \includegraphics[width=0.75\linewidth]{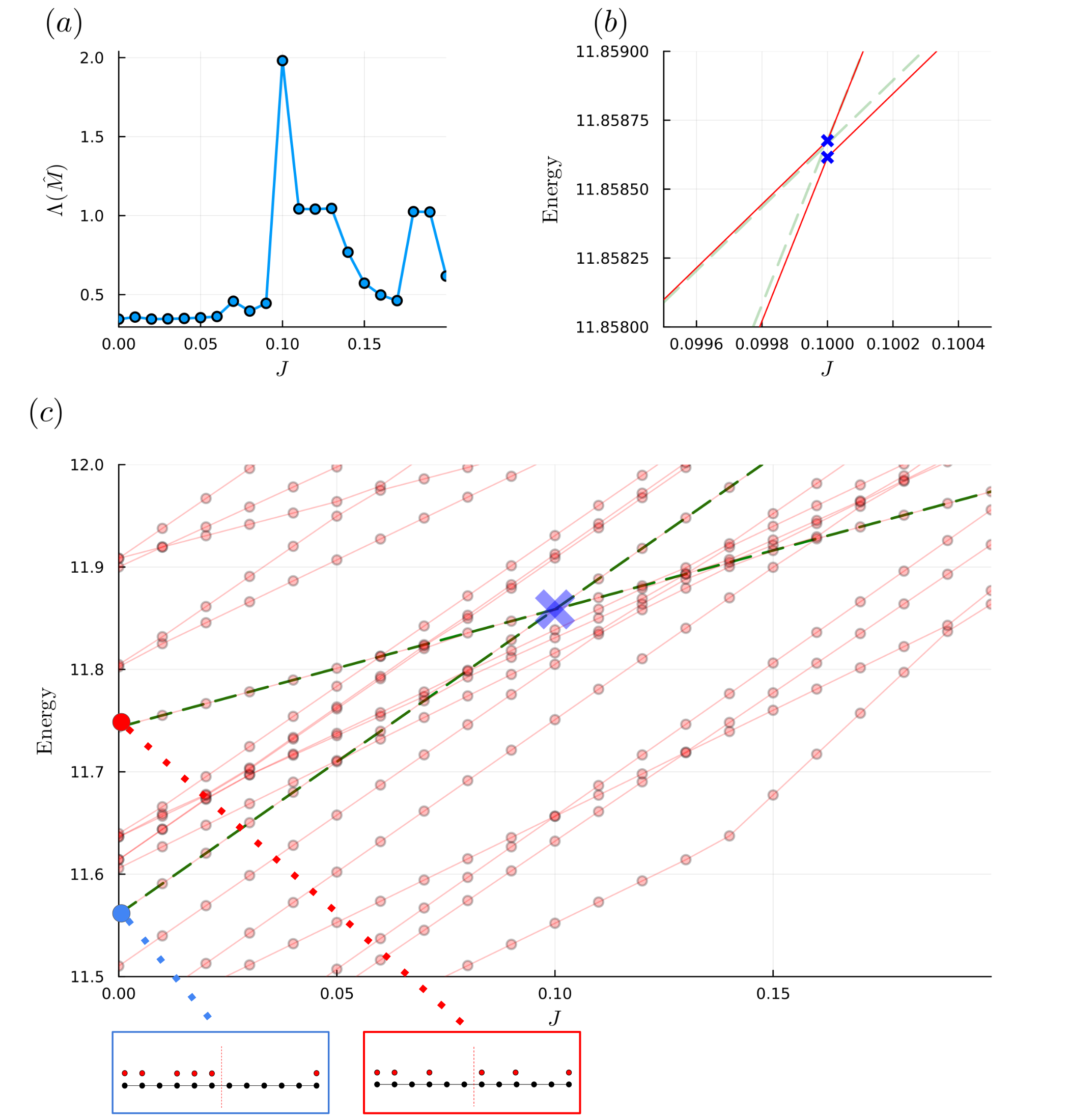}
\caption{
Summary of the connection between forbidden level crossings and charge transport resonances in the interacting Anderson model. 
In this plot, we fix the disorder realization giving rise to the example of the charge transport resonance discussed in Sec.~\ref{sec:ctr_properties}, with $\ctc = 1.98, W = 10, L = 12$, and interaction strength $J = 0.1$.
Panel (a): The CTC as a function of $J$ for this fixed disorder realization.
In the Anderson insulator ($J=0$), the CTC is small, and remains small as the interaction strength is increased until $J = 0.1$, where a CTR of $2$ particles occurs, and drops down right after the resonance (note that the window over which the resonance occurs is much smaller than the step $\Delta J=0.01$ in this plot)
Panel (b): A plot of the energy level curves as a function of the interaction strength $J$, zooming in very close to the resonance.
This CTR arises due to a forbidden level crossing between levels $\ket{804}$ and $\ket{805}$ (red lines) at the point marked by the blue crosses. 
The green lines are the estimates of the same many-body energy levels using first-order perturbation theory.
Panel (c): The full parametric level dynamics of the given disorder realization.
The light red filled circles represent many-body energy levels obtained with exact diagonalization ($y$-axis) as a function of the interaction strength $J$ ($x$-axis). 
The light red lines are guides for the eye.
The location of the resonance is marked by the blue cross.
The green dashed lines are the first-order perturbation theory estimates of the many-body energy levels, starting from the eigenstates in the Anderson insulator which are closest to the configurations involved in the CTR.
These configurations are illustrated in the red box and the blue box.
}
\label{fig:example_forbidden_crossing}
\end{figure*}

\begin{figure*}[t]
    \centering
    \includegraphics[width=\linewidth]{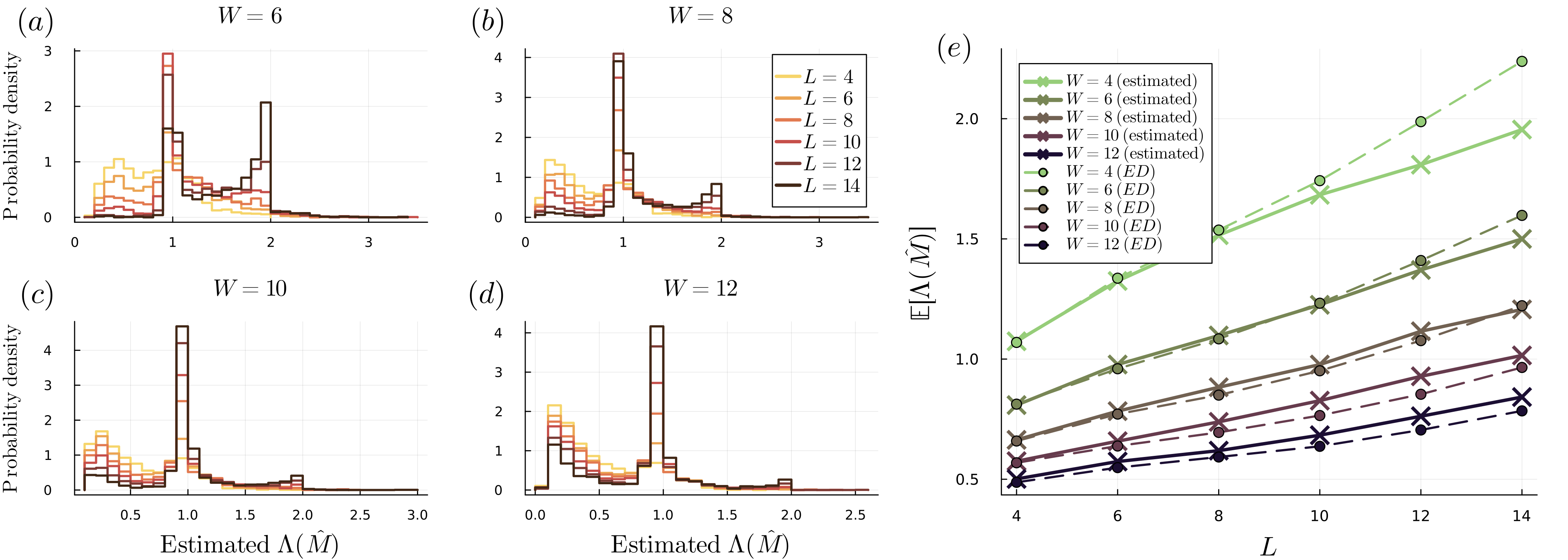}
    \caption{Calculations of the charge transport capacity $\ctc$ using the toy model, based on a diagonally-improved first-order perturbation theory, described in Sec.~\ref{sec:ctc_model}. 
    Panels (a)--(d): probability distributions of $\ctc$ over 2000 disorder realizations. 
    Each panel represents different disorder strengths, and within each panel, the colored histograms are distributions for different system sizes indicated by the legend.
    Panel (e): the disorder-averaged CTC as a function of $L$, with the toy model results overlaid, shown together with the CTC from ED. 
    Each curve is colored according to the disorder strength and marked to show if it is from the toy model (crosses, solid line) or ED (circles, dashed lines).
    }
\label{fig:predicted_ctc_vs_real}
\end{figure*}

In this section, we will show that a CTR of $m$ particles manifests as an avoided crossing between two LIOM configurations that differ by transferring $m$ particles across $i_0$.
To this end, we will analyze the many-body spectrum as we vary the interaction strength $J$.
In Fig.~\ref{fig:example_forbidden_crossing}, we show the level evolution diagram for the example of a CTR discussed in Sec.~\ref{sec:ctr_properties}. 
(For additional illustrations of the parametric level dynamics for different examples of CTRs mentioned in Sec.~\ref{subsubsec:otherCTRs}, see Appendix~\ref{app:more_ctr_examples}.)
As a quick reminder, in the present example, we found that $\ctc\approx 2$ at $J=0.1$.
Observing Fig.~\ref{fig:example_forbidden_crossing}(a), where one fixes the on-site potential disorder realization and increases interaction $J$, $\ctc$ remains close to $0$ until it suddenly peaks at 2 at $J=0.1$.
The CTC appears to drop back to near zero when we step away from $J=0.1$, but as we will see, the step size $\delta J = 0.01$ in panel (a) is much larger than the window in $J$ over which this resonance is observable; the fact that the resonance appears at $J = 0.1$ is because this value of $J$ happened to fall within that window. 
Panel (b) zooms in to reveal how narrow this window actually is.

As seen in Fig.~\ref{fig:example_forbidden_crossing}(b), this peak in $\ctc$ coincides with an avoided crossing at $J = 0.1$ between energy-ordered eigenstates $\ket{804}$ and $\ket{805}$ (the same eigenstates that dominate $\ket{\maxstate}$ and $\ket{\minstate}$). 
At $J=0.1$, these many-body eigenstates are a hybridization of two Anderson LIOM configurations $\ket{a}, \ket{b}$ that exchange two particles across $i_0$.
These LIOM configurations are illustrated in Fig.~\ref{fig:example_forbidden_crossing}(c) (which are the same ones as identified earlier in Fig.~\ref{fig:example_ctr}).
The energy levels in this figure also show an approximately linear dependence on $J$, suggesting that weak-$J$ CTRs arise from first-order effects: off-diagonal perturbation terms of specific configurations of the non-interacting LIOMs are brought to resonance by diagonal perturbation terms. 
Furthermore, this data shows that the first-order perturbation theory (linear in $J$) estimate of the energies works remarkably well, with straight energy lines exiting forbidden crossings almost untouched at small $J$.
These observations are key for formulating a simple theory of charge transport resonances that captures many aspects of the growth of the CTC, which we will describe in the next section.

\subsubsection{The model}
The overall strategy of our diagonally-improved perturbative model of CTRs is to treat interactions as perturbatively weak on top of the Anderson insulator to estimate the many-body energies at small $J$, and then to compare the energy splitting between many-body levels with the corresponding off-diagonal matrix element.
This gives rise to resonances between Anderson LIOM configurations, where such a resonance occurs when the corresponding off-diagonal matrix element is larger than the energy splitting. The described process of counting resonances is inherently non-perturbative. 
In line with Ref.~\cite{lin_slow_2020}, we refer to this method as a \emph{diagonally-improved perturbation theory}, to describe the important role that the diagonal terms of the perturbation play in counting these resonances.

To be precise, the starting point is the non-interacting Anderson model ($J = 0$) with orbitals $\phi_\alpha$ and corresponding single-particle energies $\epsilon_\alpha$ with $\alpha = 1, \dots, L$. 
Here, one can roughly interpret the labeling as the ``position'' of $\phi_\alpha$, i.e., the location where $\phi_\alpha$ is peaked.
In the half-filling sector of the non-interacting Anderson model, the eigenstates are Slater determinants formed by choosing which $L/2$ of the $L$ localized single-particle orbitals to occupy.
We label the possible orbital occupation configurations as binary representations of integers $a = 1, \dots, \binom{L}{L/2}$, 
so that the many-body energies are $E_{a} = \sum_{\alpha=1}^{L} n_\alpha[a] \epsilon_{\alpha}$, 
where $n_\alpha[a]$ is the occupation number for the orbital $\alpha$ in the configuration $a$,
with corresponding eigenstates $\ket{\Phi_a}$.

The interaction can be rewritten in terms of the localized orbital basis, $\hat{d}_\alpha= \sum_k \phi_{k\alpha}^* \hat{c}_k$:
\begin{equation}
    \hat V=\sum_{\alpha, \beta,\gamma,\delta}V_{\alpha\beta\gamma\delta}\hat{d}_\alpha^\dagger \hat{d}_\beta^\dagger \hat{d}_\gamma \hat{d}_\delta~.
\end{equation}
Here, the conveniently anti-symmetrized matrix elements are given by
\begin{multline}\label{eq:matrix_elements_perturbation}
    V_{\alpha \beta \gamma \delta}= \frac{J}{4}\sum_j(\phi_{j+1,\alpha}^* \phi_{j,\beta}^* \phi_{j,\gamma} \phi_{j+1,\delta}-\phi_{j,\alpha}^* \phi_{j+1,\beta}^* \phi_{j,\gamma} \phi_{j+1, \delta}\\
    -\phi_{j+1, \alpha}^* \phi_{j, \beta}^* \phi_{j+1, \gamma} \phi_{j, \delta}+\phi_{j, \alpha}^* \phi_{j+1,\beta}^* \phi_{j+1,\gamma} \phi_{j,\delta}) ~.
\end{multline}

Although the interaction term now appears to allow arbitrary-range terms, there is substantial local structure due to the matrix elements' dependence on the non-interacting eigenstates.
Specifically, because the Anderson model has exponentially localized orbitals, $V_{\alpha\beta\gamma\delta}$ is only substantial if orbitals labeled by $\alpha, \beta, \gamma, \delta$ are close to each other in position space.

Turning on interactions, first-order perturbation theory corrects the energy $E_a$ of the Anderson model many-body eigenstate $\ket{\Phi_a}$ (corresponding to LIOM 
configuration $a$) to
\begin{equation}
    \widetilde{E}_a = E_a + \langle \Phi_{a}|\hat{V}|\Phi_{a}\rangle~.
\end{equation}
Now, consider two LIOM configurations $a$ and $a'$ in the half-filling sector. 
If they are connected by a motion involving at most 2 particles, the interaction $\hat V$ will connect them through a nonzero off-diagonal matrix element $\langle \Phi_a|\hat V|\Phi_{a'} \rangle$ given by appropriate $4 V_{\alpha,\beta,\gamma,\delta}$ assuming $\ket{\Phi_a} = d_\alpha^\dagger d_\beta^\dagger d_\gamma d_\delta \ket{\Phi_{a'}}$).
We say that these LIOM configurations are in a charge transport resonance of $m$ particles if between the LIOM configurations $|\Phi_a\rangle$ and $|\Phi_{a'}\rangle$,  $|\langle \Phi_a| \hat N_R|\Phi_a\rangle-\langle \Phi_{a'}| \hat N_R|\Phi_{a'}\rangle|=m\leq 2$ (in the case where the single-particle orbitals are sufficiently localized, $m$ is approximately an integer), and that
\begin{equation}\label{eq:resonance_condition}
    |\widetilde E_a-\widetilde E_{a'}|\leq 2 | \langle \Phi_a|\hat V|\Phi_{a'} \rangle | ~.
\end{equation}
This resonance condition can be understood by reducing the Hamiltonian down to a 2-level system only involving $\ket{\Phi_a}$ and $\ket{\Phi_{a'}}$, and observing that the resulting approximate many-body eigenstates have significant weight on both $\ket{\Phi_a},\ket{\Phi_{a'}}$ if and only if Eq.~(\ref{eq:resonance_condition}) is true. 
At exactly the forbidden level crossing where $|\widetilde E_a-\widetilde E_{a'}|=0$, the approximate many-body eigenstates are an equal superposition of the two LIOM configurations---a perfect resonance.

In this model, one can see that interactions are necessary in inducing CTRs, both to produce mixing via off-diagonal matrix elements and also to enhance the probability of occurrence via variations in the diagonal matrix elements. 
In the non-interacting case, the energy difference between two states that differ by a few (``active'') LIOM occupations is independent of the occupations of all of the other (``background'') LIOMs. 
In the interacting case, even just counting level crossings using perturbative energy estimates, for each possible pair of states differing by the same active LIOMS, obtained by taking one out of the exponentially many configurations of the other LIOMS in the background, one gets a slightly different energy difference, so there are many more chances for the added interaction to induce a resonance.

We note that at large $W$, because the original orbitals are quasilocal with short localization lengths, the charge transfer between Anderson many-body states will be close to an integer value. 
Thus, the resonances defined this way will transfer an approximately integer amount of charge, which would explain how the peaks at integer-valued CTC arise in the interacting problem.

\subsubsection{Numerical results}\label{sec:numerical_results_ctc_model}

In this section, we use this model to provide a numerical estimation of the CTC and compare it with the full ED results.
To get a total estimate of the CTC from a given disorder realization, for every pair of LIOM configurations connected with nonzero off-diagonal elements, we find the resonances using Eq.~(\ref{eq:resonance_condition}) and, for simplicity, assume perfect transfer is allowed between the LIOM configurations if the resonance condition holds. 
Because this model builds off the non-interacting Anderson model, the charge transport resonances can be viewed approximately as ``extra'' charge transfer processes on top of those of the original Anderson insulator. 
Thus, when calculating the CTC, we can choose the maximum between the charge transferred by interaction-mediated CTRs and the $J = 0$ Anderson insulator CTC in this specific disorder realization.

We show the distributions over disorder realizations of the CTC in this model with 2000 disorder realizations, up to $L=14$, in Figs.~\ref{fig:predicted_ctc_vs_real}(a)--(d).
In Fig.~\ref{fig:predicted_ctc_vs_real}(e), we show the mean of the CTC predicted by this model compared to the ED calculation.
This model, by construction, has a finite CTC in each sample, with a maximum possible value of $\max{[\text{CTC}_{\text{Anderson}}, 2]}$, and hence a finite average in the thermodynamic limit.
Yet, the disorder-averaged CTC in this simple model seems to grow roughly linearly with $L$, with no signs of saturation at the system sizes accessible to ED.
Furthermore, this simple model is roughly quantitatively accurate at high disorder strengths (where the perturbation theory is expected to be more accurate).
We also note that the distributions of the CTC predicted by the first-order toy model appear to be qualitatively accurate, predicting changing weights in the integer-valued peaks with increasing system size.
Lastly, we numerically verified that the CTRs in this toy model do not appear to be generically caused by abnormally large off-diagonal matrix elements.
Rather, they are truly a manifestation of accidental near-degeneracies due to interaction induced level crossings.

These results show that the simple model is able to capture qualitatively and quantitatively the growth of the CTC at system sizes available to ED.
Thus, we can now use the toy model, considerably simplified from the full interacting problem, to try to understand the physical mechanisms for some of the unintuitive behaviors of the CTC at these system sizes. 
In particular, why should the CTC be so sensitive to the system size in this model, when the localization lengths in this strong disorder regime are much less than a lattice spacing?
Furthermore, how could this sensitivity be independent of the disorder strength?
Naively, one might expect these questions to have a simple answer: in the toy model, off-diagonal matrix elements are exponentially suppressed with the range of the resonance, so the resonances driving the CTC growth should primarily involve LIOMs near the cut. 
Then, one would expect that the LIOMs further out from the cut would have no ``influence'' over the active LIOMs in resonance near the cut $i_0$.
Thus, this would lead one to conclude that the ``resonance collar'' is sensitively dependent on $\xi$, and thus the CTC would grow and saturate differently for different $\xi$.
However, the observations clearly show that this naive prediction is wrong: 
Both in the toy model and the full ED data, the accessible system sizes remain well within the range that sensitively affects existing resonances, consistent with our earlier conclusion that the resonance collar must be large (cf.\ Sec.~\ref{sec:ctc_resonance_collar}). 
We offer a heuristic explanation for these unintuitive observations in the next section.

\subsubsection{Counting of the resonances}\label{sec:counting_resonances}
\begin{figure*}[t]
    \centering
    \includegraphics[width=0.85\linewidth]{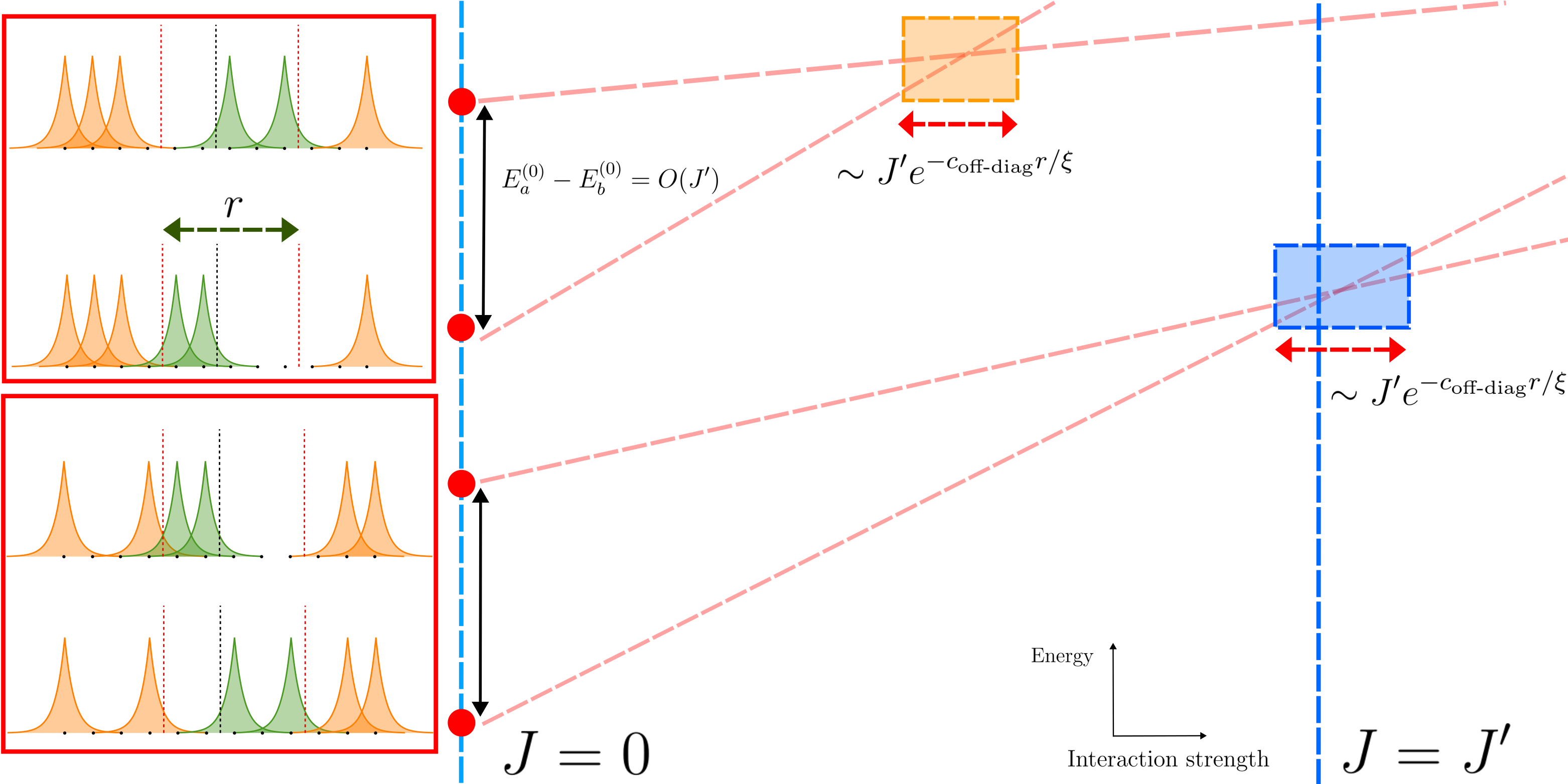}
    \caption{Illustration (not to scale) of the conditions for a given set of active LIOMs (green) with range $r=5$ to be in resonance at interaction strength $J=J'$.
    In the illustration, $L=12$, which is smaller than the resonance collar size $\sim 15$.
    The $y$-axis is meant to be understood as the energy, while the $x$-axis is the interaction strength.
    (1) The bare energy difference $\epsilon_\alpha + \epsilon_\beta - \epsilon_\gamma - \epsilon_\delta$ must be $O(J')$, which occurs with probability $\sim J'/W$. This condition is illustrated by the spacing between the red dots on the $J = 0$ axis.
    (2) 
    The dashed red lines show the first-order perturbation theory estimates of the energy levels as a function of $J$, starting from the Anderson insulator energies (red dots).
    Once the initial energy difference is small enough, we can now check all $70 = \binom{8}{4}$ 
    background configurations (for large $L$, $\sim O(2^L)$ possible configurations). 
    The background LIOMs give variation to the slopes of these curves.
    In this illustration, we show two different background configurations (orange LIOMs) out of $70$ possible ones.
    (3) The lines need to cross in a window of width $J'e^{-c_{\text{off-diag} r}/\xi}$ which falls onto $J'$.
    If this window misses $J'$ completely (orange box), we do not see those LIOM configurations in resonance.
    Alternatively, the window indeed intersects with $J'$ (blue box), and we see those LIOM configurations in resonance at $J=J'$.
    This ``box'' intersects the blue line $J=J'$ with probability $\sim J'e^{-c_{\text{off-diag}}r/\xi}/J'$.
    Thus, in total, the probability of \emph{any} of the background configurations bringing this active set of LIOMs into resonance is $\sim \frac{J'}{W} 2^L e^{-c_{\text{off-diag}}r/\xi}$, assuming that the box width is much smaller than the separation in along the horizontal axis between the crossings.
    This is the condition that the collar size exceeds $L$.
    }
\label{fig:resonance_illustration}
\end{figure*}

In this section, we provide a heuristic argument for the exponential growth of the CTC, caused by contributions from resonances whose collar exceeds the system size. 
Our argument applies in the weakly interacting regime of the interacting Anderson model, as it is based on our diagonally-improved perturbative treatment.
Specifically, we will show that $P[r; J]$,\footnote{We note that this probability distribution is over disorder realizations, and not the probability distribution within a given disorder realization, where they can be highly correlated.} the probability that a resonance of range $r$ occurs at interaction strength $J$ for a randomly sampled disorder realization, roughly doubles with each added site as long as $L$ is smaller than the resonance collar size $b_{\text{collar}} r$, where $b_{\text{collar}} \sim 3$.
This process of the probability growing exponentially with $L$ is independent of $\xi$, the localization length.
We provide an illustration and summary of this argument in Fig.~\ref{fig:resonance_illustration}.

First, let us restrict to first-order in perturbation theory and only consider resonances between LIOM configurations $\ket{a}, \ket{b}$ such that $\ket{a}, \ket{b}$ differ in the occupations of orbitals $\alpha,\beta,\gamma,\delta$, corresponding to the connection by $\hat d_\alpha^\dagger \hat d_\beta^\dagger \hat d_\gamma \hat d_\delta$.
Now, suppose that this resonance has range $r$, and that the orbitals have non-interacting energies $\epsilon_\alpha, \epsilon_\beta,\epsilon_\gamma,\epsilon_\delta$, respectively.
Again here, $\alpha$ labels the position of orbital $\alpha$, i.e., the position at which the orbital $\alpha$ peaks.
Furthermore, we will assume that the Anderson non-interacting orbital $\phi_\alpha$ has the form $\phi_\alpha(i) \sim e^{-|i - \alpha|/\xi}$, where $\xi$ is the localization length.
For a concrete example, we will follow the resonance in the main text and Fig.~\ref{fig:example_forbidden_crossing} and discuss how it came to be in resonance. 
In that example, $\alpha=5, \beta= 6$ and $\gamma=7, \delta=9$, corresponding to moving 2 particles from $\alpha, \beta$ to $\gamma, \delta$.

Now, denote the set of occupied LIOMs involved in the resonance in configuration $\ket{a}$ as $R_a$.
The \emph{active} LIOMs then refer to those in $R_a\cup R_b$.
We denote the set of \emph{background} LIOMs as $B_a$, so that by definition, $B_a = B_b \equiv B$.
For a system of size $L$, this gives us $O(2^{L})$ possible background configurations at large $L$. 
[Technically, assuming that $\alpha,\beta,\gamma,\delta$ are distinct, there are precisely $\binom{L-4}{L/2-2}$ possible backgrounds in the half-filling sector.
However, $\binom{L-4}{L/2-2} \approx 2^{L-4} \sqrt{\frac{2}{\pi(L-4)}} \sim O(2^L)$ at large $L$, ignoring numerical and polynomial in $L$ factors.]
Thus, there are $O(2^{L})$ pairs of configurations with occupations only differing on $\alpha,\beta,\gamma,\delta$.

Consider any such pair of LIOM configurations $\ket{a}, \ket{b}$.
The off-diagonal element $\langle a|\hat V|b\rangle$ is independent of the background configuration and is given by $4 V_{\alpha\beta\gamma\delta}$ in Eq.~(\ref{eq:matrix_elements_perturbation}).
Inspecting Eq.~(\ref{eq:matrix_elements_perturbation}) and assuming the Anderson orbital ansatz, we see that $\langle a|\hat V|b\rangle \sim e^{-O(r)/\xi}$.
This $O(r)$ depends specifically on the positions of $\alpha,\beta,\gamma,\delta$.
For concreteness, we will write $\langle a|\hat V|b\rangle \sim e^{-c_{\text{off-diag}}r/\xi}$. In many previous works, it is assumed that $c_{\text{off-diag}} = 1$
~\cite{garratt_local_2021,crowley_constructive_2022}.
This is an okay assumption for many active configurations.\footnote{We get $c_{\text{off-diag}} = 1$, e.g., for $\alpha = L/2-r/2, \beta = L/2, \gamma = L/2+1, \delta = L/2+r/2$; however, for some other local resonances we can have $c_{\text{off-diag}} \sim 2$, e.g., $\alpha = L/2-r/2, \beta = L/2-r/2+1, \gamma = L/2+r/2-1, \delta = L/2+r/2$.}
In the example we are following from Fig.~\ref{fig:example_forbidden_crossing}, $\langle a|\hat V|b\rangle \sim e^{-5/\xi}$.
This off-diagonal matrix element defines the energy sensitivity of the resonance: for a resonance of range $r$ to occur, the energies of the two configurations must be tuned to within a precision of  $e^{-c_{\text{off-diag}}r/\xi}$.

The corrected energy $\widetilde E_a$ as predicted by first-order perturbation theory for the LIOM configuration $\ket{a}$ can be expanded as follows:
\begin{equation}
    \widetilde E_a = E^{(0)}_a + E^{R_a-R_a} + E^{R_a-B} + E^{B-B} ~.
\end{equation}
Here, $E^{(0)}$ denotes the sum of the energy of the individual occupied LIOMs, $E^{R_a-R_a}$ denotes energy terms which encompass interactions between pairs of LIOMs in $R_a$ (e.g., $\sum_{\alpha,\beta \in R_i} V_{\alpha\beta} \widetilde n_\alpha\widetilde n_\beta$), and $E^{R_a-B}$ denotes energy terms involving interactions between background and active LIOMs, and so on.
The energy difference between the two configurations is then
\begin{align}
    \widetilde E_a - \widetilde E_b &= E^{(0)}_a - E^{(0)}_b + E^{R_a-R_a} - E^{R_b-R_b} \\
    &+ E^{R_a-B} - E^{R_b-B}~.
\end{align}
We note that the bare energy difference, $E_a^{(0)}-E_b^{(0)}=\epsilon_\alpha + \epsilon_\beta - \epsilon_\gamma - \epsilon_\delta$, is an $O(1)$ number whose distribution can be calculated, e.g., if we assume the orbital energies to be roughly independent and uniformly distributed, see Appendix D in Ref.~\cite{jiang_quasiconservation_2025}.
What is important in what follows is that this distribution has a finite probability density $\sim 1/W$ at small values of $E_a^{(0)}-E_b^{(0)}$.

With these preliminaries, we can estimate that the size of the resonance collar, as defined in Sec.~\ref{sec:ctc_resonance_collar}, is $\sim b_{\text{collar}}r$, with $b_{\text{collar}}\sim 3$ as the resonance collar spans a distance $r$ around the region containing the active LIOMs. 
This is simply because so long as an occupied background LIOM is within distance $r$ of some active LIOM, it will produce a shift $e^{-c_{\text{diag}}r/\xi}$ in the diagonal energy term $E^{R-B}$.
In the first-order perturbation theory with the diagonal terms being of the form $V_{\rho\sigma\sigma\rho} \widetilde{n}_\rho \widetilde{n}_\sigma$, we have $c_{\text{diag}} = 2$.\footnote{Note that in the above estimate we assumed for simplicity that $c_{\text{diag}} \sim c_{\text{off-diag}}$, arriving at the estimate $b_{\text{collar}} \sim 3$; we could be more precise for specific resonances, estimating that $b_{\text{collar}} \sim 1 + 2 c_{\text{off-diag}}/c_{\text{diag}}$. However, we will keep things simplified for the schematic estimates in the remaining text.}
Because this shift is comparable to the resonance sensitivity, the background LIOM is capable of facilitating or destroying the resonance.
Notably, \emph{the size of the resonance collar only depends on $r$ and not $\xi$}.
This fact is key to our later analysis of numerically accessible system sizes.

Now, we consider the conditions necessary to form a resonance between $\ket{a}$ and $\ket{b}$, i.e. $\widetilde E_a-\widetilde E_b \leq 2\langle a|\hat V|b\rangle\sim Je^{-c_{\text{off-diag}} r/\xi}$.
First, one requires that at the non-interacting level, $E_a^{(0)} - E_b^{(0)} = \epsilon_\alpha + \epsilon_\beta - \epsilon_\gamma - \epsilon_\delta = O(J)$. 
Otherwise, since the corrections to the energies are $O(J)$, these corrections cannot possibly bring $\widetilde E_a-\widetilde E_b$ down to an exponentially small number.
This occurs with probability $O(J/W)$, assuming that the non-interacting energies are distributed uniformly.
In our concrete example of Fig.~\ref{fig:example_forbidden_crossing}, in the unperturbed Anderson insulator, we already have $E_a^{(0)} -E_b^{(0)} = \epsilon_5 + \epsilon_6 - \epsilon_7 - \epsilon_9 \approx 0.2=2J$, which indeed satisfies this first condition.

Once this condition is satisfied, we then go through the $O(2^{L})$ possible backgrounds, asking the probability $P_{\alpha\beta\gamma\delta}[J]$ that \emph{any} of these configurations will bring this pair of LIOM configurations into resonance at interaction strength $J$.
This highly depends on the background LIOM configurations and how they interact with the active LIOM configurations.
In the concrete example, we have $70$ possible backgrounds. 
Note that even though $E_a^{(0)} -E_b^{(0)}\approx 0.2$ is small, it is much larger than the off-diagonal matrix element connecting $\ket{a}$ and $\ket{b}$, which is around $\lesssim 10^{-4}$ for $J=0.1$.
Thus, it is not trivial whether any of the available background configurations can bring the active LIOMs into resonance.

Now, fixing a background configuration, we ask whether the first-order corrections $\langle a|\hat V_{\text{diag}}|a\rangle$ and $\langle b|\hat V_{\text{diag}}|b\rangle$ can bring the energy levels close enough together to form a resonance. Here, $\hat V_{\text{diag}}$ is
the diagonal part of the interaction of Eq.~(\ref{eq:matrix_elements_perturbation}):
\begin{equation}
V_{\text{diag}}=\sum_{\rho \neq \sigma}2V_{\rho\sigma\sigma\rho} \tilde n_\rho^{(0)}\tilde n_{\sigma}^{(0)}~.
\end{equation}
We will refer to $\frac{1}{J}\langle a|\hat V_{\text{diag}}|a\rangle$ and $\frac{1}{J}\langle b|\hat V_{\text{diag}}|b\rangle$ as the \emph{velocities} of the level dynamics.
Whether a particular background brings the set of active LIOMs into resonance depends on the microscopic details of the configuration and the diagonal matrix elements $\langle a|\hat V_{\text{diag}}|a\rangle$.
We can understand the dependence of the velocity difference on the microscopic details through the example of Fig.~\ref{fig:example_forbidden_crossing}. Here, $E_a^{(0)}-E_b^{(0)}=0.2$, and so we require the velocity difference to be around 0.2 for a resonance to occur.
To achieve this, we note that at the crudest level, the velocity of a given configuration is given by just $\sum_{i=1}^{L-1} \tilde n_i \tilde n_{i+1}$, where $i$ and $i+1$ label the ``positions'' of the orbitals. Thus, in this example, the difference in velocities between $|a\rangle$ and $\ket{b}$ is $2$, coming from the terms $\tilde n_4 \tilde n_5 + \tilde n_5 \tilde n_6$. This means that the overall change in their energy difference is $\sim 0.2$ at $J=0.1$, bringing $\widetilde E_a - \widetilde E_b$ to near zero at $J = 0.1$ and thus facilitating a level crossing at $J=0.1$.

Now, consider system sizes $L$ which lie within the resonance collar of size $b_{\text{collar}} r$.
For simplicity, as a first approximation, we assume that the velocities are at values such that the levels $\widetilde E_a$ and $\widetilde E_b$ will participate in a forbidden level crossing at some value of $J$ (that they are not, for example, parallel to each other).
We now ask whether this crossing will occur at the specific $J$ value that we are interested in.
The resonance condition can be satisfied in a range of $J$ around the crossing with size $\sim Je^{-c_{\text{off-diag}}r/\xi}/(\frac{1}{J}\langle a|\hat V_{\text{diag}}|a\rangle-\frac{1}{J}\langle b|\hat V_{\text{diag}}|b\rangle)$.
The probability of this window containing our specific value of $J$ is then of order $\sim e^{-c_{\text{off-diag}}r/\xi}/\frac{1}{J}(\langle a|\hat V_{\text{diag}}|a\rangle-\langle b|\hat V_{\text{diag}}|b\rangle)\sim e^{-c_{\text{off-diag}}r/\xi}$.
Here, we use the fact that $\langle a|\hat V_{\text{diag}}|a\rangle- \langle b|\hat V_{\text{diag}}|b\rangle\sim O(J)$. As mentioned previously, for the example we are following, indeed $\langle a|\hat V_{\text{diag}}|a\rangle- \langle b|\hat V_{\text{diag}}|b\rangle\approx 0.2$ at $J=0.1$.
To a first approximation, now assume that the probability of each of the different background configurations bringing these two configurations into resonance is independent of that of the other background configurations.
This is justified by the condition $L \leq b_{\text{collar}} r$, so that the (somewhat random) changes in the slopes from one background to another are sufficiently large that the corresponding pairs of crossings are separated in $J$ by a distance larger than the resonance width.
Then, the total probability of at least one of the background configurations bringing this active configuration into resonance is 
\begin{equation}
    P_{\alpha\beta\gamma\delta}[J|L\leq b_{\text{collar}}r] \sim O\left(\frac{J}{W}2^Le^{-c_{\text{off-diag}}r/\xi}\right)~,
    \label{eq:P_L.leq.Rcoll}
\end{equation}
as long as the system size $L$ lies within the resonance collar of size $b_{\text{collar}} r$.

Now, if $L > b_{\text{collar}} r$, $P_{\alpha\beta\gamma\delta}[J]$ drops its $L$ dependence, saturating to value 
\begin{equation}\label{eq:probability_resonances_asymptotic}
    P_{\alpha\beta\gamma\delta}[J | L>b_{\text{collar}}r]\sim O\left(\frac{J}{W}2^{b_{\text{collar}}r}e^{-c_{\text{off-diag}}r/\xi}\right)~.
\end{equation}
This is because adding more sites outside of the resonance collar does not influence the probability of a given active configuration being in resonance for some background configuration, by definition of the resonance collar.
That is, additional crossings happen essentially inside resonance windows covered by resonances that are already accounted for with the given active sites.
Therefore, at fixed $J$ we see more CTRs with the same active sites but with different backgrounds---in fact, this number grows as $\sim 2^Lg(r)$ where $g(r)$ is some function of $r$, see Eq.~(\ref{eq:P_L.leq.Rcoll}).
However, all of them transfer exactly the same charge, and so they do not influence the probability that a resonance of range $r$ occurs (nor the value of the CTC).
The fact that, in the discussed example in Fig.~\ref{fig:example_forbidden_crossing}, the given CTR occurs only for one particular background is a very strong indication that we are far from this regime of settled resonance behavior.

Notice that when $r\rightarrow\infty$, $P_{\alpha\beta\gamma\delta}[J]$ will converge to a value smaller than 1 if $\xi$ is small enough such that $e^{-c_{\text{off-diag}}r/\xi}$ decays faster than $2^{b_{\text{collar}}r}$. On the other hand, if $\xi$ is too large, $2^{b_{\text{collar}}r}$ grows faster than $e^{-c_{\text{off-diag}}r/\xi}$, so that $P_{\alpha\beta\gamma\delta}[J]$ at large $L$ will diverge with $r$.
Thus, resonances of arbitrarily large $r$ will proliferate in an uncontrolled way.
Determining the critical $\xi$ from this theory requires careful counting of the resonances, which is beyond the scope of this work.
However, this picture of the instability of MBL to the build-up of resonances is also consistent with the avalanche instability of MBL and with previous schemes of counting MBRs in a stable MBL phase~\cite{de_roeck_stability_2017, garratt_local_2021}.

We therefore conjecture that in the simple toy model, only counting first-order resonances, for strong enough disorder strengths where $\xi$ is small, the CTC probability distribution will converge with $L$ to have non-zero probabilities below $\ctc=2$ (the maximal CTC in this treatment).
On the other hand, for small enough $W$, all disorder realizations are guaranteed to have a CTC of the maximal value $\ctc=2$, signaling that localization in this regime would likely break down with $L$.
This analysis, in principle, extends directly to resonances that can occur at higher orders in perturbation theory, and has the potential to probe the instability of the MBL phase to resonances between Anderson LIOM configurations.
A more careful treatment would be an interesting direction that we leave to future work.

\subsubsection{Implications of the toy model for numerical studies}\label{sec:implications_for_numerical_results}

In this section, we discuss the implications of our results from the diagonally-improved perturbative model for the regime of numerically accessible system sizes, giving an explanation for observations of the ED data.
These results suggest a possible physical mechanism behind the strong finite-size effects seen in numerical studies of MBL---namely, the development of the charge transport resonances which appear to transport more charge with increasing $L$ for numerically accessible $L$.
Although our focus is on explaining the behavior of resonances in the strong disorder regime, the analysis in this section, in principle, may extend to the prethermal or even the weakly ergodic regimes, where strong finite-size effects are even more pronounced~\cite{znidaric_diffusive_2016,bera_density_2017,de_tomasi_dynamics_2019,lezama_apparent_2019,panda_can_2020, abanin_distinguishing_2021}.

First, our numerical data for the toy model, which counts CTRs in a transparent way, accounts for much of the CTC growth observed in the ED data for $W \gtrsim 4-6$, even though the model only describes first-order resonant processes. 
This demonstrates that CTRs will indeed appear to transport more charge as $L$ increases in this regime.
However, we know analytically that in this toy model, the CTC must saturate at system sizes far beyond numerical reach. 
Therefore, this shows that ED data alone fundamentally cannot determine the asymptotic fate of even small resonances.

Second, our heuristic argument in Sec.~\ref{sec:counting_resonances}, which is based on this toy model, provides an understanding of why the growth of intermediately large resonances remains unsettled at numerically accessible system sizes and in a way seemingly independent of $W$.
There are two factors at play here.
The first is purely quantitative: our numerics show that the CTRs of 2 particles have a numerically nontrivial probability of occurring, i.e., in the number of disorder realizations we consider, we will find some number of these CTRs.
Second, our argument crucially showed that the size of the resonance collar only depends on $r$ as $\sim b_{\text{collar}}r$ with $b_{\text{collar}}\sim 3$, which is independent of factors like $\xi$ and $W$.
In particular, note that the resonance collar of a CTR of 2 particles spans at least $12$ sites around $i_0$, and is likely larger ($\geq 15$ sites), since most 2-particle CTRs are of range 5 or higher.
This collar size for 2-particle resonances is thus on the boundary of or larger than numerically accessible $L$.
Then, numerically accessible systems will lie \emph{inside of} the resonance collar for CTRs of 2 particles. Thus, as we argued previously, the probability of a CTR of 2 charges occurring somewhere in the spectrum then grows exponentially with $L$ at the accessible system sizes, at a rate independent of $\xi$.
We see these two factors come together and manifest as the observed nonsaturating, disorder-independent rate of exponential growth of the CTC at these system sizes.\footnote{One can understand the growth of the CTC at finite sizes in the following sense.
Suppose that either the system does not transfer charge or it has a CTR of 2 particles.
The CTC in this simple scenario is therefore either $0$ or $2$; the latter is true even if $L > b_{\text{collar}} r$ and there are $\sim 2^{L-b_{\text{collar}} r}$ CTRs in such a sample that are very correlated (differing only by inactive backgrounds), since all of them allow roughly the same charge transfer.
For large enough disorder strengths, the sum of appropriate probabilities over all possible active configurations of resonances will converge, and the CTC in this model will saturate to a value smaller than 2 at system sizes larger than the resonance collar.
In the full ED data, the CTC aggregates information about \emph{all other resonances} in a complex way, but the CTC is always upper bounded by $\sum_r rP[r; J]$.}

In the above argument, $\xi$ enters the picture by determining the asymptotic fate of this probability at a large enough system size $L^*(\xi)\propto \xi$.
One may wonder then why we do not observe in the ED data that $\doubleE[\ctc]$ saturates, as one would presume that $\xi$ is small and thus $L^*(\xi)$ is small. 
There are multiple factors that are in play here.
First, the $L\sim O(\xi)$ at which the CTC will saturate has a prefactor which sensitively depends on $c_{\text{diag}}$ and $b_{\text{collar}}$.
We do not attempt to determine such numbers precisely in this paper. 
Furthermore, the dependence of $\xi$ on $W$ is very weak in the Anderson insulator [see Eq.~(\ref{eq:anderson_ll})], and $\xi$ is nearly constant across $8 \leq W \leq 12$. 
Thus, if one of the $W$ values in this range has a CTC that saturates at $L$ values much larger than what is accessible to ED, then most of the $W$ values investigated in numerical studies will also not saturate.
We see this quantitatively, where it is apparent that the current system sizes are not close to $L^*(\xi)$ at those $W$ values. 
Consequently, one will not see the CTC saturate for this range of $W$ considered to be deep in the MBL regime.
However, if one cranks up the disorder strength to be exceedingly large, then the CTC is indeed expected to saturate with $L$ at numerically accessible system sizes. 
However, it is not desirable to perform numerics in a regime where the disorder is so large that the interactions effectively have no effect at numerically accessible system sizes.

It is worth discussing the differences between the toy model and the ED data at these numerically accessible system sizes.
One may first notice that the exponential growth of the CTC with $L$ in the full ED data of Fig.~\ref{fig:ctcb_main_results_J_01} contrasts with the apparent linear growth of the CTC in the toy model. 
We numerically verified that in our toy model, the probability of occurrence of a CTR transferring 1 particle across $i_0$ is already significant and grows approximately linearly in $L$, while for CTRs of 2 particles it grows exponentially with $L$.
We thus speculate that the linear growth of the CTC in the toy model reflects a delicate balancing act between quantitatively important resonances which are beginning to ``settle,'' and other resonances which are still growing. 
On the other hand, in the full ED data, we suspect that the exponential growth of the CTC highlights the presence of many quantitatively important CTRs not accounted for in the first-order perturbation theory, whose resonance collars are much larger than the accessible $L$.
Finally, it is also noteworthy that at large $W$, the toy model appears to overestimate the number of CTRs of 1 or 2 particles in the system, which can likely be attributed to the model's loose application of the resonance condition Eq.~(\ref{eq:resonance_condition}) (i.e. that any pair of LIOM configurations satisfying Eq.~(\ref{eq:resonance_condition}) are assumed to undergo \emph{perfect} particle transfer).

To summarize, the implications of our toy model's results are that at numerically accessible system sizes $L$, even if one tunes the disorder strength to be very strong into what is presumed to be the MBL regime, there will apparently to be larger resonances appearing as one increases $L$.
Even though they may have a very small probability, once the number of disorder realizations studied is enough to see them even just a few times, one will see their numbers growing with $L$.
We suggest that these resonances of growing size contribute to what appears to be finite-size drifts towards thermalization observed at these system sizes.
We note that this drift at finite sizes could still be consistent with the resonances eventually having a suppressed presence at large enough $L$ and a true MBL phase, but this suppression would not be visible at numerically accessible system sizes.
On the other hand, in the finite-size MBL regime, which eventually thermalizes at large $L$, such growing resonances may be harbingers of an eventual flow to true thermalization on long length scales.
These growing short-ranged resonances are an important factor to consider for numerical studies of MBL limited to system sizes that cannot see their asymptotic behavior.
We leave it to future work to study the direct consequences of these growing resonances on the physical properties of the finite-size MBL systems.

\subsection{Relation to many-body resonances}\label{sec:mbr_relation_to_ctr}
In this section, we briefly review the current understanding of many-body resonances (MBRs) in the literature and establish the relationship between MBRs and the CTRs identified in this work.
We then explain how the results in this section fit into the current accepted picture of MBRs.

The simplest form of a many-body resonance of range $r$ is an (approximately) equal-weight superposition of two or more LIOM configurations which differ in a spatial region of range $r$---a pair of eigenstates that are cat-like states.
The consequence of the presence of these types of simple many-body resonances is that if the system evolves under quench dynamics starting from one of the LIOM configurations, then the system can time evolve to the other LIOM configuration.
If one assumes the simple LIOM model with no resonances at some given disorder landscape, MBRs between such LIOM configurations unavoidably arise as one adiabatically varies the disorder, which induces forbidden level crossings between LIOM configurations~\cite{kjall_many-body_2018,villalonga_eigenstates_2020,garratt_local_2021}.
We note that in thermodynamically large systems, a given eigenstate in an MBL phase can plausibly be involved with many such simple resonances, and we can view a single pair of states discussed above (and often used below) as a building block in such a broader picture~\cite{garratt_local_2021}.
This is consistent with MBL as long as in a given eigenstate, the spatial density of resonances is low enough that the whole state is approximately a tensor product of such sufficiently isolated local resonances, i.e., that the resonances are not so spatially dense such that they begin to overlap~\cite{garratt_local_2021, crowley_constructive_2022}.
The CTC quantity specifically detects resonances that move charges across $i_0$, and therefore probes local resonances in a specific region of the chain.

This general definition of MBRs encompasses CTRs as defined in this work.
As we saw in Sec.~\ref{sec:ctc_model}, a CTR also arises from forbidden level crossings that occur when one turns on the interaction strength.
Specifically, a charge transport resonance of $m$ particles is simply an MBR of range $r$ with $r\geq 2 m$, and between the two product states, charge $m$ is transferred across the cut $i_0$.
On the other hand, not all long-ranged MBRs are necessarily CTRs that transfer charge.
For example, an MBR between the product states $|001100\rangle$ and $|010010\rangle$ is an MBR of range $r=4$, but would be a CTR of 0 particles, since no net charge was moved across the cut.
In some sense, CTRs can be thought of as categorizing MBRs in terms of the charge transfer across the cut $i_0$.

It must be clarified that the MBRs that have been extensively studied in much of numerical studies are \emph{system-wide} MBRs, which are MBRs with range $r$ of at least $L/2$ (only very few studies more carefully considered local resonances, see~\cite{garratt_local_2021, crowley_constructive_2022, garratt_resonant_2022}).
Typically, these system-wide resonances are detected by considering measures such as the mutual information and correlations between far-away spins.
Alternatively, the Jacobi method identifies resonances at different energy scales via large off-diagonal matrix elements which emerge from the RG-like flow under Jacobi rotations~\cite{morningstar_avalanches_2022,long_phenomenology_2023, colbois_statistics_2024, laflorencie_cat_2025, padhan_long-range_2025}.
Numerical results in the literature showed that system-wide resonances occur with negligible probability in a strong disorder regime for the system sizes accessible to ED.
This was used to conclude that the \emph{number} of system-wide resonances in the spectrum decreases with system size at very strong disorders~\cite{morningstar_avalanches_2022,laflorencie_cat_2025}.
Many of the charge transport resonances studied in this work would not be considered system-wide MBRs, as transporting 1 or 2 particles across the cut over a range of 2 or 4--5, for example, is a localized process around $i_0$ (we do discuss an example of a system-wide CTR of 2 particles in Appendix \ref{app:more_ctr_examples}, illustrated in Fig.~\ref{fig:example_3_ctr}).

Our work thus resolved microscopic details of \emph{short-ranged} MBRs and how they evolve with system size.
This allowed us to find that even though the resonances that transfer the most particles are indeed small at large disorder strengths (i.e., not system-wide), they systematically become \emph{larger} as the system size is increased. 
Unfortunately, as discussed earlier, numerical studies are fundamentally unable to see asymptotic resonance behavior due to the large sizes of resonance collars of quantitatively important resonances. 
We argued that this limitation is consistent with scenarios where MBL exists in the $L\rightarrow\infty$ limit, but it must be kept in mind that this is also consistent with no MBL phase.
From the ED data alone, one cannot fundamentally tell the difference between these two scenarios---one where the resonance growth is a transient feature of small system sizes, or whether these resonances will eventually grow to be system-wide at a large enough system size.
Such is the curse of using numerics on small system sizes in this very challenging problem.

\subsection{The growth of the CTC in an MBL phase}
With this background on resonances, we now close this section by discussing different possibilities on how the CTC could scale with $L$ in the MBL phase as $L\rightarrow\infty$.
We argue that in the MBL phase, the CTC will be $O(1)$ with the system size, just like in the Anderson insulator. We combine two separate facts: one is that in the naive LIOM model, where LIOMs have strictly exponentially decaying tails, the CTC will be finite with $L$.
Furthermore, if one modifies the LIOM model to include MBRs which occur with probability suppressed exponentially in their size (i.e., the model of resonances resulting from the argument in Sec.~\ref{sec:counting_resonances} and in Ref.~\cite{garratt_local_2021}), then the CTC would still be finite with $L$.

\subsubsection{The naive LIOM model}\label{sec:l-bit-model-results}
First, we discuss how the LIOM model of MBL with truly quasilocal LIOMs (truly quasilocal in the sense of Lieb-Robinson bounds) implies that the CTC should be finite in the $L\rightarrow\infty$ limit in the MBL phase.
Specifically, it is conjectured that in the MBL phase, one can rewrite $\hat H$ in the form of the LIOM model~\cite{serbyn_local_2013, huse_phenomenology_2014}:
\begin{equation}\label{eq:liom_equation}
    \hat H= \sum_{\alpha} \widetilde \epsilon_\alpha \widetilde n_\alpha + \sum_{\alpha>\beta} J_{\alpha,\beta} \widetilde n_\alpha \widetilde n_\beta+\sum_{\alpha> \beta>\gamma} J_{\alpha, \beta, \gamma} \widetilde n_\alpha \widetilde n_\beta \widetilde n_\gamma + ... ~.
\end{equation}
Here, $\widetilde n_\alpha$ are referred to as the (quasi)-local integrals of motion (LIOMs), otherwise known as $\ell$-bits.
These LIOMs are conserved quantities (i.e., $[\widetilde n_{\alpha}, \hat H]=0$), and it is assumed that they can be constructed from the local operators $\hat n_i$ as $\widetilde n_{i}=\hat U^\dagger \hat n_i\hat U$, using a quasilocal unitary $\hat U$.
Here, $\widetilde n_\alpha$ are quasilocal, and were thought to be ``dressed'' versions of the non-interacting LIOMs (which are simply occupation numbers of the original Anderson-localized orbitals)~\cite{bauer_area_2013}.

The LIOM model constrains the behavior of the CTC in the thermodynamic limit, and in particular would be incompatible with an $O(L)$ CTC. 
When the LIOMs are strictly finite-ranged, we proved in Appendix~\ref{app:derivation_local_m_lioms} that the CTC operator is also finite-ranged around $i_0$, meaning that the disorder-averaged CTC is finite in the thermodynamic limit. 
In the case of quasilocal LIOMs, it was argued in our previous work Ref.~\cite{jiang_quasiconservation_2025} that $\hat{M}$ should likewise be quasilocal around $i_0$, though a rigorous analytical proof remains beyond the scope of this work. 
Instead, we verified this claim numerically for LIOMs that are quasilocal in the sense of the Lieb--Robinson bound~\cite{lieb_finite_1972}. We refer the reader to Appendix~\ref{app:LIOM_model_results} and Fig.~\ref{fig:l-bit-model-results} for details of our CTC calculations in an artificial LIOM model, in which the LIOMs are constructed using short-time evolution under a local Hamiltonian with disorder.
We find that the distributions of the CTC in this artificial LIOM model are Anderson-like, and that the convergence of this distribution with $L$---or, equivalently, the saturation of the disorder-averaged $\ctc$---is visible at \emph{numerically accessible system sizes} (i.e. $L\leq 16$). 

Despite the finite-size limitations, our data for the CTC in Fig.~\ref{fig:ctcb_main_results_J_01} already rules out the naive truly quasilocal LIOM description for the XXZ model where the LIOMs are simply smooth rotations from trivial local operators.
This is because if the naive effective model were true, then we should observe saturation of the CTC at these system sizes, which we do not.
A more accurate LIOM model in the MBL phase thus must encode some degree of nonlocality into the LIOMs--- in particular, the presence of CTRs.

\subsubsection{CTC in the presence of resonances}\label{sec:ctcb_with_resonances}
Here, we briefly discuss the possible ways the CTC can scale with $L$ when the truly quasilocal LIOM model is modified to take into account the presence of CTRs, similar to what was done in Ref.~\cite{garratt_local_2021}.
By this, we specifically mean the following: we assume that we start with the truly quasilocal LIOM model, and now CTRs are induced between configurations of the truly quasilocal LIOMs due to some physical mechanism (for instance, by varying the disorder landscape/strength as in Refs.~\cite{villalonga_eigenstates_2020,garratt_local_2021}, or via local perturbations as in Ref.~\cite{crowley_constructive_2022}).
Now, the probability that two configurations come into resonance is controlled by some underlying probability density function which could depend on many variables, including the system size, the range of the resonance, parameters of the model, etc. 
One would imagine that this probability distribution must be controlled, i.e., that arbitrarily long-ranged resonances are sufficiently suppressed. 
One such scenario is the one described in Sec.~\ref{sec:counting_resonances}, where the probability of some disorder realization $\{h_i\}$ having a resonance of range $r$ straddling $i_0$ is proportional to
\begin{equation}\label{eq:resonances_probability_all_ranges}
    P[r;J, W]\sim \frac{J}{W} r2^re^{-r/\xi}2^{b_{\text{collar}}r}
\end{equation}
Here, $r2^r$ provides an order-of-magnitude estimate of the number of distinct active resonance configurations of range $r$ which intersect with $i_0$, while the factor $e^{-r/\xi}2^{b_{\text{collar}}r}$ is what we derived previously to be the probability of a given active configuration of range $r$ occurring in a randomly chosen disorder realization.
In this case, for small enough $\xi$, $P[r; J, W]$ is exponentially small in $r$ as $P[r; J, W]\sim e^{-cr}$.
In this case, the disorder-averaged CTC at system size $L$ is upper bounded by (assuming $L$ is very large)
\begin{equation}
\lim_{L\rightarrow\infty}\doubleE[\ctc]\leq \sum_{r=2}^{\infty} rP[r; J, W] ~.
\end{equation}
The right-hand side is a finite value as long as $\xi$ is small enough, and so the CTC will be $O(1)$ in the limit $L\rightarrow\infty$.

At the MBL-ETH transition, we speculate from Eq.~(\ref{eq:resonances_probability_all_ranges}) that a form of scale invariance emerges---resonances of any range $r$ are equally probable, in the sense that there is no exponential dependence on $r$, while there may still be polynomial dependence on $r$. 
(Such scale-invariance of the resonances was the claim of Ref.~\cite{villalonga_eigenstates_2020}.) 
At the transition, then, the CTC will \emph{grow} with $L$ due to the presence of system-wide resonances.

We discussed one scenario of the presence of many-body resonances that arises from our diagonally-improved perturbative model in which the CTC is finite.
However, we cannot exclude scenarios that have a larger presence of resonances, which could still be consistent with stable MBL.
One possibility, for instance, is for the average \emph{number} of system-wide resonances to grow with the system size, while remaining negligible compared to the Hilbert space size $2^L$.
In this case, the CTC will grow at least as $O(\sqrt{L})$, arising from the fluctuation of charges to the right of $i_0$ between two random charge configurations of length $O(L)$.
Although our model does not suggest that there is such a regime, we cannot rule it out as a scenario near the MBL-ETH transition.

Taken together with our discussion of the LIOM model, we speculate that the CTC is $O(1)$ with the system size in the MBL phase, and $O(L)$ in the critical and ETH phases. 
The ED data cannot see this saturation at the presently accessible sizes due to the continued growth of resonances.
\section{Charge Transport in Average States}\label{sec:charge_transport_typical_states}
The biggest unanswered question from the last section is whether or not these unsettled CTRs will eventually \emph{proliferate}, becoming system-wide and taking up a nonvanishing fraction of the Hilbert space. 
As mentioned previously, the fraction of resonances which transport $O(L)$ charge across the cut in the Hilbert space, of dimension $2^L$, should vanish in the $L \to \infty$ limit in order for MBL to be stable.
Specifically, the remaining questions are as follows:
Firstly, is there some way to quantify how important the expanding CTRs are for destabilizing MBL? 
Secondly, we now understand the worst-case scenario charge transport from studying the CTC; how can we probe the \emph{average} scenario for transport?

We will not be able to provide a definitive answer in this work, but we make a step towards this direction using two separate measures:
First, we introduce and study the normalized Frobenius norm of the CTC operator, which serves as a measure of \emph{average} transport.
Second, we introduce a quantity we named the \emph{state charge transfer bound} (SCTB), which serves as an upper bound for the transported charges across $i_0$ for a fixed initial state, and study it for all product initial states. 
Using both of these measures, we find evidence that the disorder average of these quantities is $O(1)$ in system size for strong enough disorder, in contrast to the expected $O(\sqrt{L})$ scaling in ergodic systems.
This provides some evidence that the presence of growing resonances at accessible system sizes may be consistent with a stable asymptotic MBL phase.

\subsection{Normalized Frobenius Norm of $\hat M$}\label{sec:ctcb_frobenius_norm}
\begin{figure*}[t]
    \centering
    \includegraphics[width=0.85\linewidth]{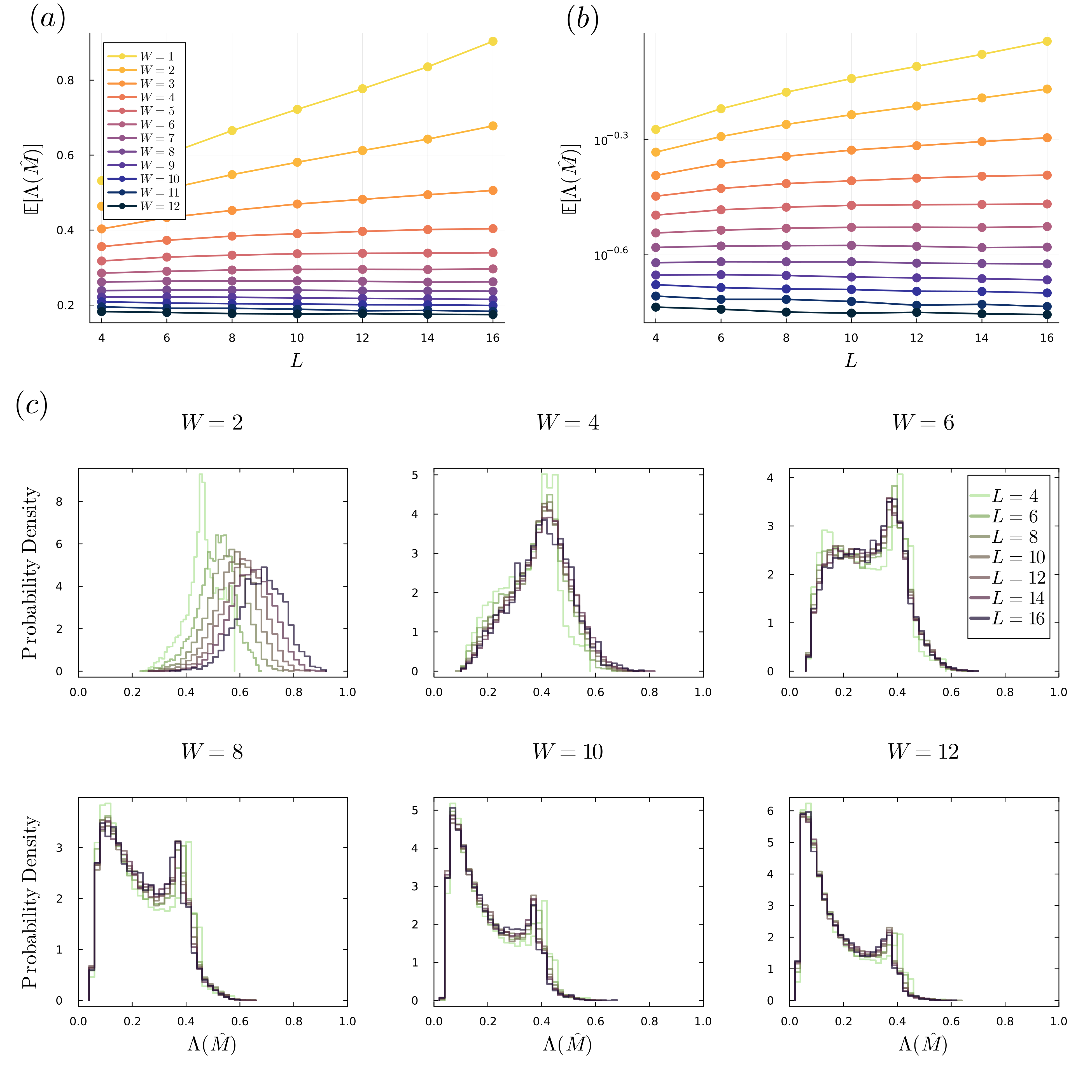}
    \caption{ 
    Analysis of the normalized Frobenius norm of the charge transport capacity operator in the half-filling sector, $\|\hat M\|_F^{\text{half-filling}}$, for the interacting Anderson model with $J=0.1$.
    Panel (a): the disorder-averaged $\|\hat M\|_F^{\text{half-filling}}$ ($y$-axis) as a function of the system size $L$.
    Panel (b): the same data as in (a), but with a log-log scale.
    Each curve represents a different disorder strength.
    Panel (c): the probability density functions of $\|\hat M\|_F^{\text{half-filling}}$, with each panel representing different disorder strengths.
    In a given panel, each colored curve represents the system size indicated in the legend.
    For these plots and at $L\leq 14$, 18000 disorder realizations were used for $W\leq 4$, and 22000 disorder realizations were used for $W\geq 5$.
    For $L=16$, $8000$ disorder realizations were used per disorder strength for $W 
    \leq 4$, whereas between $16000$ and $18000$ disorder realizations were used for $W\geq 5$.
    }
\label{fig:frobenius_norm_j_01}
\end{figure*}

The CTC exclusively contains information about the extremal eigenvalues of $\hat M$, thus making it a probe of extremal cases of transport. 
However, there is useful information contained in other eigenvalues in the spectrum of $\hat M$ about other transport processes in the system.
An aggregate measure of all of the eigenvalues of $\hat M$ is its normalized Frobenius norm:
\begin{equation}
    \|\hat M\|_F=\sqrt{\frac{1}{\mathcal{D}}\mathrm{Tr}[\hat M^{\dagger}\hat M]}~.
\end{equation}
Here, $\mathcal{D}$ is the total Hilbert space dimension.
We are interested in the above {\it normalized} Frobenius norm, as this results in localized operators having $O(1)$ norm. 
For ergodic systems, we conjecture that the Frobenius norm of $\hat M$ generally scales as:
\begin{equation}\label{eq:ergodic_system_scaling}
    \|\hat M\|_F \sim \frac{\sqrt{L}}{4} ~.
\end{equation}
We proved this result in the clean free fermion model~\cite{jiang_quasiconservation_2025}.
Numerically, it appears to hold for systems obeying the ETH, which we verified numerically by considering eigenstates drawn from the GOE and also by studying a separate local 1D model with charge conservation, c.f.\ Appendix~\ref{app:goe_results}.
This result has an intuitive understanding (once we appreciate the relation of the Frobenius norm to average transportable charge, which we will discuss later):
Recall that an average product state at half-filling will have $\frac{L}{4} + O(\sqrt{L})$ particle number in each half, and in an ETH system we expect $O(\sqrt{L})$ fluctuations in this number at subsequent times.
A scaling of $\|\hat M\|_F$ differing from the typical scaling, therefore, is a strong indication of an ETH violation.
In particular, if $\hat M$ is quasilocal, its normalized Frobenius norm is expected to be finite as $L\rightarrow\infty$, which would be a strong signature of ergodicity breaking.
A relevant and closely related quantity is the ``half-filling'' normalized Frobenius norm of $\hat M$:
\begin{equation}
    \|\hat M\|_F^{\text{half-filling}}=\sqrt{\frac{1}{\mathcal{D_{\text{half-filling}}}}\mathrm{Tr}[\hat P_{\text{half-filling}}\hat  M^{\dagger}\hat M\hat P_{\text{half-filling}}]}.
\end{equation}
Here, $\hat P_{\text{half-filling}}$ is the projector onto the half-filling sector, and $\mathcal{D_{\text{half-filling}}}$ is the Hilbert space dimension of this sector $\binom{L}{L/2}$.
We observe numerically that $\|\hat M\|_F\leq \|\hat M\|_F^{\text{half-filling}}$, at least for the system sizes accessible to ED.
Thus, we expect that if $\|\hat M\|_F^{\text{half-filling}}$ shows signs of saturation, then so will the full Frobenius norm.
Regardless, since the majority of states lie in the half-filling sector, we expect $\|\hat M\|_F^{\text{half-filling}}$ to be representative of the full-Frobenius norm.

The physical operational meaning of the Frobenius norm is less clear-cut than that of the CTC, but it can be interpreted as an upper bound on an averaged charge transport over all states, in a sense made precise as follows.
Consider any orthonormal basis $\{|n\rangle \}$, and fix a time $t$.
Now, we compute the absolute value of the expectation value of $\Delta\hat N_R(t)=e^{i\hat H t}\hat N_Re^{-i\hat H t}-\hat N_R$, averaged over all basis states.
Observe that
\begin{align}
    &\frac{1}{2^{L}}\sum_n |\langle n|\Delta\hat N_R(t)|n\rangle|=\frac{1}{2^{L}}\sum_n \left|\langle n|\Delta\hat M(t)| n\rangle\right|\nonumber \\
    &\leq \frac{1}{2^{L}}\sum_n \left(\left|\langle n|\hat M(t)| n\rangle\right| + \left|\langle n|\hat M(0)| n\rangle\right|\right) \leq 2\|\hat M\|_F~.
\end{align}
Here, we used simple inequalities between arithmetic and quadratic means, $\frac{1}{\mathcal{D}} \sum_{n=1}^\mathcal{D} |A_{nn}| \leq \sqrt{\frac{1}{\mathcal{D}} \sum_{n=1}^\mathcal{D} |A_{nn}|^2} \leq \|A\|_F$, and the fact that the time-evolved orthonormal states $\{ e^{-i\hat H t}|n\rangle \}$ still form an orthonormal basis.
The crucial thing here is that all the basis states are time-evolved up to the same time. Thus, the Frobenius norm of the CTC operator can be interpreted as a measure of the charge transport in an average state.\footnote{One may be tempted to interpret this as an infinite-temperature ensemble average, but this is not true since this quantity is basis-dependent because of the absolute values in the definition.
Nevertheless, one can imagine running an experimental protocol measuring such a quantity.}

Finally, but most importantly, the behavior of the Frobenius norm of $\hat M$ as $L\rightarrow\infty$ also provides information about the fraction of states involved in resonances. 
The Frobenius norm of an operator is basis independent, but for the sake of this argument, consider $\hat M$ as a matrix written in the Hamiltonian eigenstate basis $\{ |n\rangle \}$, so that
\begin{equation}\label{eq:frobenius_norm_n_r}
    \|\hat M\|_F=\sqrt{\frac{1}{\mathcal{D}}\sum_{n\neq m}|\langle n|\hat N_R|m\rangle|^2}~.
\end{equation}
For a pair of eigenstates $|n\rangle, |m\rangle$ involved in a charge transport resonance of $\Delta N\sim O(L)$ particles as in Eq.~(\ref{eq:resonance_structure}), one can calculate that their corresponding off-diagonal matrix element $\langle n|\hat N_R|m\rangle$ will be of order $\Delta N$. 
To see this, recall that in a perfect resonance, the eigenstates take the form $|n\rangle = \frac{1}{\sqrt{2}}(|a\rangle + |b\rangle)$ and $|m\rangle = \frac{1}{\sqrt{2}}(|a\rangle - |b\rangle)$, where $|a\rangle$ and $|b\rangle$ are approximate computational basis states with $\hat{N}_R |a\rangle = a|a\rangle$ and $\hat{N}_R |b\rangle = b|b\rangle$. A direct calculation then gives
\begin{equation}
    \langle n|\hat{N}_R|m\rangle  = \frac{1}{2}(a - b) \sim \Delta N,
\end{equation}
where $\Delta N = |a - b|$ is the number of particles transported across the cut. More generally, for an imperfect resonance, the prefactor will differ from $1/2$, but the matrix element remains $O(\Delta N)$.
If the fraction of CTRs in the spectrum (out of $2^L$ states) with $\Delta N\sim O(L)$ was nonvanishing as $L\rightarrow\infty$, then they will contribute a $O(L)$ term in the sum of Eq.~(\ref{eq:frobenius_norm_n_r}).
Thus, this would cause the Frobenius norm of $\hat M$ to grow with $L$. Thus, an $O(1)$ Frobenius norm of $\hat M$ as $L\rightarrow\infty$ would imply that the fraction of resonating states transferring $O(L)$ charge vanishes with $L$.

Here, due to computational ease, we compute $\|\hat M\|_F^{\text{half-filling}}$ instead of the full Frobenius norm. 
Figure~\ref{fig:frobenius_norm_j_01} shows the results for $\|\hat M\|_F^{\text{half-filling}}$ for the interacting Anderson model at $J=0.1$.
For calculations of $\|\hat M\|_F^{\text{half-filling}}$, but with parameters equivalent to the literature-studied XXZ chain, see Appendix~\ref{app:large_interaction_strength}.
For small disorder strengths $W\leq 4$, in what is clearly the ergodic regime, the disorder-averaged Frobenius norm appears to grow with $L$ with a power-law, as expected. 
However, for very large disorder strengths ($W\geq 8$), one can see that the distributions of $\|\hat M\|_F^{\text{half-filling}}$ appear to converge with $L$, which is further supported by the disorder-average of $\|\hat M\|_F^{\text{half-filling}}$ saturating very quickly with $L$. There is an intermediate regime, roughly $6<W<8$, where the normalized Frobenius norm is seen to grow with $L$ at these system sizes, but at a very slow rate. 

If the trends for $\|\hat M\|_F$ continue to larger $L$, that would indicate ergodicity breaking for strong enough disorder strength $W$. 
In this work, we do not claim to observe a specific $W^*$, as it is impossible to rule out a very slowly increasing $\|\hat M\|_F$ due to noise from statistical fluctuations.
Furthermore, as mentioned beforehand, one must be cautious about claiming that the behavior of resonances is settled at our finite system sizes, since one cannot access system sizes that would rule out resonances building up to become system-wide.

\subsection{Charge transport in average product states}\label{sec:sctb}
In this section, we introduce a new measure that upper bounds the charge transfer in the setting where a specific initial condition is fixed.
Here, charge transport refers to possible changes in $\langle\hat N_R\rangle$ at any later (or prior) time after the system is initiated in this state.
Instead of placing an upper bound on $\mathcal{Q}$ over all possible initial states as with the CTC, we can instead restrict ourselves to a particular initial condition $\ket{\psi_0}$ and ask for its maximum transportable charge.
We call this upper bound the \emph{state charge transfer bound} (SCTB).

\subsubsection{The state charge transfer bound}\label{sec:ctcb}
Consider a Hamiltonian $\hat H$ with energy eigenvalues $E_n$ and eigenstates $|n\rangle$.
Now, prepare the initial quantum state $\ket{\psi(t=0)} \equiv\ket{\psi_0}$ and time evolve it under the dynamics of the Hamiltonian $\hat H$.
Let us expand the state in the eigenbasis of the Hamiltonian, $\ket{\psi_0} = \sum_n c_n \ket{n}$.
We can calculate the change in the expectation value of $\hat N_R$ at time $t$ and then upper-bound it as
\begin{align}
\label{eq:time_evolve_n_r}
\langle \Delta\hat N_R(t) \rangle &= \sum_{n,m} c_n^* c_m e^{i(E_n-E_m)t} \langle n|\hat N_R|m\rangle - \langle \hat N_R(0)\rangle \nonumber \\
&\leq \sum_{n,m} |c_n^* c_m| |\langle n|\hat N_R|m\rangle| - \langle \hat N_R(0)\rangle =: B_R ~.
\end{align}
Similarly for $\hat N_L$,
\begin{equation}
\langle \Delta\hat N_L(t) \rangle \leq \sum_{n,m} |c_n^* c_m| |\langle n|\hat N_L|m\rangle| - \langle \hat N_L(0)\rangle =: B_L ~.
\end{equation}
Since $\langle \Delta\hat N_L(t) \rangle = -\langle \Delta\hat N_R(t) \rangle$, we have
\begin{align}\label{eq:dctc_equation}
& |\langle \Delta\hat N_R(t)\rangle| = \max\{\langle \Delta\hat N_R(t)\rangle, \langle \Delta\hat N_L(t)\rangle\} \nonumber \\
& \leq \max\{ B_R, B_L \} =: \eta(\ket{\psi_0})~,
\end{align}
which we call the state charge transfer bound.
(Note that both $B_R$ and $B_L$ are non-negative, but need not be the same:
They arose in {\it upper-bounding} the change in $N_R$ and $N_L$ respectively, and such bounds need not be related even though $N_R + N_L = L/2$ is fixed.)

For a generic $\ket{\psi_0}$ and many-body Hamiltonian, $\eta(\ket{\psi_0})$ appears to be an extremely large quantity, as it is a double sum over exponentially many positive terms. 
Thus, there is no reason that one should expect the SCTB to be a good bound in these cases.
Indeed, for a system obeying ETH, it is generically the case that $\eta(\ket{\psi_0})$ increases exponentially with $L$ for an initial condition which is a generic product state.
A rough argument goes as follows: Assuming the ETH ansatz, and that $\ket{\psi_0}$ has a finite energy density, one expects that typical matrix elements of a local observable are of order $\sim O(1/\sqrt{\mathcal{D}})$. Thus, as $\hat N_R$ is a sum of local terms, $\langle n|\hat N_R|m\rangle\sim O(1/\sqrt{\mathcal{D}})$, here and below ignoring powers of $L$ compared to $\mathcal{D} = 2^L$. Furthermore, for a general product state expanded in the eigenstate basis, $c_n\sim O(1/\sqrt{\mathcal{D}})$. 
Thus, each member of the sum Eq.~(\ref{eq:dctc_equation}) is of order $O(\frac{1}{\mathcal{D}^{3/2}})$. 
There are $\mathcal{D}^2$ such terms in the sum, so overall we expect $\eta(\ket{\psi_0})\sim O(\mathcal{D}^{1/2})$. 
Thus, this rough estimate tells us that $\eta(\ket{\psi_0})\sim 2^{L/2}$ for large $L$. 
We verified numerically in Appendix~\ref{app:goe_results} that the SCTB grows exponentially with $L$ in thermalizing systems, through the examples of the Gaussian Orthogonal Ensemble and a local thermalizing chain of spinless fermions.
We expect this scaling to hold generically for all ergodic systems. 

It is impossible in many situations for $\eta(\ket{\psi_0})$ to be a tight bound (i.e., that at some time $t$, the initial state moves an amount of charge across $i_0$ equal to its SCTB). This is because this requires that $e^{i(E_n - E_m)t} = \text{sign}(c_n c_m \langle n | N_R |m \rangle)$ (assuming a real-valued problem), for all $n, m$ such that $c_n c_m \langle n | N_R |m \rangle > 0$.
It is easy to come up with simple situations where these conditions can never be achieved.\footnote{One such example is if there is a triple $n,m,k$ such that the product of $\langle n|\hat N_R|m\rangle, \langle m|\hat N_R|k\rangle, \langle k|\hat N_R|n\rangle$ negative while $c_n, c_m, c_k$ are non-zero, since in this case $g_{nm}(t) g_{mk}(t) g_{kn}(t) = |c_n|^2 |c_m|^2 |c_k|^2 \langle n|\hat N_R|m\rangle \langle m|\hat N_R|k\rangle  \langle k|\hat N_R|n\rangle < 0$ at any $t$ and cannot be $|g_{nm}(t)| |g_{mn}(t)| |g_{nk}(t)|$. 
Then, it is clear that the SCTB in this case can never be tight. }
The usefulness of the SCTB would therefore be in a scenario in which the SCTB is constant with $L$ for average product states as initial conditions. This can only occur if the initial condition $\ket{\psi_0}$ is such that only an $O(1)$ number of energy eigenstates have significant $O(1)$ weight on $\ket{\psi_0}$. This is the first hint that this quantity may be useful to probe the structure of the energy eigenstates in the MBL regime, which are thought to have only significant weight on a few product states. 
We will demonstrate its behavior in the interacting Anderson model in the next few sections and use it as a probe of the density of resonances in the spectrum.

\subsubsection{Melting of the domain-wall state}\label{sec:domain_wall_melting}
\begin{figure}[t]
    \centering
    \includegraphics[width=\linewidth]{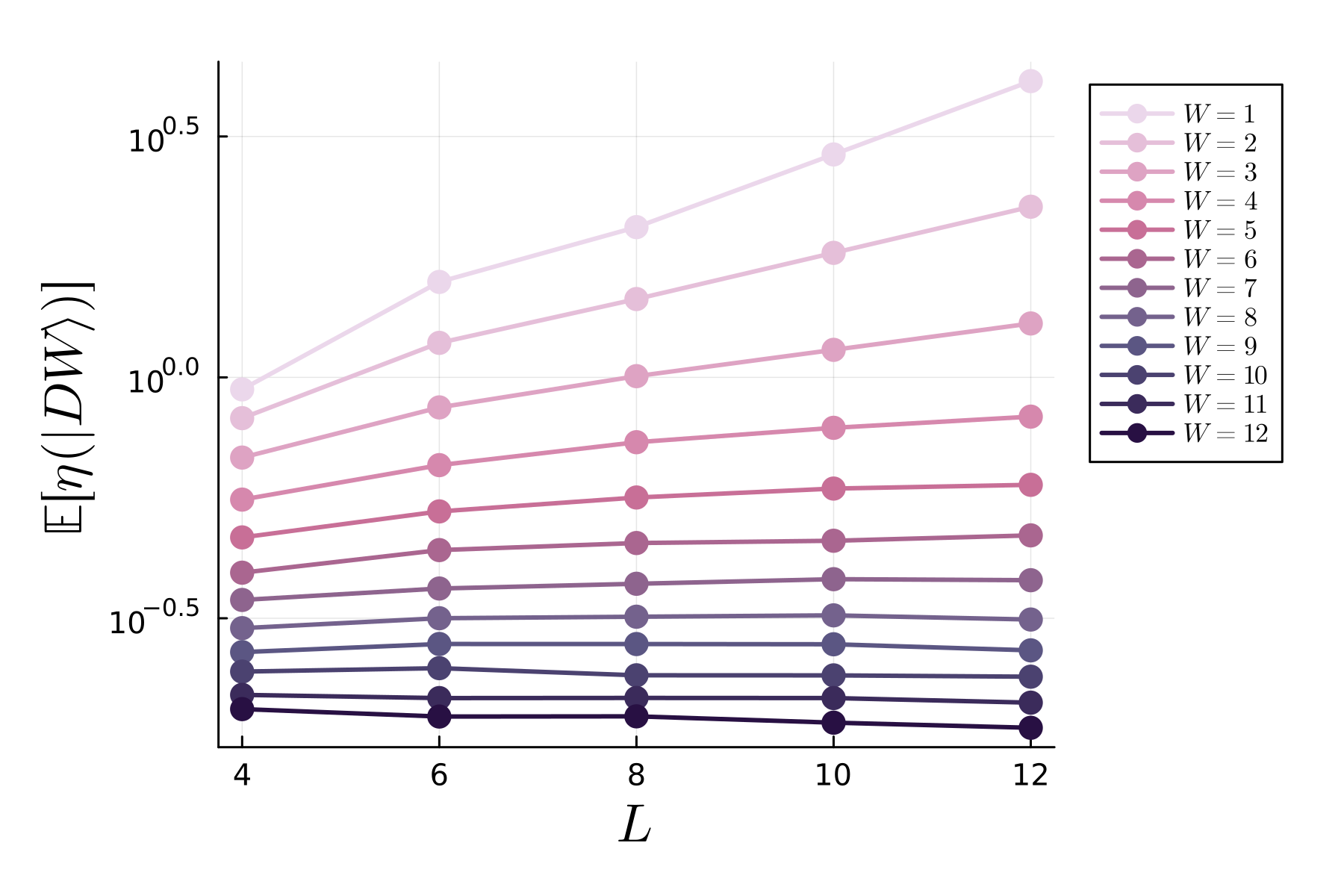}
    \caption{The disorder averaged $\eta(\ket{\text{DW}})$ as a function of the system size $L$ for the interacting Anderson model with $J=0.1$. Here, $\ket{\text{DW}}$ is the domain wall state.
    The $y$-axis is on a log scale. Each colored curve represents a different disorder strength, ranging from $W=1$ to $W=12$ in steps of 1.
    Each data point is the average over 2000 disorder realizations for $W<7$, and $12000$ disorder realizations for $W\geq 7$.} 
    \label{fig:dctc_DW_state.png}
\end{figure}
To gain some intuition for the SCTB, let us first consider the SCTB calculated for a specific product state which might naively transport the most charge---the ``domain-wall state'' $\ket{\text{DW}}$:
\begin{equation}
    \ket{\text{DW}} = \ket{11 \dots 10\dots 00}~,
\end{equation}
where all sites to the left of position $i_0=\lfloor L/2\rfloor$ are 1s, and the rest 0s.
This denotes a state where all $\lfloor L/2\rfloor$ particles are placed in the left half of the system.

In ergodic systems, this state should time-evolve to one where the expectation value of $\hat N_R$ is around $L/4$ due to their memoryless property
(note that the initial melting of the domain wall may be slow since the hopping terms act trivially away from the domain wall; however, our bounds are valid for all times including infinitely long, where this initial slow melting is not important).
In contrast, if one is in the MBL regime, all of the eigenstates are close to product states. Thus, the state above would be only a superposition of a few eigenstates, causing the system to retain its memory of the initial state $\ket{\text{DW}}$, leaving only $O(1)$ charge transferred across $i_0$.
This makes the SCTB an interesting quantity to study in this context.

We note that the problem of domain wall melting has been studied previously in Refs.~\cite{hauschild_domain-wall_2016, panda_can_2020}, and is particularly relevant for the context of cold atom experiments, where preparing such a state is natural~\cite{choi_exploring_2016}.
The SCTB approach is a different perspective on the same problem, where we bypass the need for long-time evolution by placing an absolute upper bound on $\langle\Delta N_R(t)\rangle$ valid up to infinite time, at the cost of requiring full exact diagonalization.

We show the analysis of the disorder-averaged SCTB for $\ket{\text{DW}}$ in Fig.~\ref{fig:dctc_DW_state.png}.
As mentioned before, deep in the ergodic regime, the SCTB is a poor bound, and we correspondingly observe that $\mathbb{E}[\eta(\ket{\text{DW}})]$ scales exponentially with the system size.
On the other hand, one can see that $\mathbb{E}[\eta(\ket{\text{DW}})]$ appears to be independent of $L$ (for the system sizes considered) for large disorder strengths. 
Assuming that the disorder-averaged SCTB for this product state stays saturated with $L$ for large enough disorder strengths, this would show that this domain wall state will never melt, i.e., it can only transfer $O(1)$ charge even in the infinite system size limit, which is a strongly nonergodic behavior.
In principle, a quench experiment starting from $\ket{\text{DW}}$ tracking the change in the amount of charge to the right using imaging is feasible in state-of-the-art cold atom or ion trap experiments~\cite{lukin_probing_2019, rispoli_quantum_2019, leonard_probing_2023}.
It would be interesting to see how this upper bound holds at much larger system sizes in experimental platforms that are inaccessible to numerics.

\subsubsection{State charge transfer bound for all product states}
\begin{figure*}[t]
    \centering
    \includegraphics[width=\linewidth]{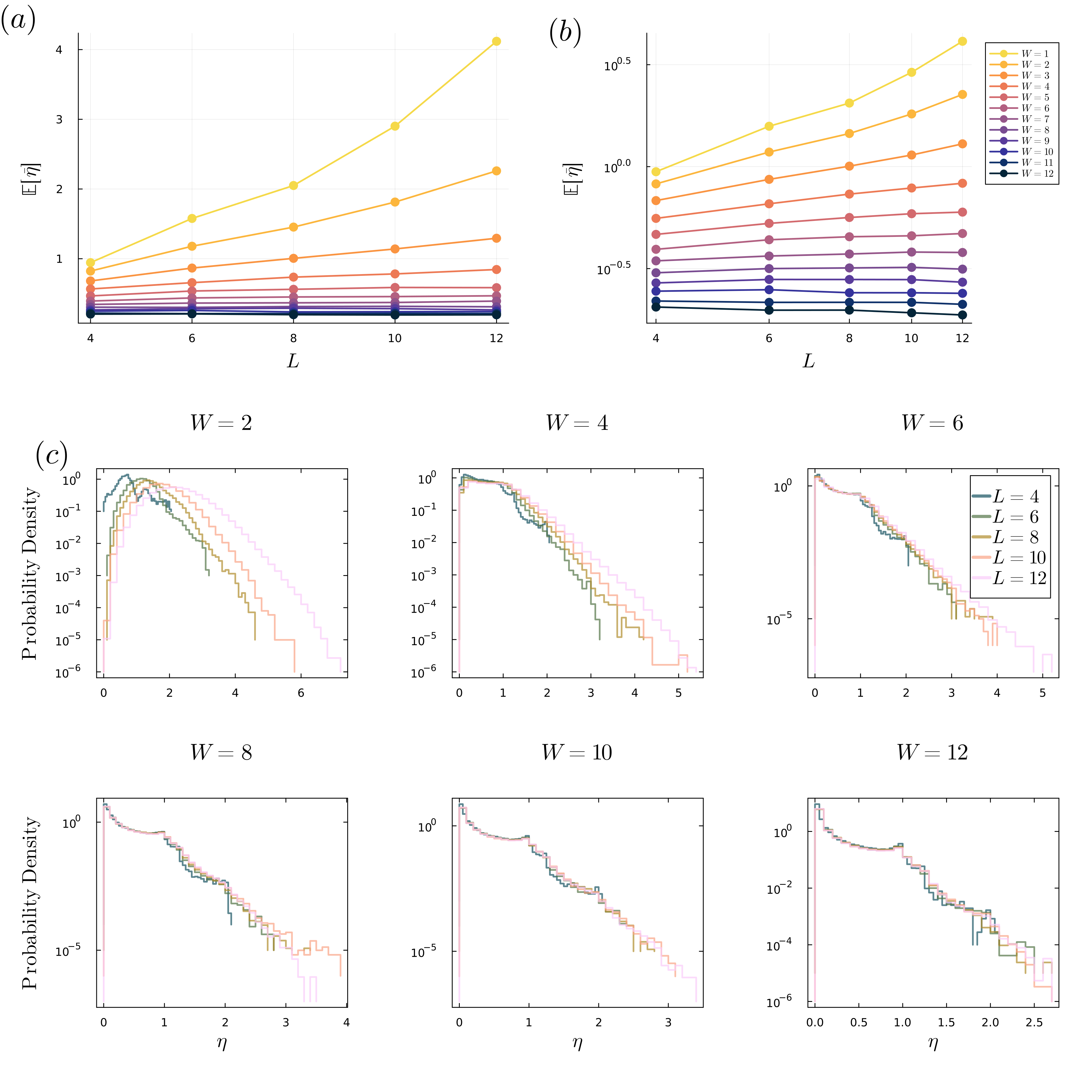}
    \caption{Analysis of the SCTB $\eta(\ket{a})$ for every product state $|a\rangle$ (i.e., computational basis state) in the interacting Anderson model with $J=0.1$.
    All of these plots were generated using 2000 disorder realizations for $W<4$, and $12000$ disorder realizations for $W\geq 4$.
    Panel (a): the SCTB averaged over all disorder realizations and $\binom{L}{L/2}$ product states in the half-filling sector as a function of $L$.
    Panel (b): the same data as in (a), but with a logarithmic scale on the $y$-axis.
    Panel (c): probability distributions of the SCTB over all product states and disorder realizations, each curve representing a different system size ranging from $L=4$ to $L=12$ in steps of 2, and each panel representing a different disorder strength.
    Note the logarithmic scale on the $y$-axis.}
    \label{fig:dctc_all_product_states.png}
\end{figure*}
Now, we are interested in an aggregate measure of how the SCTB behaves for all product states as $L\rightarrow\infty$. 
Let $|i\rangle$ with $i = 0, \dots, \binom{L}{\lfloor L/2\rfloor}-1$ index computational basis states in the half-filling sector, e.g., with occupations labeled by the binary representation of index $i$. 
We are interested in the SCTB for all of these product states under the Hamiltonian Eq.~(\ref{eq:interacting_hamiltonian}), the distribution of which is shown over 12000 disorder realizations in Fig.~\ref{fig:dctc_all_product_states.png}(a)--(d), for $J=0.1$.
We note that these distributions are cumulative distributions of $\eta$ over $\{h_i\}$ and $\ket{\psi_0}$, where $\{h_i\}$ is the disorder realization, and $\ket{\psi_0}$ is the starting product state. 
In particular, we do not have information about the distributions of $\eta$ for a given disorder realization, as the $\eta$ for different product states are highly correlated within a given sample.

In Fig.~\ref{fig:dctc_all_product_states.png}(e), we consider the SCTB averaged over all product states:
\begin{equation}\label{eq:sctb_averaged}
    \overline \eta = \frac{1}{\binom{L}{\lfloor L/2\rfloor}} \sum_{i=0}^{\binom{L}{\lfloor L/2\rfloor}-1} \eta(|i\rangle)~,
\end{equation}
and study how $\doubleE[\overline \eta]$ ($\doubleE[\cdot]$ denotes disorder averaging) grows with $L$ in Fig.~\ref{fig:dctc_all_product_states.png}(a) and (b). 
Here, panel (a) shows the data on a linear-linear scale, while panel (b) shows the same data with the y-axis on a log scale. 
At low disorder strengths ($W=2, 4$ as pictured), it is of no surprise that $\mathbb{E}[\overline \eta]$ grows exponentially with $L$, and the corresponding distributions shift consistently towards the right.
For an intermediate regime roughly corresponding to $6 \leq W \leq 8$, $\doubleE[\overline \eta]$ seems to increase very slowly with $L$ for the system sizes accessible to ED, but at a rate much slower than exponential growth. This suggests that the distributions of $\overline \eta$ slowly drift rightwards, even though it is not immediately visible just by inspecting them. For $W>8$, $\doubleE[\overline \eta]$ seems to be $O(1)$ with system size starting from $W=9$, with no apparent drift observed at the accessible system sizes for $12000$ disorder realizations. 

The saturation of $\doubleE[\overline \eta]$, if it persists to the thermodynamic limit, would have a lot of implications for the properties of an MBL system.
Firstly, this scaling contains information about the structure of the eigenstates. 
In particular, if $\doubleE[\overline \eta]$ remains $O(1)$ in the $L\rightarrow\infty$ limit, that would imply that the energy eigenstates must have only significant weight on only $O(1)$ computational basis states that only transfer $O(1)$ charge across the link between each other.
Otherwise, the averaged SCTB cannot be $O(1)$.
Secondly, $\doubleE[\overline\eta]$ also gives an aggregate measure of how many CTRs there are in the spectrum that transfer extensive amounts of charge. 
This is straightforward to see: consider a pair of product states $|a\rangle$ and $|b\rangle$ with eigenvalues of $\hat N_R$ being $a, b$.
Now, suppose there are a pair of energy eigenstates of the following form: $|E_1\rangle=\frac{1}{\sqrt{2}}(|a\rangle+|b\rangle),|E_2\rangle=\frac{1}{\sqrt{2}}(|a\rangle-|b\rangle)$, i.e. there is a CTR of $|b-a|$ particles.
It is easy to verify that:
\begin{align}
    \eta(\ket{a})=\eta(\ket{b})=|b-a|~.
\end{align}
Thus, a product state involved in a CTR that transports extensively many charges will contribute a $O(L)$ term to the sum in Eq.~(\ref{eq:sctb_averaged}).
If the fraction of CTRs in the spectrum that transport extensively many charges does not vanish as $L \to \infty$, then Eq.~(\ref{eq:sctb_averaged}) should grow with the system size.
On the other hand, if a vanishing fraction of CTRs transfers extensively many charges, then Eq.~(\ref{eq:sctb_averaged}) should be an $O(1)$ value.
These observations about $\doubleE[\overline \eta]$ are consistent with our picture of the occurrence of resonances in Sec.~\ref{sec:counting_resonances}.

We also observe from the data in Fig.~\ref{fig:dctc_all_product_states.png}(c) that the distributions of $\eta(\ket{a})$ over $\{h_i\}$ and $\{\ket{a}\}$ in the strong disorder regime appear to be converged to a distribution with an exponentially decaying tail. 
This implies that a random product state, for a system sampled from a random disorder realization, will transfer an amount of charge $m$ that is exponentially suppressed in $m$ across $i_0$.
The existence of $P[\eta]$ as a continuous function defined in the thermodynamic limit, with exponential decay at large $\eta$, is consistent with the phenomenological description of MBRs/CTRs in the stable MBL phase, even though at first glance it may appear that it would allow for a large number of resonances (scaling $\sim 2^L$).
To illustrate this point, note that the distribution of $\eta$ for $W=12$ appears to have a significant ``bump'' in the tail.
The ``bump'' represents a single disorder realization with effectively significantly smaller disorder (i.e., a ``rare instance''), causing all of the product states to have very large SCTB in that realization.
This is an illustration of an important point mentioned earlier, cautioning about the interpretation of the distributions of $\eta(\ket{\psi})$, which is accumulated over different initial states $\ket{\psi}$ and different realizations of disorder.
Namely, that instances of large $\eta(\ket{\psi})$ are strongly correlated:
If a given disorder realization does not have a local resonance near $i_0$, then for all initial states $\ket{\psi}$ the corresponding $\eta(\ket{\psi})$ will be small.
On the other hand, if there is a local resonance, then the same resonance will be present for any possible LIOM background states outside of the resonance collar, i.e., there will be $\sim 2^{L-|C|}$ initial states that will produce the same largish $\eta(\ket{\psi})$, where $|C|$ is the size of the collar (assumed to include the active sites as well).
Thus, these calculated distributions of $\eta$ are consistent with our analysis of the probability of resonance occurrence in Eq.~(\ref{eq:probability_resonances_asymptotic}) of Sec.~\ref{sec:counting_resonances}.

These points together emphasize that $\overline \eta$ is similar to the Frobenius norm in that it provides some measure for the presence of system-wide resonances. 
However, unlike $\|\hat M\|_F$, $\overline \eta$ has a nice feature in that it has a very clear physical operational meaning---it upper bounds the charge transport for all times for a fixed initial state, whereas this is less clear-cut for the Frobenius norm of $\hat M$. 
The data here provides some evidence for the existence of a nonergodic phase for strong enough disorder, in which the fraction of CTR pairs in the spectrum that transport extensive amounts of charge vanish for $L\rightarrow\infty$.
We once again emphasize that we do not attempt to find or even lower bound $W^*$.
At very large disorder strengths, resonances become very rare, and so one becomes more susceptible to noise in the data.
Thus, we cannot rule out a very slow growth of the SCTB, especially because we know from the previous results on the CTC that the behavior of the resonances is far from settled.

\section{Conclusions, Discussion, and Outlook}\label{sec:conclusion}
In this work, we introduced the charge transport capacity (CTC), which focuses on the worst-case (i.e., largest transfer) charge transport process across the central cut in the lattice.
We performed an ED study of the CTC in the interacting Anderson model, which is the prototypical model for MBL.
The results show that, in stark contrast to its non-interacting counterpart, the interacting model becomes more capable of transferring charge as the system size increases---even at very large disorder strengths presumed to be deep in the MBL regime.
We showed that this trend arises from interaction-induced charge transport resonances (CTRs).
These resonances are pairs of nearly degenerate eigenstates that are cat-like superpositions of product states that can transfer a net charge across the cut in oscillations between them.
Even at very strong disorder, charge transport resonances that transfer larger amounts of charge become increasingly more probable with increasing system size.

We then introduced a diagonally-improved first-order perturbation theory model of resonances.
This model reproduces the growth of these CTRs at numerically accessible $L$ in the strong disorder regime.
However, in the toy model, the CTC must saturate since it is limited to first-order processes.
The buildup observed at our system sizes, therefore, manifests strong finite-size effects as far as the CTC measure is concerned.
Consequently, the settled behavior of resonances remains inaccessible to exact diagonalization due to the large disorder-strength-independent resonance collars.
We suggest that this buildup of larger CTRs likely contributes to the numerically observed ``crossover'' regime, where MBL systems in some measures appear to drift towards thermalization at larger and larger system sizes.
On the other hand, we also argued that at large enough disorder strengths, the CTRs become exponentially suppressed in their range in this toy model.
We then suggest that such behavior could persist also when higher-order processes are taken into account.
This would be consistent with the current phenomenological picture of how resonances can coexist with MBL~\cite{garratt_local_2021, garratt_resonant_2022}.

We also study quantities which probe the \emph{average}, as opposed to \emph{worst-case}, charge transport, giving aggregate measures on the prevalence of these resonances and on their relative weight in the Hilbert space.
To this end, we studied the scaling of two quantities with $L$: the Frobenius norm of the CTC operator and the state charge transfer bound (SCTB) for an ensemble of initial states.
The former is an aggregate measure over all resonances rather than the single one maximizing charge transfer in a given sample, while the latter is a direct dynamical probe that is also sensitive to CTRs.
Our numerical evidence suggests that both these quantities are $O(1)$ with the system size past some critical disorder strength. 
In contrast, for weak disorder, these measured properties grow with system size and suggest the interpretation that the fraction of CTRs with extensive charge transfers in the spectrum does not vanish as $L \to \infty$, destabilizing localization at large enough system sizes.
In the strong disorder regime, these results are consistent with previous conclusions about system-wide resonances at these system sizes, and with a phenomenological description of stable MBL with local resonances, whose probability is exponentially suppressed with the resonance size.

Our work expands our understanding of many-body resonances and finite-size effects.
By working in the framework of the worst-case (i.e., largest) charge transfer, we found a rich zoo of physics in the growth of short-ranged CTRs at these finite system sizes. 
This expands on past numerical studies, which tended to focus on system-wide MBRs, and contributes to the understanding of local resonances beyond their cumulative effects.
Our diagonally-improved perturbation theory model provides quantitative and qualitative reasons for why the probabilities of larger CTRs grow with $L$ at these system sizes, even in the strong disorder regime.
Although the focus of our model is to explain observations at numerically accessible system sizes, it shows that there is an instability emerging from the proliferation of arbitrarily long-ranged resonances for large enough Anderson localization length $\xi$.
This can also be regarded as a starting point for a theory that describes the instability of the MBL phase to many-body resonances at both small $L$ and large $L$.
(We note that the statistics of resonances in the regime deep in the MBL phase at large $L$ were studied in Ref.~\cite{garratt_local_2021}. 
Also, similar to our work, Ref.~\cite{crowley_constructive_2022} focused on the numerically accessible regime; however,
it does not explicitly describe the transition at large $L$.)
We point out that it was previously assumed that finite-size effects are more suppressed as $W$ increases~\cite{abanin_distinguishing_2021}, but we showed that this is not the case for measures that are sensitive to MBRs.
Unfortunately, the large sizes of resonance collars of most quantitatively significant resonances mean that ED studies cannot probe the asymptotic behavior of these resonances.
Thus, our results suggest that the growing short-ranged resonances contribute to the finite-size effects in MBL studies, which we argue is an \emph{inherent} feature of a system featuring an asymptotic MBL phase.

Our work emphasizes that numerical studies should be cautious about extracting phase behavior from only system-wide resonances at finite sizes, and that characterizations of the profile of short-ranged resonances are equally important for the destabilization of MBL.
We showed that, even at very large disorder strengths, the system hosts CTRs of increasing charge as $L$ increases.
Two scenarios are possible for the fate of these resonances at larger system sizes.
In the first, the buildup of resonances never saturates, and system-wide resonances eventually emerge.
In this scenario, at numerically available system sizes, the small resonances have not had a chance to develop into those that can transfer extensive charges across the cut due to the large resonance collar of long-ranged resonances.
In this case, the absence of system-wide resonances at these system sizes is a transient (finite-size) feature.
In the second scenario, the resonance buildup saturates at sufficiently large system sizes---consistent with the phenomenological picture of stable MBL, in which resonances of any size are permitted but occur with probabilities exponentially suppressed in the resonance size~\cite{garratt_local_2021, crowley_constructive_2022}.
Here, the absence of system-wide resonances at numerically accessible sizes is an expected feature of the theory due to their very low probabilities.
We do not make any claim as to which scenario for resonances is realized, as the system sizes required to resolve this question remain inaccessible. 
Instead, we highlight the importance of accounting for the buildup of short-range resonances in the potential destabilization of MBL, as these may seed system-wide resonances at sufficiently large system sizes.
We also urge that their effects be taken into consideration when analyzing numerical data of small system sizes.
This applies to both the strong disorder regime and the intermediate regime with prethermal behavior.

We further point out that the CTC operator $\hat{M}$ offers additional advantages in detecting resonances.
In previous approaches, the system-wide resonances are detected through probes such as system-wide spin-spin correlations~\cite{laflorencie_cat_2025}, the end-to-end mutual information~\cite{morningstar_avalanches_2022}, or large off-diagonal elements from Jacobi rotations on the Hamiltonian~\cite{long_phenomenology_2023}.
In our work, MBRs arose naturally as the microscopic mechanism that can move the most charge across the cut, easily identified by analyzing the eigenstates of $\hat M$.
This provides a physically transparent way to categorize MBRs, which can help our understanding of them.
Thus, the CTC operator or its analogues can be useful for future studies of other nonergodic systems.
A natural extension would be to study the CTC in other potential candidates for MBL, such as Stark-Bloch (tilted lattice)~\cite{van_nieuwenburg_bloch_2019, schulz_stark_2019} and quasiperiodic models of MBL~\cite{iyer_many-body_2013}, which are plagued by finite-size effects in ED studies similar to those in the disordered model.
We particularly point to the quasiperiodic model as a possibly interesting case~\cite{iyer_many-body_2013, khemani_two_2017, znidaric_interaction_2018, padhan_long-range_2025}.
As mentioned previously, the CTC operator's eigenstates are useful for detecting resonances in the spectrum and can be used to probe the structure of resonances in these models of MBL.
Another immediate interesting future direction is to extend our studies of systems with $U(1)$ charge conservation to Hamiltonian systems with no symmetries.
In such systems, since energy is the only conserved quantity, the worst-case energy transport would be an interesting quantity to study~\cite{varma_energy_2017, maksymov_energy_2019, de_roeck_absence_2024}.
On the other hand, for Floquet systems, where there are no such conserved quantities, it is not clear if this type of analysis will be directly applicable. 
However, our work suggests that finding a systematic way to study short-ranged resonances in such models is valuable for understanding the finite-size effects present in these systems.

Finally, as briefly mentioned in Sec.~\ref{sec:domain_wall_melting}, our SCTB data suggests that the maximal charge transfer starting from the domain wall state $\ket{\text{DW}}$ (or from any typical product state) should saturate with the system size.
This saturation is a strong signature of MBL, which can plausibly be observed in current ion trap or cold atom platforms, at much larger system sizes than what is possible with numerics~\cite{schreiber_observation_2015, rispoli_quantum_2019, lukin_probing_2019, leonard_probing_2023}.
This provides a new signature in experiments for MBL beyond the decay of autocorrelation functions, which leaves open questions about the asymptotic fate of MBL due to strong finite-size effects.

To summarize, our work is a step towards understanding the mechanisms behind finite-size effects affecting studies of many-body localization. 
This will help inform future analysis of ED studies, and provides an interesting microscopic perspective on why interacting quantum many-body systems trend towards thermalization.
Our analysis of the probabilities of resonances in Sec.~\ref{sec:counting_resonances} is also an exciting starting point for understanding the possibility of a stable MBL phase description incorporating many-body resonances.
This perspective provides a framework that can encompass the MBL phenomenology in both the thermodynamic limit at large $L$, and in the regime of numerically accessible system sizes.

\begin{acknowledgements}
We thank Anushya Chandran, Philip Crowley, Sam Garratt, Sarang Gopalakrishnan, Hyunsoo Ha, David Huse, Vedika Khemani, David Long, David Pekker, and Gil Refael for very useful discussions.
We are particularly grateful to Anushya Chandran, Philip Crowley, Sam Garratt, Sarang Gopalakrishnan, Hyunsoo Ha, and David Huse for extensively discussing and clarifying the picture of MBRs and resonance collars, and to Gil Refael for many discussions of MBL physics and renormalization group approaches to it.
J.K.J.~is supported by the U.S.\ Department of Energy, Office of Science, Office of Advanced Scientific Computing Research, Department of Energy Computational Science Graduate Fellowship under Award Number(s) DE-SC0025528.
FMS acknowledges support provided by the U.S.~Department of Energy Office of Science, Office of Advanced Scientific Computing Research, (DE-SC0020290); DOE National Quantum Information Science Research Centers, Quantum Systems Accelerator; and by Amazon Web Services, AWS Quantum Program. This research was supported in part by grant NSF PHY-2309135 to the Kavli Institute for Theoretical Physics (KITP).
This work was also supported by the National Science Foundation through grant DMR-2001186.
This research used resources of the National Energy Research Scientific Computing Center (NERSC), a Department of Energy User Facility using NERSC award ERCAP0036065.

This report was prepared as an account of work sponsored by an agency of the United States Government. Neither the United States Government nor any agency thereof, nor any of their employees, makes any warranty, express or implied, or assumes any legal liability or responsibility for the accuracy, completeness, or usefulness of any information, apparatus, product, or process disclosed, or represents that its use would not infringe privately owned rights. Reference herein to any specific commercial product, process, or service by trade name, trademark, manufacturer, or otherwise does not necessarily constitute or imply its endorsement, recommendation, or favoring by the United States Government or any agency thereof.
The views and opinions of authors expressed herein do not necessarily state or reflect those of the United States Government or any agency thereof.
\end{acknowledgements}

\appendix

\section{Properties of the Charge Transport Capacity Operator}
In this Appendix, we derive, show numerical results for, and discuss various important properties of the charge transport capacity operator $\hat M$, and the CTC $\ctc$.

\subsection{An upper bound on the charge transport capacity}
Here, we prove that $\ctc\leq L$, providing a strict upper bound on how the CTC can grow in the thermodynamic limit. 
Using simply that both the spectra of $\hat N_R$ and $[\hat N_R]_\text{diag}$ are bounded between $0$ and $L/2$, it can be derived that the spectrum of the difference $\hat M =\hat N_R-[\hat N_R]_\text{diag}$ is bounded between $-L/2$ and $L/2$. More precisely, we have that for any state $\ket{\psi}$
\begin{equation}
    0\le \braket{\psi|\hat N_R|\psi}\le L/2,
\end{equation}
\begin{equation}
    0\le \braket{\psi|[\hat N_R]_\text{diag}|\psi}\le L/2.
\end{equation}
We then obtain:
\begin{equation}
   \braket{\psi|\hat M|\psi}= \braket{\psi|\hat N_R|\psi} -\braket{\psi|[\hat N_R]_\text{diag}|\psi}\le L/2, 
\end{equation}
\begin{equation}
   \braket{\psi|\hat M|\psi}= \braket{\psi|\hat N_R|\psi} -\braket{\psi|[\hat N_R]_\text{diag}|\psi}\ge -L/2 .
\end{equation}
Since these bounds on $\braket{\psi|\hat M|\psi}$ hold for any state $\ket{\psi}$, we have in particular  
\begin{equation}
  \lambda_\mathrm{max}(\hat M)\le L/2, \qquad    \lambda_\mathrm{min}(\hat M)\ge -L/2,
\end{equation}
and therefore the CTC is bounded by $\lambda_\mathrm{max}(\hat M)-\lambda_\mathrm{min}(\hat M)\le L$.

\subsection{Charge transport capacity in inversion symmetric systems}\label{app:derivation_inversion_symmetric}
We will show that for any $\hat H$ with inversion and $U(1)$ symmetry, we have $\ctc = L/2$ in the half-filling sector, and that the eigenvalues of $\hat{M}$ across all filling sectors are the half-integers in $[-L/4,L/4]$.

Let $\hat U$ be the inversion operator. Consider a Hamiltonian with inversion symmetry, such that $[\hat H, \hat U]=0$. Then, $\hat H$ and $\hat U$ share a set of eigenstates, $\hat U |n\rangle= \pm|n\rangle$. Now, observe that by the definition of $\hat U$,
\begin{equation}
    \hat U^\dagger \hat N_R \hat U=\hat N_L,
\end{equation}
it follows that
\begin{equation}\label{eq:relation_inv_symm}
    \langle n|\hat N_R |n\rangle=\langle n|\hat U^\dagger \hat N_R \hat U|n\rangle = \langle n| \hat N_L |n\rangle.
\end{equation}
In the $N$ particle sector, $\langle n|\hat N_R |n\rangle+\langle n|\hat N_L |n\rangle=N$. 
From Eq.~(\ref{eq:relation_inv_symm}), we obtain $\langle n|\hat N_R |n\rangle = N/2$.
Thus:
\begin{equation}
    \hat M = \hat N_R-\sum_n \langle n|\hat N_R|n\rangle |n\rangle\langle n|=\hat N_R-\frac{N}{2}\sum_n  |n\rangle\langle n|.
\end{equation}
The maximum eigenvalue of $\hat M$ is $L/4$, which occurs when $N=L/2$ and $\langle\hat N_R\rangle=L/2$.
The minimum eigenvalue of $\hat M$ is $-L/4$ and occurs when $N=L/2$ and $\langle\hat N_R\rangle=0$.
This gives $\ctc = L/2$. 
In the half-filling sector, the eigenvalues of $\hat M$ range from $-L/4$ to $L/4$ in integer steps, whereas across all sectors, the eigenvalues range from $-L/4$ to $L/4$ in \emph{half}-integer steps.

\subsection{Properties of $\hat M$ and related quantities in thermalizing systems}\label{app:goe_results}
\begin{figure*}[t]
    \centering
\includegraphics[width=0.85\linewidth]{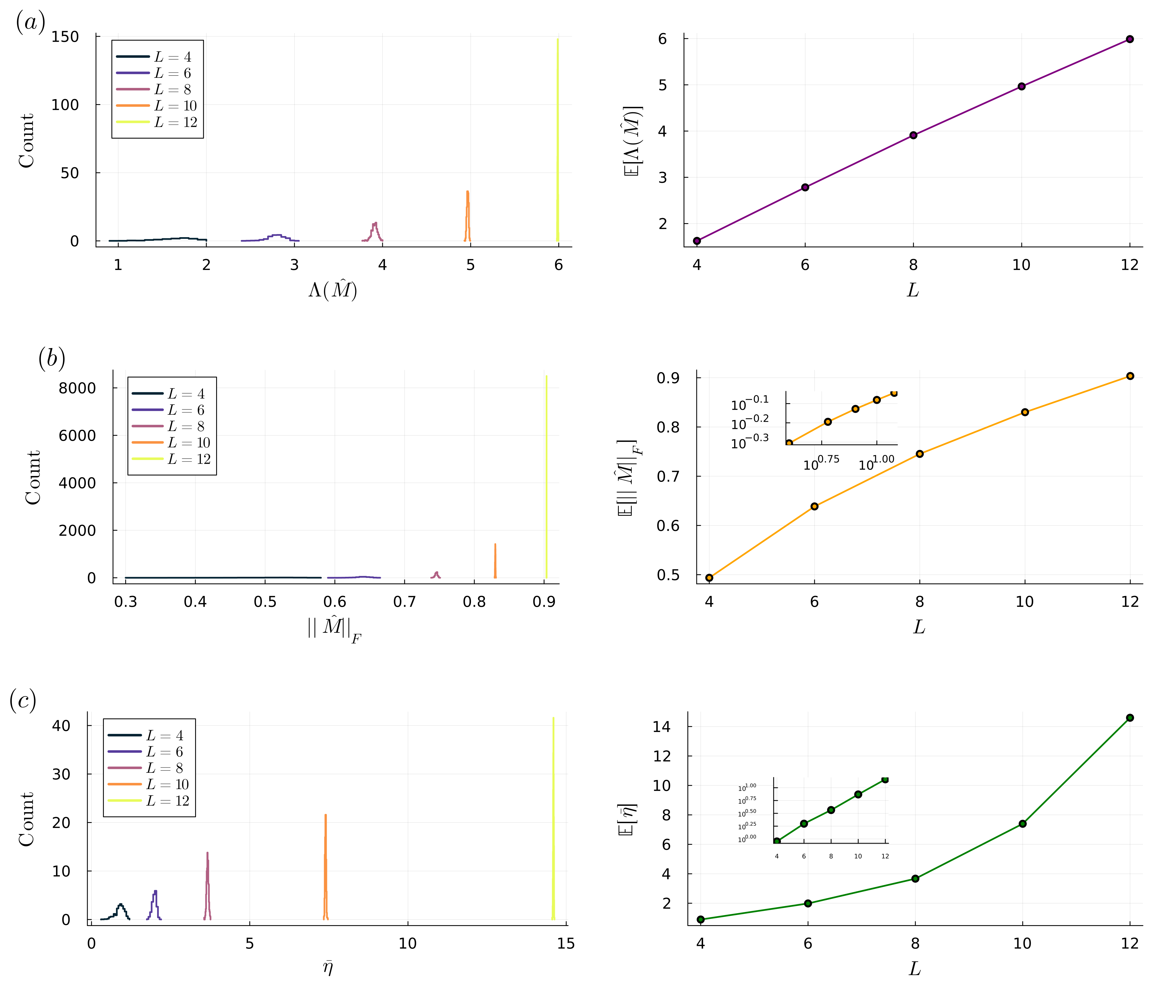}
    \caption{Results for the CTC operator $\hat M$ and related quantities for presumed Hamiltonian eigenvectors drawn from the Gaussian Orthogonal Ensemble (GOE). 
    Panel (a): analysis of the CTC for the GOE. 
    The left panel shows normalized probability distributions of the CTC in the GOE, each colored curve representing different system sizes.
    The right panel shows the CTC averaged over GOE realizations ($y$-axis) as a function of the system size $L$ ($x$-axis).
    Panel (b): analysis of the normalized Frobenius norm of the CTC operator $\hat M$, $\|\hat M\|_F$ for the GOE.
    The left panel similarly shows normalized probability distributions of $\|\hat M\|_F$ over GOE realizations, each colored curve representing a different system size.
    The right panel shows $\|\hat M\|_F$ averaged over GOE realizations ($y$-axis) as a function of $L$ ($x$-axis). 
    The inset shows the same plot but on a log-log scale.
    The best-fit power-law exponent $\alpha$ to the curve $\sim  L^\alpha$ is $\alpha \approx 0.54$.
    Panel (c): analysis of the SCTB $\eta(\ket{\psi_0})$, Eq.~(\ref{eq:dctc_equation}), as a function of $L$, with probability density functions on the left over both product states $\ket{\psi_0}$ and in the GOE, and the mean of $\eta(\ket{\psi_0})$ over all product states and GOE realizations as a function of $L$ on the right hand side.
    The inset in the right plot is the same data but with the $y$-axis on a logarithmic scale, showing the exponential scaling of the averaged SCTB with $L$ for GOE systems.
    }
\label{fig:GOE_results}
\end{figure*}

\begin{figure*}[t]
    \centering
    \includegraphics[width=\linewidth]{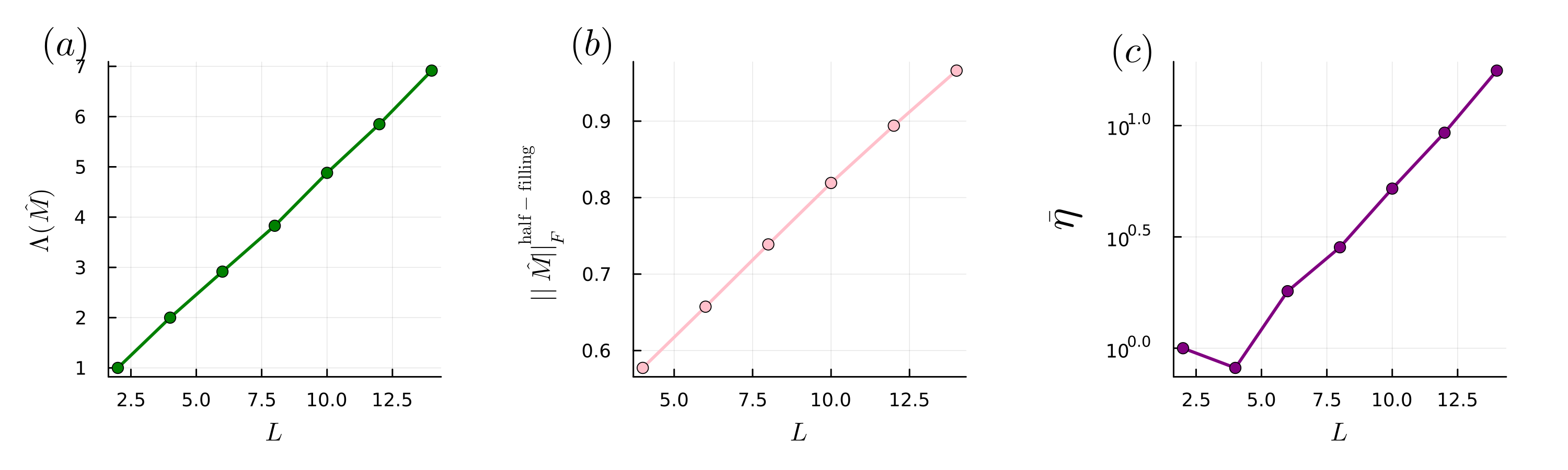}
    \caption{Analysis of quantities associated with $\hat M$ for the non-disordered Hamiltonian in Eq.~(\ref{eq:next_nearest_neighbor}), which is nonintegrable, non-inversion-symmetric, and expected to thermalize.
    Panel (a): The charge transport capacity $\ctc$ as a function of $L$, where we see the expected scaling $\approx L/2$.
    Panel (b): The normalized Frobenius norm of $\hat M$ in the half-filling sector, $\|\hat M\|_F^{\text{half-filling}}$, as a function of $L$. 
    Here, $\|\hat M\|_F^{\text{half-filling}}\sim L^\alpha$ with best-fit $\alpha \approx 0.414$.
    Beyond small $L$, the data is well described by $\approx \sqrt{L}/4$.
    Panel (c): The state charge transfer bound (SCTB) averaged over all product states, $\overline\eta$, ($y$-axis, logarithmic scale) versus the system size on the $x$-axis.
    }
\label{fig:next_nearest_neighbor_results}
\end{figure*}
Here, we provide numerical evidence of the following two properties of the charge transport capacity operator $\hat M$ in thermalizing systems: that $\ctc$ generically scales as $\approx L/2$, and that $\|\hat M\|_F \sim \sqrt{L}$. 

We first test these properties for a Hamiltonian sampled from the GOE with Hilbert space dimension $\binom{L}{L/2}$, restricting to the half-filling sector. 
The eigenvectors are constructed by drawing i.i.d.\ $\mathcal{N}(0,1)$ entries and orthonormalizing via the Gram-Schmidt process. (Note that both the CTC operator $\hat{M}$ and the SCTB for any initial state depend only on the assumed eigenstates of the Hamiltonian, not on the corresponding eigenvalues.)
We show results for the CTC in Fig.~\ref{fig:GOE_results}(a), and results for the Frobenius norm of $\hat M$ in Fig.~\ref{fig:GOE_results}(b).
We also study the SCTB [Eq.~(\ref{eq:dctc_equation})] over all computational basis states in Fig.~\ref{fig:GOE_results}(c).
These figures affirm the expected scaling of these quantities in an ergodic phase.
We briefly note that the distributions of these quantities appear to narrow with $L$, reflecting the fact that the sampling process decreases the variance of the off-diagonal matrix elements by a factor of the Hilbert space dimension.

While GOE eigenstates are known to satisfy ETH, the GOE Hamiltonian lacks locality, conservation laws, and a physical thermodynamic limit, making it an overly idealized model.
Thus, we also compute $\hat M$ for a 1D non-disordered model with nearest-neighbor hopping and a six-fermion interaction term:
\begin{equation}\label{eq:next_nearest_neighbor}
\hat H = -t \sum_{i=1}^{L-1} \big[ \hat c_i^\dagger \hat c_{i+1} + \hat c_{i+1}^\dagger \hat c_i \big]
+ J \sum_{i=1}^{L-2} \hat n_i\hat n_{i+1}\hat n_{i+3} ~,
\end{equation}
with $J$, $t$ being constants. 
We expect that this model is nonintegrable and will thermalize.
Furthermore, with open boundary conditions, this model does not have inversion symmetry, so the proof from App.~\ref{app:derivation_inversion_symmetric} does not apply.
In Fig.~\ref{fig:next_nearest_neighbor_results}, we show an analysis of $\hat M$ for this model with $J=1, t=1$ in the half-filling sector.
One can see in panel (a) that the results align with the predictions for $\ctc \approx L/2$. Furthermore, in panel (b) we observe that $\|\hat M\|_F^{\text{half-filling}}\sim L^\alpha$ where the best power-law coefficient is $\alpha\sim{0.42}$.
We note that it appears that this power-law scaling appears to increase slightly with $L$, and it is not clear what the asymptotic scaling will be.
This model serves as a physical example confirming both the expected scaling $\ctc \sim L/2$ and the power-law growth $\|\hat{M}\|_F^{\text{half-filling}} \sim L^\alpha$ with $\alpha \approx 1/2$, similar to the free-fermion result.

Finally, we also confirm that the state charge transfer bound (SCTB), defined in Eq.~(\ref{eq:dctc_equation}), averaged over all product states, grows exponentially with system size in Fig.~\ref{fig:next_nearest_neighbor_results}(c).

\subsection{Degenerate systems and $\hat M$}\label{app:degenerate_hamiltonians}
There are subtleties which arise when one defines $\hat M=\hat N_R-[\hat N_R]_{\text{diag}}$ for a Hamiltonian in a system with degeneracies.
In particular, $[\hat N_R]_{\text{diag}}$ will depend on the chosen basis for the degenerate subspace, which can affect $\hat M$.

One simple thought experiment is to take the Hamiltonian $\hat H = \sum_{i=1}^L \hat n_i$.
This Hamiltonian clearly has product states as eigenstates. 
Thus, $\hat N_R$ will be diagonal in this eigenstate basis, and so then $\hat N_R-[\hat N_R]_{\text{diag}}$ will be exactly the 0 matrix.
Thus, the CTC will be zero when defined using these energy eigenstates.
At the same time, we have shown in Sec.~\ref{app:derivation_inversion_symmetric} that because $\hat H$ is inversion symmetric, $\ctc$ will scale as $L/4$ when the energy eigenstates are chosen to be also inversion eigenstates.
This appears to be a major contradiction, so what is the problem here?

The resolution here is that $\hat H$ is highly degenerate, and $\hat M$ can have a $\ctc$ of $L/4$ or 0 depending on which degenerate basis is picked.
Specifically, for this thought experiment, if one works explicitly in the computational basis, then for the reason described $\ctc$ will be 0.
On the other hand, consider the two product states $|a\rangle$ and $\hat U_I|a\rangle$ such that $|a\rangle$ is the inverted mirror image of $\hat U_I|a\rangle$ (e.g., $|010110\rangle$ and $|011010\rangle$). 
These two states are degenerate, with total energy being the total occupation of the product states.
We can consider working in a new eigenbasis where for every product state $|a\rangle$ and its inverted counterpart $\hat U_I\ket{a}$, the new eigenstates are $\frac{1}{\sqrt{2}}(|a\rangle\pm \hat U_I\ket{a})$.
In this basis, then the argument in Sec.~\ref{app:derivation_inversion_symmetric} applies, as now the eigenstates of the Hamiltonian (in this basis) are also shared eigenstates of the inversion symmetric operator $\hat U$, and now indeed $\ctc=L/2$.
Both of these results are technically correct: $\ctc=0$ is as much as an \emph{upper bound} on the transportable charge across $i_0$ as $\ctc=L/2$ is, it is just that $\ctc=0$ is a much better upper bound and thus is more representative of the charge dynamics of the $\hat H$.

This is a potential concern when working with the CTC operator in systems that are expected to have a degenerate eigenspectrum.
One might have to pick the degenerate subspace basis in a way that minimizes the CTC in some way.
In this case, we suspect that the basis closest to the eigenbasis of $\hat N_R$ will minimize the CTC.
In the context of this paper, this particular point is not a concern, as the Hamiltonians we work with are disordered and have no symmetries (other than charge conservation) and thus explicitly have no degeneracies.

\subsection{Charge transport capacity of other subregions and in inversion symmetric models}
Here, we discuss generalizing the charge transport capacity operator to an arbitrary geometry.
We then discuss in detail a possible application of this generalization to problems of inversion-symmetric localization.

Consider a lattice $\mathcal{L}$ and a subsystem $A \subset \mathcal{L}$ and its complement $A^c$. 
Define $\hat N_A = \sum_{i\in A}\hat n_i$ and likewise $\hat N_{A^c} = \sum_{i\in A^c}\hat n_i$.
Now we consider the change in the amount of charge between these subregions, i.e. $\Delta \hat N_A(t) := \hat{N}_A(t) - \hat{N}_A$ and similarly for $\Delta \hat N_{A^c}(t)$,
we can put an upper bound on $\|\Delta \hat N_A(t)\| = \|\Delta \hat N_{A^c}(t)\|$ just as before by defining the corresponding CTC operator:
\begin{equation}
    \hat M_{A, A^c} = \hat N_{A} - [\hat N_A]_{\text{diag}} = -\hat N_{A^c} + [\hat N_{A^c}]_{\text{diag}} ~,
\end{equation}
so that $\|\Delta \hat N_A(t)\| = \|\Delta \hat N_{A^c}(t)\| \leq \lambda_{\text{max}}(\hat M_{A, A^c}) - \lambda_{\text{min}}(\hat M_{A, A^c})$.
The CTC for this partition is then $\lambda_{\text{max}}(\hat M_{A, A^c}) - \lambda_{\text{min}}(\hat M_{A, A^c})$.
Here, we will refer to the CTC in the main text as the half-cut CTC, while the \emph{generalized} CTC can be defined on other subsystems.

The generalized CTC can be useful in many scenarios of MBL studies.
One straightforward case is the analog of the 1D CTC in higher-dimensional lattices.
Here, one can, for instance, study the maximal charge transfer between one half of the square/cubic lattice.
Although numerics in higher-dimensional systems are out of reach, analytical questions about the generalized CTC here may be of interest.

Another very interesting application of the generalized CTC is in the context of inversion-symmetric versions of localized problems.
This will be the focus of the rest of this section.

In App.~\ref{app:derivation_inversion_symmetric}, we showed that a model with inversion symmetry in the middle of the bond $[i_0, i_0+1]$, $i_0 = L/2$ (assuming $L$ is even), will always have a half-cut CTC of $L/2$.
While the proof in App.~\ref{app:derivation_inversion_symmetric} was motivated by thinking about systems without disorder and their half-cut CTC as a representative of ergodic systems, the following question naturally arises.
Suppose we construct an artificial system with strong disorder and inversion symmetry, e.g., by generating random on-site potentials $\{h_1, h_2, \dots, h_{L/2}\}$ on the left half and copying them over using the above inversion onto the right half, \{$h_i = h_{L-i+1}, i = L/2+1, \dots, L\}$.
If the localization exists as a stable phase in a generic strongly disordered model (i.e., without inversion symmetry), one would expect that some physics of localization would also persist in this inversion-symmetric model.
However, the localization is clearly not captured by the half-cut CTC measure, since the half-cut CTC is $L/2$ for any inversion-symmetric disorder realization in this case.
This type of tension was noted in Ref.~\cite{kloss_absence_2023} in a somewhat different context.
Here, we discuss a potential reconciliation using the generalized CTC defined on an inversion symmetric subregion.

First, we note that even the half-cut CTC of $L/2$ could be consistent with some notion of suppressed transport.
Let us start from the limit of zero hopping and, for the sake of simplicity, ignore the physics of resonances within the non-symmetric disordered system.
At zero hopping, the computational product states $\{\ket{a}\}$ are exact eigenstates; however, in the inversion-symmetric system we would consider the symmetric/antisymmetric states $\frac{1}{\sqrt{2}} (\ket{a} \pm \ket{a'})$, where $\ket{a'}$ is obtained by applying the inversion transformation on $\ket{a}$.
In the presence of very small tunneling, these would be true eigenstates in the inversion symmetric problem; in some sense, the addition of inversion symmetry induces a very specific set of resonances.
These resonances can clearly transfer large amounts of charge between the two halves of the system. 
However, the associated energy splittings are exponentially small, so such transfers only happen on exponentially long timescales. 
Any notion of transport that forbids exponentially long waiting times is therefore still compatible with localization in this setting. 
Our half-cut CTC, by contrast, is an infinite-time measure by construction, and hence sensitive to precisely these exponentially slow processes --- it will not detect the suppressed transport here.
In summary, the worst-case character of the CTC is useful when the CTC is finite, since it then bounds any notion of transport; its utility is more limited when the CTC diverges, as in this example with disorder and inversion symmetry.

Nevertheless, in this example, we can invoke the generalized CTC, defined on a clever choice of subregions, which would signal sharp notions of localization.
Specifically, instead of dividing the system into left and right halves, we can partition it into inversion-symmetric regions, e.g., the inner region $\Omega_{\text{inner}} = [L/4+1, L/4+2, \dots, 3L/4-1, 3L/4]$ and the outer region $\Omega_{\text{outer}} = \Omega_{\text{inner}}^c$.
For simplicity, we assumed that $L$ is divisible by $4$ and took the inner region to be half of the total size.
The idea for considering this setup is that the inversion-symmetry-induced resonances can transport charges only within these regions but not between them.
Thus, we can consider the generalized CTC defined on the subregions $\Omega_{\text{inner}}$ and $\Omega_{\text{outer}}$.
In the non-interacting inversion-symmetric Anderson model, we expect this generalized CTC quantity to be finite.
This shows that localization can still be precisely defined here, despite the exponentially slow, arbitrarily long-range transport processes induced by the inversion symmetry at finite sizes.
We then conjecture that, in the interacting version of this problem, if stable MBL exists at strong disorder in the generic (non-inversion-symmetric) case, then localization, in the sense of the generalized CTC proposed above, also holds once inversion symmetry is imposed.
We leave detailed explorations of such ideas to future work.

\section{Charge Transport Capacity in Free Fermion Models}\label{app:free_fermion_ctcb}
In this Appendix, we numerically study the charge transport capacity measure in standard non-interacting models of localization.
All of the models considered in this Appendix describe non-interacting spinless fermions hopping on a one-dimensional (1D) and finite lattice with sites $i = 1, 2, \dots, L$, with open boundary conditions:
\begin{equation}\label{eq:noninteracting_hamiltonian}
    \hat H=-t\sum_{j=1}^{L-1} \left(\hat{c}_j^\dagger \hat{c}_{j+1}+\hat{c}_{j+1}^
    \dagger \hat{c}_{j}\right)+\sum_{j=1}^L h_j \hat{c}_j^\dagger \hat{c}_j .
\end{equation}
Here, the dependence of the onsite potentials $h_j$ on $j$ varies between the models that we consider.

For the above Hamiltonian, we can write an exact expression for $\hat M$ in the form of a fermion bilinear operator:
\begin{equation}
    \hat M = \sum_{\alpha, \beta} M_{\alpha\beta} \hat{d}^\dagger_{\alpha} \hat{d}_{\beta}.
\end{equation}
Here, we define
\begin{equation}\label{eq:occupation_of_orbitals}
d_\alpha = \sum_i \phi_\alpha^{*}(i)\hat c_i 
\end{equation} 
as the annihilation operator for the $\alpha$th eigenstate $\phi_\alpha(x)$ with energy $\epsilon_\alpha$, and $x$ is the position.
The original fermions/hardcore bosons are $c_j = \sum_\alpha \phi_\alpha(j) d_\alpha$, and we can easily express $N_R(i_0)$ and identify $[N_R(i_0)]_{\text{diag}}$ with respect to $H = \sum_\alpha \epsilon_\alpha \hat{d}_\alpha^\dagger \hat{d}_\alpha$.
We denote $\phi_\alpha(x)\equiv \phi_{x, \alpha}$ using matrix notation.
Then, the (off-diagonal) matrix elements $M_{\alpha\beta}$ are
\begin{equation}\label{eq:no_denominator_rewriting_m0}
    M_{\alpha\beta} = -\sum_{j=1}^{i_0} \phi^{*}_{j\alpha} \phi_{j\beta}=\sum_{j=i_0+1}^{L} \phi^{*}_{j\alpha} \phi_{j\beta}~. 
\end{equation}
By the definition of $\hat M$, the diagonal elements are zero: $M_{\alpha\alpha}=0$. 
A full discussion of these results can be found in the previous work~\cite{jiang_quasiconservation_2025}.
Equation (\ref{eq:no_denominator_rewriting_m0}) makes effects of localization manifest: if the orbitals $\phi$ are localized, then $M_{\alpha\beta}$ is appreciable only when $\alpha$ and $\beta$ lie near $i_0$. If $\alpha$ and $\beta$ are far apart, their overlap is exponentially small, so $M_{\alpha\beta}$ is exponentially suppressed. If they are close to each other but both far from $i_0$ (say, to the right), then they have exponentially small weight on the sites $j\le i_0$ that appear in the sum in Eq.~(\ref{eq:no_denominator_rewriting_m0}); each term is suppressed, hence the sum is as well. The same holds on the left. Therefore, $\hat M$ is quasilocal around $i_0$.

From its analytical expression, it is clear that $\hat M$ is easy to compute numerically for large sizes for the non-interacting problem. Its calculation only depends on the energy eigenstates, which are easy to compute numerically. In the following sections, we will show the results of these calculations for various non-interacting models that exhibit localization phenomena. 
We only consider 1D lattices with open boundary conditions, and $\hat M$ is always calculated through the middle of the chain, with $i_0=\lfloor L/2\rfloor$. As a result, we only focus on even lattice sizes to mitigate fluctuations in the charge transport capacity due to the placement of the cut.

\subsection{Non-interacting Anderson Model}\label{app:anderson_model_ctc}

\begin{figure*}[t]
    \centering
    \includegraphics[width=\linewidth]{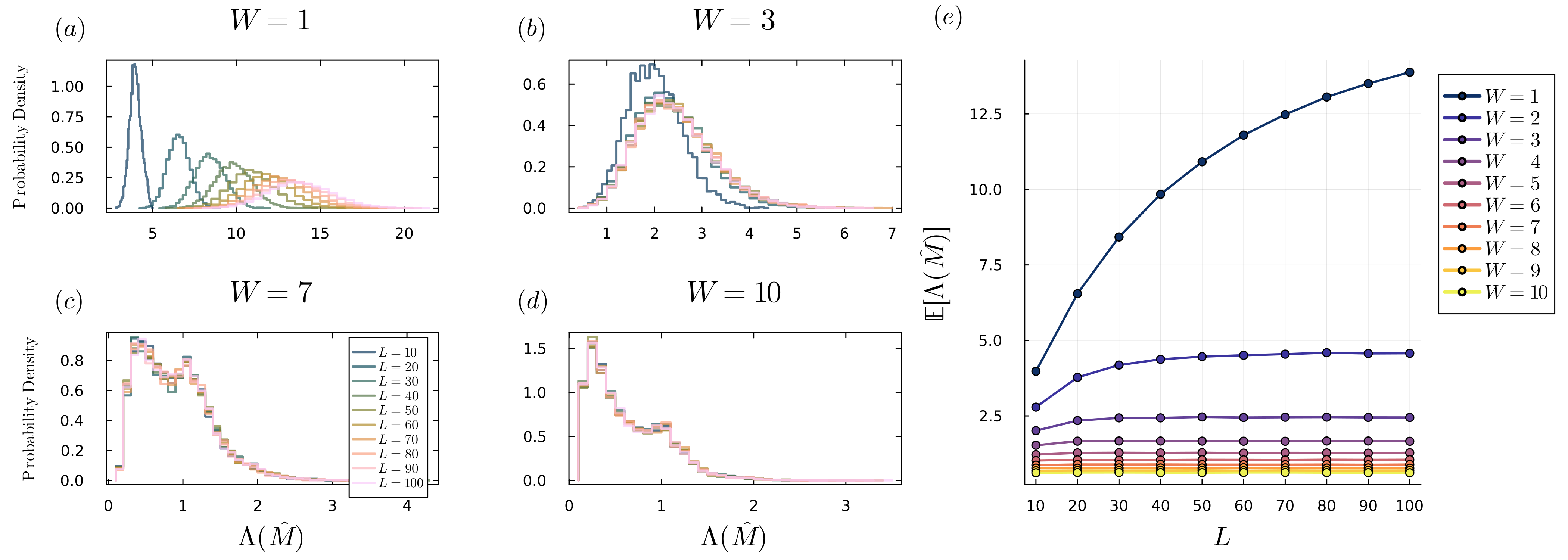}
    \caption{Analysis of the charge transport capacity $\ctc$ in the Anderson model.
    Panels (a)-(d) show the probability density functions of $\ctc$ over disorder realizations, with each panel representing the labeled disorder strength $W$, and each colored curve representing a different system size $L$, ranging from $L=10$ to $L=100$ in steps of $10$. 
    Panel (e) shows the disorder-averaged $\ctc$, $\doubleE[\ctc]$, as a function of $L$, with each curve representing a different disorder strength $W=1\dots 10$ in steps of 1.
    }
\label{fig:anderson_CTC.png}
\end{figure*}
Now we discuss the behavior of $\hat M$ in the 1D Anderson model, the paradigmatic setting for single-particle localization. In this model, the on-site potentials $h_j$ are chosen independently from a uniform distribution $\mathrm{Unif}[-W, W]$, where $W \in \mathbb{R}$ denotes the disorder strength.
It is a well-known result that in 1D, for any finite disorder strength $W$, all eigenstates of the Anderson model are exponentially localized~\cite{anderson_absence_1958}. The consequence of this is that Eq.~(\ref{eq:noninteracting_hamiltonian}) can be diagonalized in terms of \emph{quasilocal} integrals of motion:
\begin{equation}
    \hat H = \sum_{\alpha}\epsilon_\alpha \hat d^{\dagger}_\alpha\hat d_\alpha~.
\end{equation}
Here, the \emph{quasilocal} integrals of motion are $\hat d^{\dagger}_\alpha\hat d_\alpha$, whose localization follows naturally from the fact that the eigenstates are localized, since $\hat d^{\dagger}_\alpha\hat d_\alpha$ represent occupations of the localized orbitals. 

Because of the nature of localization, we expect an absence of diffusion and therefore, only a finite number of charges can pass through $i_0$ even in the thermodynamic limit. 
Indeed, we see this behavior clearly reflected in the distributions of $\ctc$ for the Anderson model in Fig.~\ref{fig:anderson_CTC.png}, which can be calculated numerically easily through exact diagonalization and using Eq.~(\ref{eq:no_denominator_rewriting_m0}).
Here, all calculations of $\ctc$ involve 10000 disorder realizations. 
First, we note that the distributions of $\ctc$ converge towards a characteristic form with exponential tails.
At large $W$, a separate, smaller peak appears at 1, meaning that a significant fraction of disorder realizations allow one particle to cross the link at most. 
Intuitively, this peak is explained by the fact that a particle is free to move roughly on the length scale of a localization length $\xi$.
If a LIOM happens to be positioned right on top of the cut $i_0$, it could facilitate the process of a particle moving a distance of $\xi$ near the cut.
The net result of this is a single particle hopping back and forth near the cut, a process which is captured by the charge transport capacity.

Next, we consider the growth of the mean of those distributions as a proxy for how well they are converged, i.e., the disorder averaged $\ctc$, which we denote as $\mathbb{E}[\ctc]$.
One can see clearly that $\mathbb{E}[\ctc]$ quickly converges to a finite value for large enough $L$, once $L$ becomes larger than the localization length $\xi$.
The CTC that it converges to is dependent on $\xi$. 
One understanding is that we expect that this saturated value should be of order $\xi/2$.
This is because in a localized system, charges can still move on an effective lattice size of order of the localization length $\xi$, which has CTC $\xi/2$.

\subsection{Non-interacting quasiperiodic model}\label{app:quasiperiodic_noninteracting_ctcb}

\begin{figure*}[t]
    \centering
    \includegraphics[width=\linewidth]{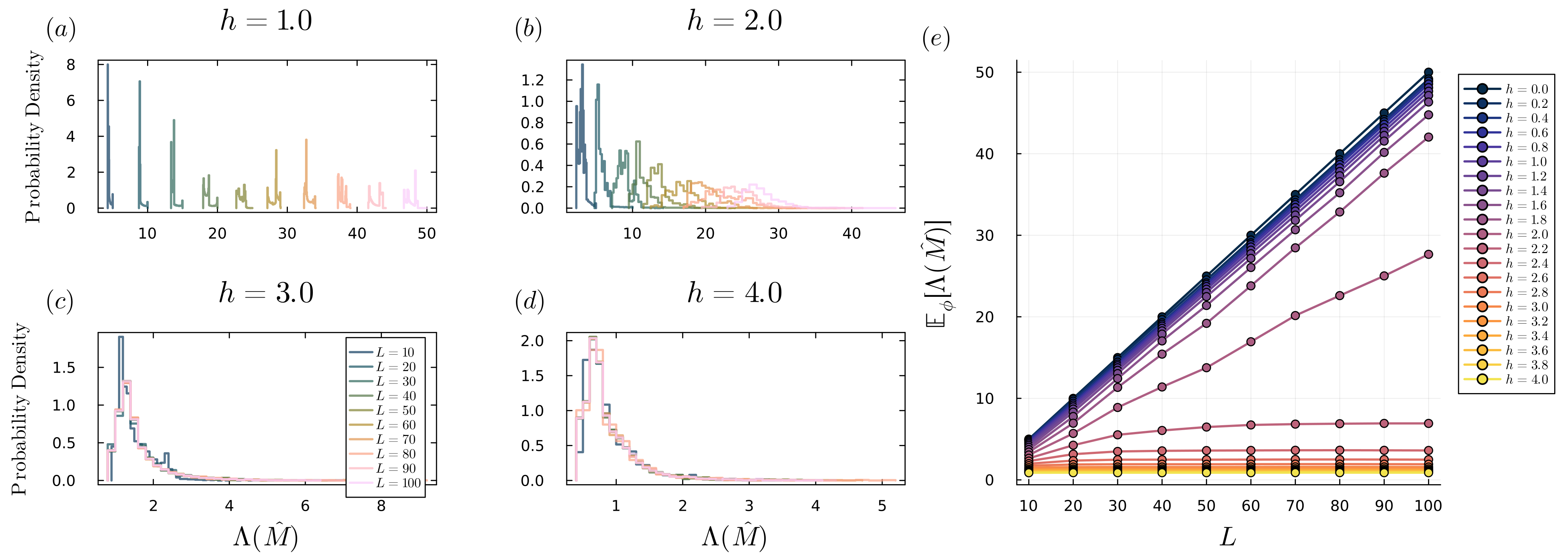}
    \caption{Analysis of the charge transport capacity $\ctc$ in the quasiperiodic Aubry-André model.
    Panels (a)--(d): distributions of $\ctc$ over phase shifts $\phi$, with each panel representing a different potential strength $h$, and each colored curve representing a distribution at a different system size $L$, from $10$ to $100$ in steps of 10.
    Panel (e): $\doubleE_{\phi}[\ctc]$ as a function of $L$, where $\doubleE_{\phi}[\cdot]$ denotes averaging over phases $\phi$. Each colored curve represents a different potential strength $h$.
    The system transitions at $h=2$ from fully delocalized (all curves showing scaling $\approx L/2$) to fully localized (all curves saturating with $L$).
    }
    \label{fig:AA_CTC.png}
\end{figure*}
The scaling of the charge transport capacity with $L$ is also useful as a diagnostic of localization transitions.
We will demonstrate this by studying $\hat M$ in the quasiperiodic Aubry-André model~\cite{aubry_analyticity_1980}, with onsite potentials
\begin{equation}\label{eq:quasiperiodic_potential}
    h_j=h\cos{\left[2\pi k\left( j+\delta\right)\right]}.
\end{equation}
Here, $k$ is some irrational number, and $\delta$ is a phase shift.
For this model, fixing $t=1$, there is a transition at $h=2$ from a phase where all of the eigenstates are delocalized to a phase where all of the eigenstates are localized in real space.
There is a duality of this model in real space and momentum space~\cite{aubry_analyticity_1980}.
Adding a perturbation which breaks this duality can lead to the existence of mobility edges, which we will probe later in the generalized Aubry-André model. 

In our numerics in the quasiperiodic model, we fix $k=\sqrt{2}$.
This choice does not affect the behavior of $\ctc$ as long as $k$ is irrational.
Furthermore, to avoid commensuration issues, we adopt the standard practice and perform calculations for different phase shifts between $0\leq \delta\leq 1/k$, in uniform steps, and compute the distributions of $\ctc$ over these phase shifts.
Here, we use $\mathbb{E}_{\phi}[\cdot]$ to denote averaging over phase shifts.

We show the results of the calculation of the charge transport capacity of the quasiperiodic model in Fig.~\ref{fig:AA_CTC.png}.
In the delocalized phase $h<2$, the distributions of the transport capacity shift rightwards, with the phase averaged CTC growing linearly with $L$ with a slope of roughly $\sim 1/2$, consistent with the free-particle case.
Interestingly, at exactly $h=2$, it appears that the mean of the charge transport capacity scales linearly with a slope of $\sim 0.27$. 
Curves in the delocalized regime with $h<2$ have slopes larger than $\sim 0.27$.
We again note that the slope of the $\mathrm{mean}(\ctc)$ curve being above $1/4$ is consistent with ergodicity.
For the localized phase $h>2$, the distributions instead quickly converge to a fixed distribution with increasing $L$. 
Different from the Anderson case, the converged distributions appear to have fast power-law tails, i.e., $P[\ctc] \sim 1/\mathbb{E}[\ctc]^4$ for large $\ctc$, seemingly independent of the choice of the rational number inside the potential.
The emergence of a power-law tail here highlights a difference between Anderson and quasiperiodic localization and deserves further investigation.
Furthermore, in this regime, the mean of the transport capacity converges quickly with $L$ for large enough $h$, consistent with the distributions converging. 

The scaling of the disorder-averaged transport capacity is strikingly different in the delocalized and localized phases.
This highlights the power of $\ctc$ in probing a delocalization-localization transition.

\subsection{Non-interacting Generalized Aubry-André Model}\label{sec:ctc_gaa_model}

\begin{figure*}[t]
    \centering
    \includegraphics[width=\linewidth]{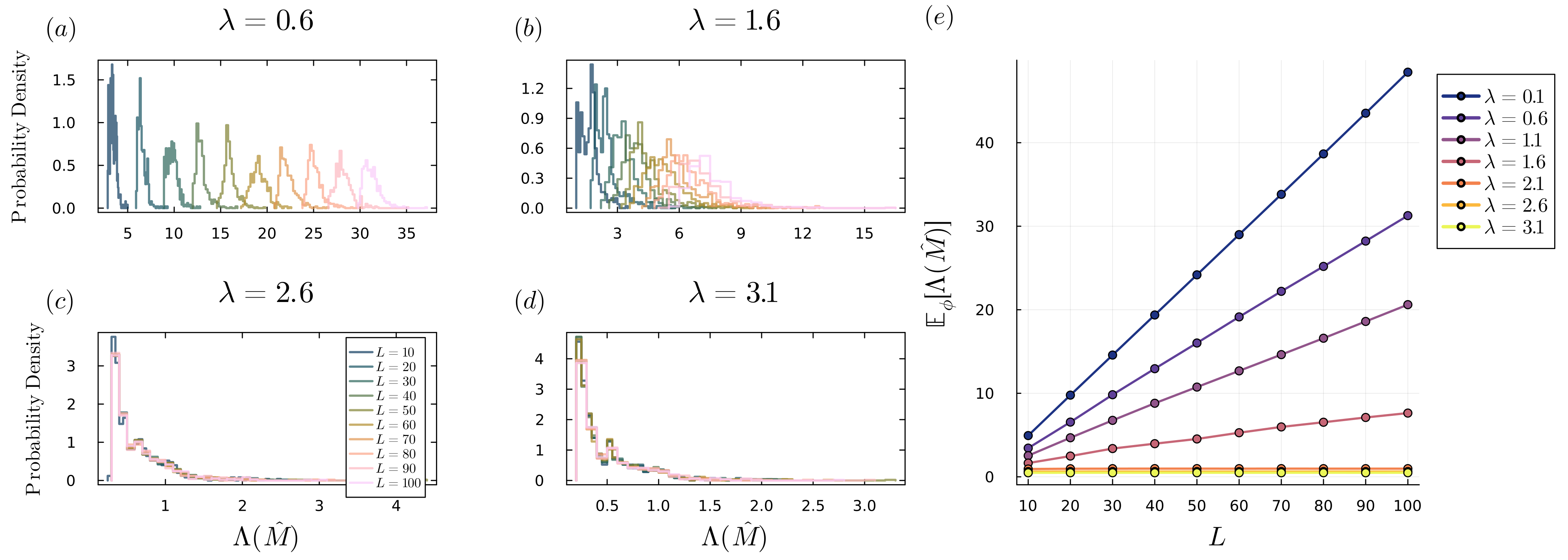}
    \caption{Analysis of the charge transport capacity $\ctc$ in the Generalized Aubry-André (GAA) model for $\alpha=0.5$. Panels (a)-(d): distributions of $\ctc$, with each panel representing a different potential strength $\lambda$, and each colored curve representing a distribution at different system size $L$, from $10$ to $100$ in steps of 10.
    Panel (e): the $\ctc$ averaged over phases $\phi$ as a function of $L$, with each curve representing a different potential strength $\lambda$.
    The most striking feature of the curves is that in the regime of $\lambda$ where there is a nontrivial mobility edge (for this plot, that is $\lambda=0.6, 1.1, 1.6$), the slope of the linear curve varies with $\lambda$ and provides information about the fraction of delocalized states in the system.
    }
\label{fig:GAA_CTC.png}
\end{figure*}

Finally, we study the CTC numerically in the Generalized Aubry-André (GAA) model, a model realizing a mobility edge in 1D. Here, we take Eq.~(\ref{eq:noninteracting_hamiltonian}) with the onsite potentials
\begin{equation}\label{eq:gaa_model}
    h_j=2\lambda \frac{\cos(2\pi k(j+\phi))}{1-\alpha\cos(2\pi k(j+\phi))}~.
\end{equation}
This model has a mobility edge $E$ given by
\begin{equation}
    \alpha E = \mathrm{sgn}(\lambda)\left||t|-|\lambda|\right|~,
\end{equation}
which separates localized from extended eigenstates, where the localized (extended) states lie above (below) $E$ for $|\lambda| >|t|$, with the roles reversed for $|\lambda| <|t|$.
Here, $\alpha$ can be thought of as a tuning parameter for a duality-breaking perturbation: when $\alpha=0$, the model reduces to the regular Aubry-André model where the ``mobility edge" ceases to exist~\cite{ganeshan_nearest_2015}. 

We show numerical calculations of the charge transport capacity of the GAA model for $\alpha=0.5$ in Fig.~\ref{fig:GAA_CTC.png}.
Just like with the quasiperiodic model, we use $k=\sqrt{2}$ and calculate the transport capacity over 500 evenly spaced phase shifts $\phi$ between $0$ and $1/k$. For $\lambda=0.1$, all of the eigenstates are delocalized, and we see the typical ergodic scaling $\mathbb{E}_{\{h_i\}} \sim L/2$. 
For $0.6\leq\lambda\leq 2.1$, there exists a nontrivial mobility edge. 
In this regime, we verify that indeed, $\mathbb{E}_{\phi}[\ctc] \sim \alpha L/2$, as conjectured.
Intuitively, as long as an extensive number of delocalized states exist (which occurs when there exists a mobility edge), the system will always be able to transport an extensive number of charges via those delocalized states through an effectively smaller system, and this is reflected in the linear scaling of $\ctc$. 
Thus, the scaling of the charge transport capacity with $L$ provides a clear signature of the presence of a mobility edge that also provides some information about where the mobility edge is located in the spectrum.

One can numerically verify that, for the $\lambda$ investigated in Fig.~\ref{fig:GAA_CTC.png}, all of the eigenstates in the spectrum are localized starting from $\lambda=2.6$, and we see that the transport capacity saturates quickly with $L$, reflecting the lack of ergodicity in this regime just as in the Anderson model and the localized regime of the quasiperiodic model.

\section{Proof of Locality of $\hat M$ in the Finite Range LIOM Model}\label{app:derivation_local_m_lioms}

In this section, we prove that the finite-range LIOM model implies that the resulting $\hat M$ will also be finite-ranged and localized near the cut at $i_0$.
Our set up in this section is inspired by the LIOM model proposed in~\cite{aceituno_chavez_ultraslow_2024}.
Here, for clarity and convenience, we elect to work in the spin language rather than the particle occupation language used in this paper.

We consider a chain of spinless fermions, with $U$(1) charge conservation. In particular, a model of this type can be obtained from our fermionic chain, Eq.~(\ref{eq:interacting_hamiltonian}), using a Jordan-Wigner transformation:
\begin{equation}
    \hat H= -\frac{t}{2}\sum_{i=1}^{L-1} [\hat \sigma^1 _{i} \hat \sigma^1_{i+1} + \hat \sigma^2 _{i} \hat \sigma^2_{i+1}]+\frac{J}{4}\sum_{i}^{L-1}\hat \sigma^3_{i} \hat \sigma^3_{i+1} + \sum_{i=1}^{L} \frac{h_i}{2}\hat \sigma^3_i + C~.
\end{equation}
Here, the upper index $\alpha=0,1,2,3$ denotes the 4 Pauli matrices, and the lower index denotes the position that the Pauli operator acts on, so that $\sigma ^0=I, \sigma^1=X, \sigma^2=Y, \sigma^3=Z$, and the operators $\hat \sigma_j^\alpha$ act on site $j$.

Let us now assume that the Hamiltonian can be rewritten in the form:
\begin{equation}
    \hat H= \sum_{i}  \epsilon_i \hat \tau^3_i + \sum_{i>j} J_{i,j} \hat \tau^3_i \hat \tau^3_j+\sum_{i> j>k} J_{i,j,k} \hat \tau^3_i \hat \tau^3_j \hat \tau^3_k + \dots , 
\end{equation}
where $\hat \tau_i^3$, called local integrals of motion (LIOMs), are deformations of the single-site Pauli operators that can be constructed from $\hat\sigma^3_i$ using a finite-depth unitary (hence, the unitary transforms $\hat H$ into a form that is diagonal in the computational basis):
\begin{equation}
    \hat \tau_i^3=\hat U^\dagger \hat \sigma^3_i \hat U~.
\end{equation}
Clearly, $\hat \tau_i^3$ are conserved quantities, and they mutually commute. One can define corresponding $X$ and $Y$ components of the LIOMs:
\begin{align}
    \hat \tau_i^1 = \hat U^\dagger \hat \sigma_i^1 \hat U~,\\
    \hat \tau_i^2 = \hat U^\dagger \hat \sigma_i^2 \hat U~.
\end{align}
Because $\hat U$ is a finite-depth circuit, $\hat \tau_i^3$ can be decomposed into Pauli strings with support on $[i-r_{\text{max}}, i+r_{\text{max}}]$ for some integer $r_{\text{max}}$, i.e., 
\begin{equation}\label{eq:finite_range_tau}
    \hat \tau_i^\alpha=\sum_{\substack{ \{I\}\ni i:\\\text{range}(\{I\})\leq r_{\text{max}}}  } \sum_{\{\mu_I\}} h^{\{\mu_I\}}_{\{I\}} \hat \sigma^{\{\mu_I\}}_{\{I\}}
\end{equation}
for $h^{\{\mu_I\}}_{\{I\}}\in\mathbb{R}$.
Here, $\{I\}$ and $\{\mu_I\}$ denote ordered sets, so that $\hat \sigma^{\{\mu_I\}}_{\{I\}}=\hat \sigma^{\mu_1}_{j_1}\hat \sigma^{\mu_2}_{j_2}\dots\hat \sigma^{\mu_{\{I\}}}_{j_{|\{I\}|}}$ for $j_1< j_2<\dots\in\{I\}$ and $\mu_1,\mu_2,\dots\mu_{|\{I\}|}\in\{\mu_I\}$, where the first and last indices $\mu_1, \mu_{\{I\}}$ can take values $1,2$ or $3$, while the others can be  $\mu_j=0,1,2,3$. 
We define $\text{range}(\{I\})=|j_{|\{I\}|}-j_1|$. 
Together, $I, \tau^1, \tau^2, \tau^3$ and strings of them form an orthonormal basis (with respect to the Hilbert–Schmidt inner product) for the real vector space of traceless Hermitian operators.

With these assumptions, we want to show that the charge transport capacity operator, $\hat M = \hat N_R-[\hat N_R]_{\text{diag}}$ is of the form
\begin{equation}
    \hat M = \sum_{\substack{ \{I\}\ni i:\\\text{range}(\{I\})\leq r}  } \sum_{\{\mu_I\}} k^{\{\mu_I\}}_{\{I\}} \hat \sigma^{\{\mu_I\}}_{\{I\}}~.
\end{equation}
For some finite $r$ so that $\hat M$ is an operator with finite range $r$ around $i_0$. 

To prove this, we will make use of the following lemma.
\begin{lemma}\label{lem:lemma_finite_range_ctc}
    Given Eq.~(\ref{eq:finite_range_tau}), and that $U$ is a finite-depth circuit, one can rewrite $\sigma_i^3$ in terms of finite-range strings of $\tau^\mu_i$
\begin{equation}
    \hat \sigma^3_i=
    \sum_{\substack{ \{I\}\ni i:\\\text{range}(\{I\})\leq r_\text{max}'}  }
    \sum_{\{\mu_I\}} 
    k^{\{\mu_I\}}_{\{I\}} \hat \tau^{\{\mu_I\}}_{\{I\}}~.
\end{equation}
Here, $r_{\text{max}}'$ is not necessarily equal to $r_{\text{max}}$.
\end{lemma}
We now prove Lemma~\ref{lem:lemma_finite_range_ctc}. First, we note that the strings $\tau^{\{\mu_I\}}_{\{I\}}$ form an orthonormal basis (with respect to the Hilbert–Schmidt inner product) for the real vector space of traceless Hermitian operators. 
Thus, we can expand $\hat \sigma ^3_i$ in terms of $\hat \tau^{\{\mu_I\}}_{\{I\}}$, as so:
\begin{equation}
    \hat \sigma ^3_i=\sum_{\{I\},\{\mu_I\}} \left\langle \hat \tau^{\{\mu_I\}}_{\{I\}}|\hat \sigma ^3_i\right\rangle\hat \tau^{\{\mu_I\}}_{\{I\}}~,
\end{equation}
where
\begin{equation}
    \left\langle \tau^{\{\mu_I\}}_{\{I\}}|\sigma ^3_i\right\rangle = \mathcal D^{-1} \Tr[\tau^{\{\mu_I\}}_{\{I\}} \hat \sigma ^3_i].
\end{equation}
By construction, $\tau^{\{\mu_I\}}_{\{I\}}=\hat \tau^{\mu_1}_{i_1}\hat \tau^{\mu_2}_{i_2}\dots=U\hat \sigma^{\mu_1}_{i_1}\hat \sigma^{\mu_2}_{i_2}\dots U^\dagger, \mu_1, \mu_2\dots \in \{\mu_I\}$, so we have
\begin{align}\label{eq:inner_product_sigma_z}
    \Tr\left[\hat \sigma ^3_i\hat \tau^{\{\mu_I\}}_{\{I\}}\right] = \Tr[\hat \sigma ^3_iU\hat \sigma^{\mu_1}_{i_1}\hat\sigma^{\mu_2}_{i_2}\dots U^\dagger]\nonumber\\
    =\Tr[U^\dagger\hat \sigma ^3_iU\hat \sigma^{\mu_1}_{i_1}\hat \sigma^{\mu_2}_{i_2}\dots]
\end{align}
Crucially, we invoke the fact that $U$ is a finite-depth circuit here so that $U^\dagger\hat \sigma ^3_iU$ is an operator with at most range $[i-r_{\text{max}}', i+r_{\text{max}}']$, where $r_{\text{max}}'$ is not necessarily equal to $r_{\text{max}}$, but is finite.
Thus, in order for Eq.~(\ref{eq:inner_product_sigma_z}) to be nonzero (note that Pauli strings form an orthonormal basis in the linear space of traceless Hermitian operators), $\{I\}$ must be of range at most $r_{\text{max}}'$ around $i$.
Thus, $\hat \sigma ^3_i$ is a sum of strings of $\hat \tau^\mu_j$ such that $j\in [i-r_{\text{max}}', i+r_{\text{max}}']$.

The correctness of the lemma implies the desired result. 
From the lemma, we can rewrite $\hat N_L$ in terms of $\hat \tau_i$:
\begin{align}
    \hat N_L&=\frac{1}{2}\lp\sum_{i=1}^{i_0} \hat \sigma^3_i\rp+\frac{L}{4}\nonumber \\
    &=\frac{1}{2}\lp\sum_{i=1}^{i_0}\sum_{\text{range}(\{I\})\leq r_{\text{max}'}, i\in\{I\}, \{\mu_I\}} k^{\{\mu_I\}}_{\{I\}} \hat \tau^{\{\mu_I\}}_{\{I\}}\rp+\frac{L}{4}~.
\end{align}
Similarly, for $\hat N_R$:
\begin{align}\label{eq:n_r_lioms}
    \hat N_R&=\frac{1}{2}\lp\sum_{i=i_0+1}^{L} \hat \sigma^3_i\rp+\frac{L}{4}\nonumber\\
    &=\frac{1}{2}\lp\sum_{i=i_0+1}^{L}\sum_{\substack{\text{range}(\{I\})\leq r_{\text{max}}'\\ i\in\{I\}, \{\mu_I\}}} k^{\{\mu_I\}}_{\{I\}} \hat \tau^{\{\mu_I\}}_{\{I\}}\rp+\frac{L}{4}~.
\end{align}
Now, recall that one can rewrite $\hat M$ in terms of either $\hat N_L$ or $\hat N_R$. i.e.
\begin{equation}\label{eq:m_n_r_n_l}
    \hat M=\hat N_R-[\hat N_R]_{\text{diag}}=-\hat N_L+[\hat N_L]_{\text{diag}}~.
\end{equation}
Using the lemma, we rewrote $[\hat N_R]_{\text{diag}}$ as a sum of strings of $\hat \tau^3$ only. 
These terms are subtracted out from the sum in Eq.~(\ref{eq:n_r_lioms}), leaving only strings containing $\hat \tau^1, \hat \tau^2$ in them.
By construction, then, $\hat N_R-[\hat N_R]_{\text{diag}}$ at most only has support between the positions of range $[i_0+1-2r_{\text{max}}', L]$.
By a similar argument, $\hat N_L-[\hat N_L]_{\text{diag}}$ at most has support in the spatial interval $[1, i_0+2r'_{\text{max}}]$.
In order for the equality in Eq.~(\ref{eq:m_n_r_n_l}) to hold, $\hat M$ must only have support between $[i_0+1-2r_{\text{max}}', i_0+2r_{\text{max}}']$. 
Therefore, $\hat M$ is a finite-ranged operator, concluding the proof.
A corollary to this result is that the CTC $\ctc$ is finite as $L\rightarrow\infty$.

Although its proof is beyond the scope of this paper, we suspect that a similar argument will hold for the case where the LIOMs have strict exponential tails, in the Lieb-Robinson sense. 
We verify this claim numerically in the next Appendix~\ref{app:LIOM_model_results}, where we observe that the artificial LIOM model (with LIOMs that decay exponentially in the sense of Lieb-Robinson bounds) indeed shows numerically observable saturation at system sizes accessible to ED. 

\section{Charge transport capacity in the artificial LIOM model}\label{app:LIOM_model_results}
In this Appendix, we provide further numerical evidence for the argument that the presence of truly quasilocal LIOMs, quasilocal in the Lieb-Robinson sense, implies that the CTC will be finite in the thermodynamic limit.
Notably, a truly quasilocal LIOM model would show signatures of the saturation of the CTC even at the numerically accessible system sizes to ED. 
We also show that the distributions of the CTC for the truly quasilocal LIOM model are similar in spirit to those of Anderson localization.
To this end, we numerically computed the CTC in an artificial LIOM model, inspired in spirit by the artificial LIOM model in~\cite{aceituno_chavez_ultraslow_2024}.

In this artificial model, we construct the LIOMs $\widetilde n_{j}$ through the short-``time'' evolution of the number operators $\hat n_j$ under an extensive local and hermitian operator $\hat X$ with disorder and $U(1)$ symmetry:
\begin{equation}\label{eq:artificial_l_bits}
    \widetilde n_{j}(t) = e^{i\hat X t}\hat n_j e^{-i\hat X t} ~. 
\end{equation}
Here, $t$ is a real-valued parameter.
If $\hat X$ is an extensive local Hermitian operator, the Lieb--Robinson bound states that:
\begin{equation}
   \| [\widetilde n_{j}(t), \hat B]\|\leq C\|\hat{n}_j\| \|\hat B\|e^{-\mu [d(n_{j}, B) - v_{\text{LB}}t]}~.
\end{equation}
Here, $\hat B$ is some operator that has disjoint support from $n_{j}$, and $v_{\text{LB}}, \mu, C$ are constants, with $v_{\text{LB}}$ known as the Lieb-Robinson velocity, and $d(n_{j}, B)$ is the distance (in lattice spacings) between the supports of $n_{j}$ and $B$.
This property shows that $\widetilde n_j(t)$ develops exponentially small tails outside the light cone $d \gtrsim v_\text{LB}|t|$ set by the Lieb--Robinson velocity~\cite{lieb_finite_1972}.

For $\hat X$, we use the following extensive local operator:
\begin{equation}\label{eq:liom_model_hamiltonian}
\hat X = -x \sum_{i=1}^{L-1} \big[ \hat c_i^\dagger \hat c_{i+1} + \hat c_{i+1}^\dagger \hat c_i \big] 
+ \sum_{i=1}^{L} y_i \, \hat c_i^\dagger \hat c_i 
+ z \sum_{i=1}^{L-1} \hat c_i^\dagger \hat c_{i+1}^\dagger \hat c_{i+1} \hat c_i .
\end{equation}
Here, $y_i\sim \mathrm{Unif[-Y, Y]}$ for $Y\in\mathbb{R}$, where the presence of disorder allows the resulting LIOMs to have variation. The parameters $x=1$ and $z=0.1$ are fixed. One can tune the ``localization length'' of the LIOMs by changing the range of the random variable $X$ of the onsite potentials and the length of the time evolution $t$. 

These artificially quasilocal LIOM models successfully capture even surprising aspects of MBL phenomenology---such as the double-logarithmic growth of the number entropy and the logarithmic growth of the entanglement entropy with time~\cite{aceituno_chavez_ultraslow_2024}.
In contrast to Ref.~\cite{aceituno_chavez_ultraslow_2024}, our LIOMs are constructed from continuous time-evolution, taking advantage of the Lieb-Robinson bound. 
We once again emphasize that we do not believe that LIOMs in a true MBL phase will be of precisely this structure.
Rather, our goal is to study how the CTC behaves in the presence of quasilocal conserved quantities for an interacting system at the system sizes accessible to ED, with the goal of guiding the analysis of how the CTC should behave in the true MBL problem.
\begin{figure*}[t]
    \centering
    \includegraphics[width=0.85\linewidth]{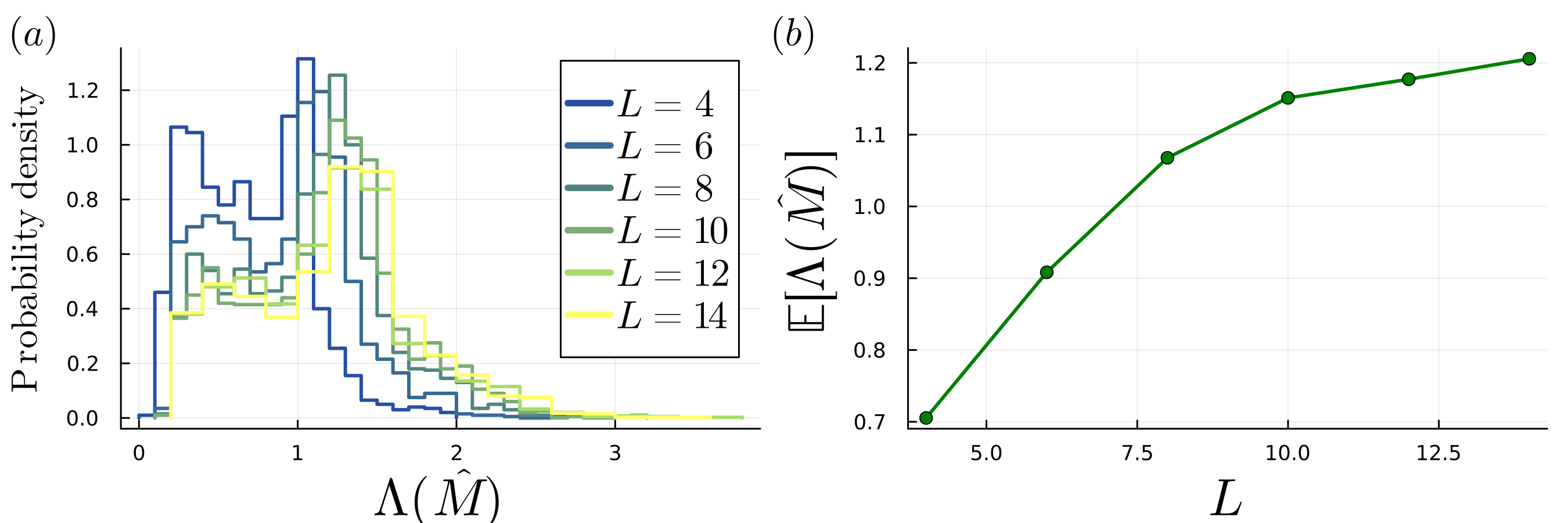}
    \caption{Analysis of the CTC behavior in the artificial LIOM model as described in Sec.~\ref{app:LIOM_model_results}, where LIOMs are constructed through short-time evolution of single-site Pauli operators according to Eqs.~(\ref{eq:artificial_l_bits}) and (\ref{eq:liom_model_hamiltonian}) with parameters $Y=10$ and $t=20$.
    Panel (a): probability distributions of the CTC over 2000 disorder realizations of the LIOMs. Each colored curve represents a different system size as indicated by the legend. Panel (b): the disorder averaged CTC as a function of $L$ as calculated in this LIOM model.}
\label{fig:l-bit-model-results}
\end{figure*}

We show the results of the CTC in the LIOM model in Fig.~\ref{fig:l-bit-model-results}. 
Observing the distributions of the CTC as a function of $L$ in Fig.~\ref{fig:l-bit-model-results}(a), we see that the probability distribution functions (PDFs) show two peaks, one at a small nonzero value, and the other at the integer value 1, with exponentially decaying tails at large CTC.
Furthermore, in Fig.~\ref{fig:l-bit-model-results}(b), we plot the disorder-averaged CTC as a function of $L$. We note that even at these small system sizes, the disorder-averaged CTC already appears to show signs of saturation, with the growth of the curve visibly slowing down.
We conclude here that the properties of the CTC in this interacting LIOM model look similar to those of the Anderson insulator, and that if there were truly quasilocal LIOMs present in Eq.~(\ref{eq:interacting_hamiltonian}) with exponentially decaying tails in the Lieb-Robinson sense, then there would be signatures of the saturation of $\ctc$ with $L$ even at the sizes available to ED.

\section{Additional data on the absolute-value-state overlap of $|\ket{\maxstate}| \cdot |\ket{\minstate}|$ and related quantities}\label{app:overlap}
\begin{figure*}[t]
    \centering
    \includegraphics[width=\linewidth]{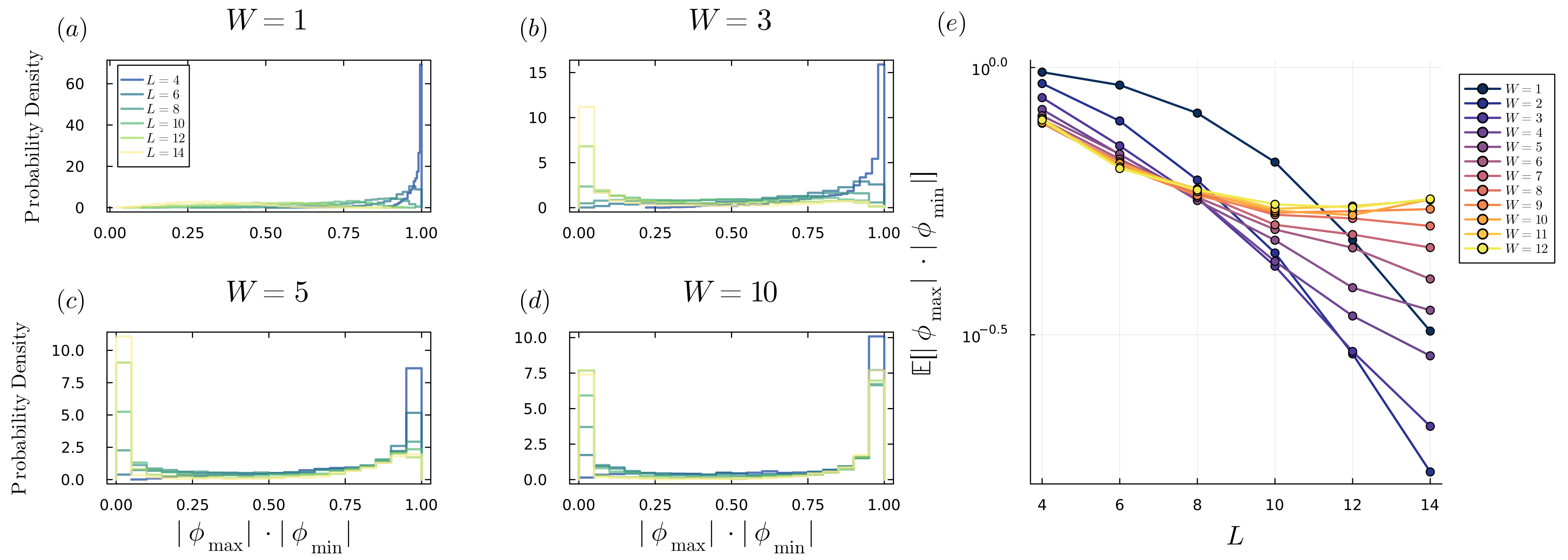}
\caption{Analysis of the absolute-value-state overlap $|\ket{\phi_\text{max}}| \cdot |\ket{\phi_\text{min}}|$ for $J=0.1$ in the interacting Anderson model, where $\ket{\maxstate}$ and $\ket{\minstate}$ are eigenstates of the CTC operator $\hat M$ with the maximum and minimum eigenvalues respectively.
Panels (a)--(d): Calculated probability density functions of $|\ket{\phi_\text{max}}| \cdot |\ket{\phi_\text{min}}|$ over $4000$ disorder realizations.
Each panel represents a different disorder strength indicated in its title, with each colored curve representing a different system size labeled by the legend.
Panel (e): the disorder-averaged $|\ket{\phi_\text{max}}| \cdot |\ket{\phi_\text{min}}|$ as a function of $L$, with each curve representing a different disorder strength.}
\label{fig:overlap_j_01.png}
\end{figure*}

\begin{figure*}[t]
    \centering
    \includegraphics[width=0.85\linewidth]{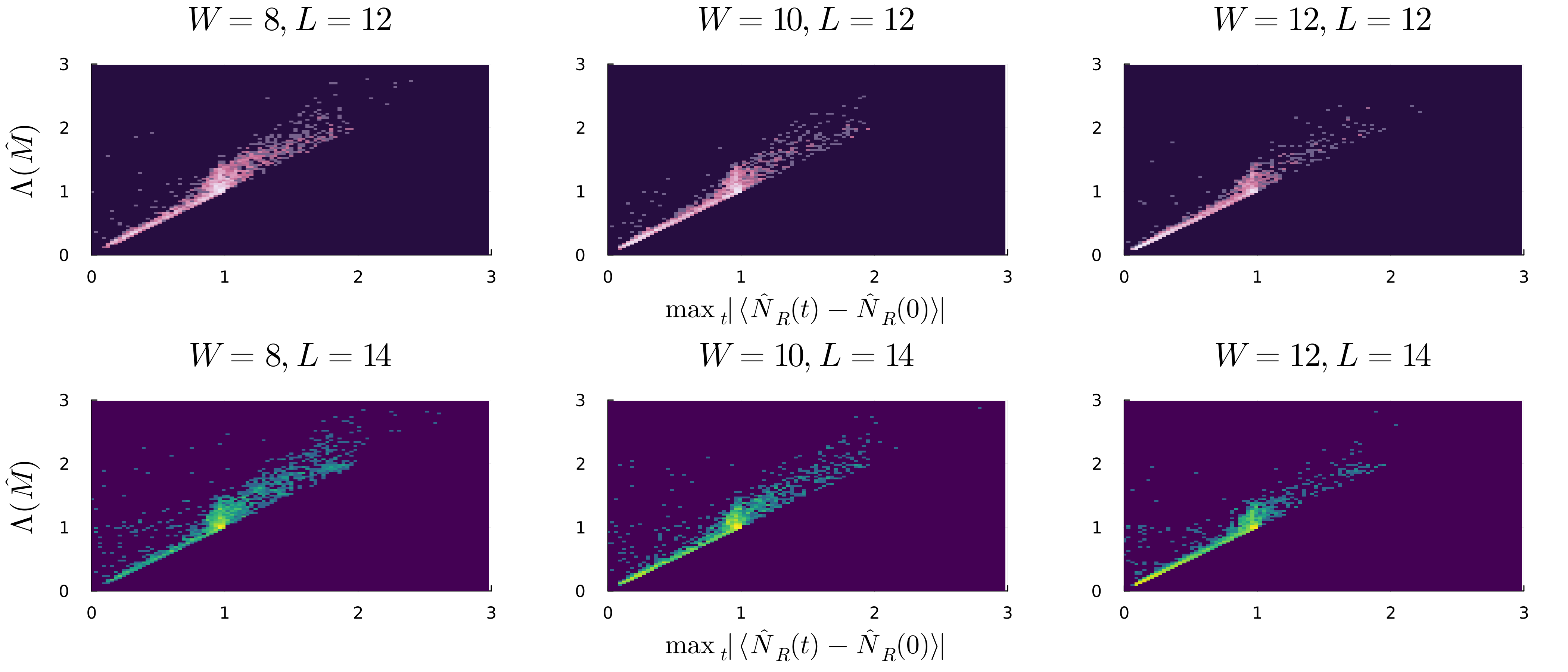} 
    \caption{Density heat maps comparing the CTC ($y$-axis) with $\max_t{|\langle\hat{N}_R(t)-\hat{N}_R(0)\rangle|}$ in a quench dynamics simulation ($x$-axis) in the strong disorder regime and for $J=0.1$.
    The quench evolves $|\phi_{\min}\rangle$ as an initial state, to times $t\in [0,10^5]$ in steps of $\Delta t=10$.
    Each panel corresponds to a different disorder strength $W$ and system sizes $L=12$ or $14$, with 2000 disorder realizations per panel.
    The color scale, shown on a logarithmic scale, ranges from dark (low density of instances) to light (high density).
    }
\label{fig:scatter_quench_ctcb_vs_ctcb}
\end{figure*}

\begin{figure*}[t]
    \centering
    \includegraphics[width=\linewidth]{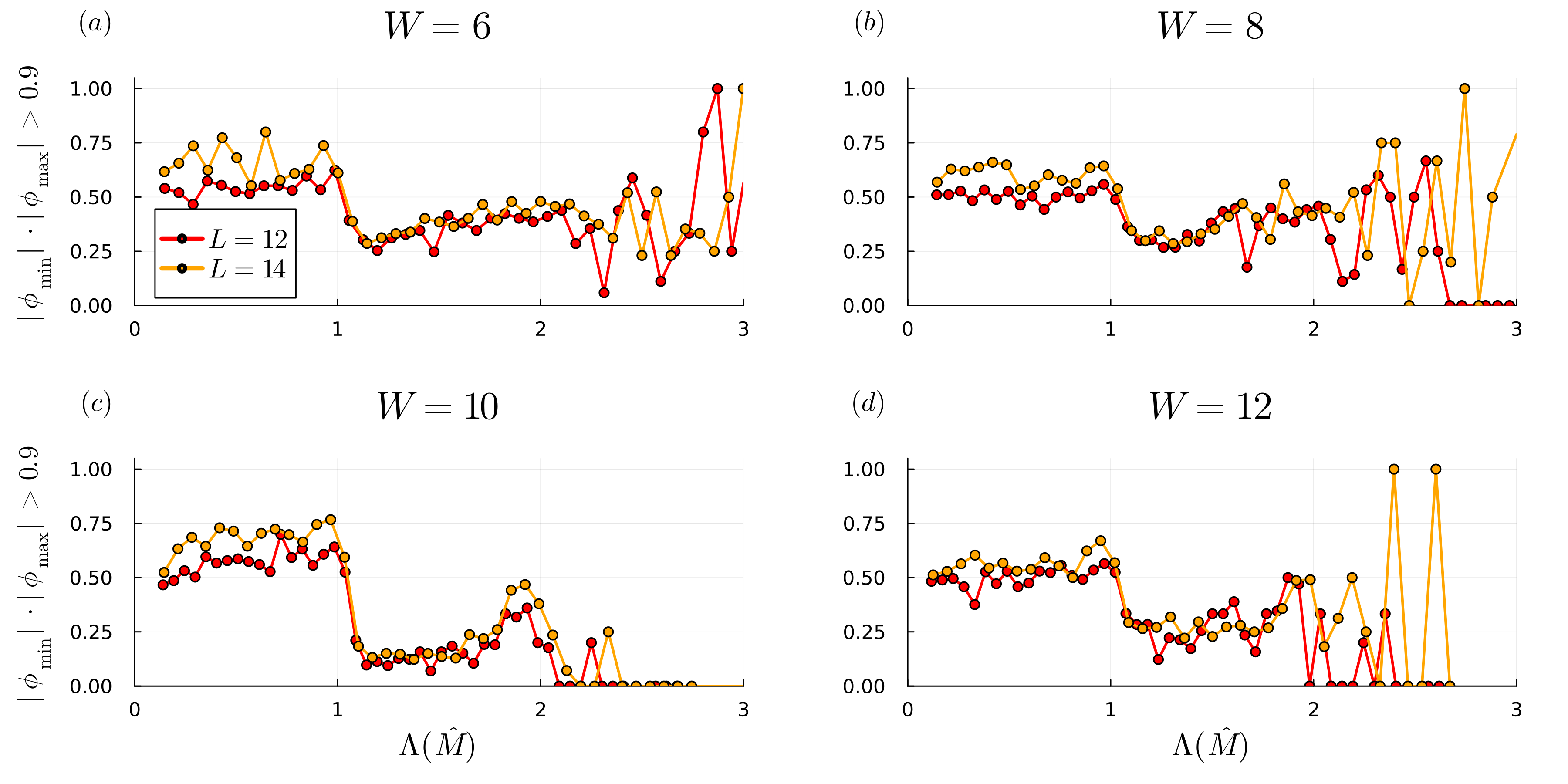} 
    \caption{Fraction of disorder realizations ($y$-axis) as a function of the value of $\ctc$ ($x$-axis) which have the absolute-value-state overlap $|\ket{\maxstate}|\cdot|\ket{\minstate}|>0.9$, indicating a charge transport resonance-like structure of $\ket{\minstate}$ and $\ket{\maxstate}$. Each panel represents a different disorder strength, with $W=6, 8, 10, 12$ selected. Within each panel, there are curves representing $L=12$ (red) and $L=14$ (orange). }
\label{fig:overlap_fraction}
\end{figure*}

In this section, we present additional data on the following quantities related to Sec.~\ref{sec:ctc_is_a_tight_bound}:
\begin{enumerate}
    \item In Fig.~\ref{fig:overlap_j_01.png}, we show the analysis of $|\ket{\maxstate}|\cdot|\ket{\minstate}|$.
    In panels (a)-(d) we have the probability distributions of $|\ket{\maxstate}|\cdot|\ket{\minstate}|$ over 2000 disorder realizations. 
    In panel (e), we show the disorder averaged $|\ket{\maxstate}|\cdot|\ket{\minstate}|$ as a function of $L$.
    \item In Fig.~\ref{fig:scatter_quench_ctcb_vs_ctcb}, we show a comparison of the CTC with predictions of quench dynamics starting from $\ket{\minstate}$ as an initial state.
    This is the counterpart to Fig.~\ref{fig:ctcb_is_tight_bound}(a) of the main text, while here, we present data for additional disorder strengths and system sizes $L=12$ and $L=14$.
    \item In Fig.~\ref{fig:overlap_fraction}, we show the fraction of disorder realizations with $|\ket{\maxstate}|\cdot|\ket{\minstate}|>0.9$ as a function of the CTC. 
    Notably, we observe that at $\ctc\approx2$, the fraction of instances with $|\ket{\maxstate}|\cdot|\ket{\minstate}|>0.9$ is more than half of the total disorder realizations for $W=9, 12$, and close to $1/2$ for $W=8, 10$.
    As discussed in Sec.~\ref{sec:ctr_properties}, observing $|\ket{\maxstate}|\cdot|\ket{\minstate}|>0.9$ in a given disorder realization is a strong indication that the large CTC is caused by a charge transport resonance involved in a nearly perfect cat state, such as the example CTR presented in the main text of the paper in Sec.~\ref{sec:ctr_properties}.
    Thus, our conclusion here is that the example presented in Sec.~\ref{sec:ctr_properties} and the majority of the main text is representative of samples that lie inside the integer peaks of the CTC distributions of Fig.~\ref{fig:ctcb_main_results_J_01}(c).
    We also briefly note that interestingly, there seem to be anomalous peaks at large values of $\ctc$, indicating that these instances mostly have very large overlap. This could indicate that at these disorder strengths, a very large CTC is usually caused by a charge transport resonance. 
    This is consistent with our understanding of charge transport resonances in this regime.
\end{enumerate}

\section{Additional examples of charge transport resonances}\label{app:more_ctr_examples}

\begin{figure*}[t]
    \centering
    \includegraphics[width=0.65\linewidth]{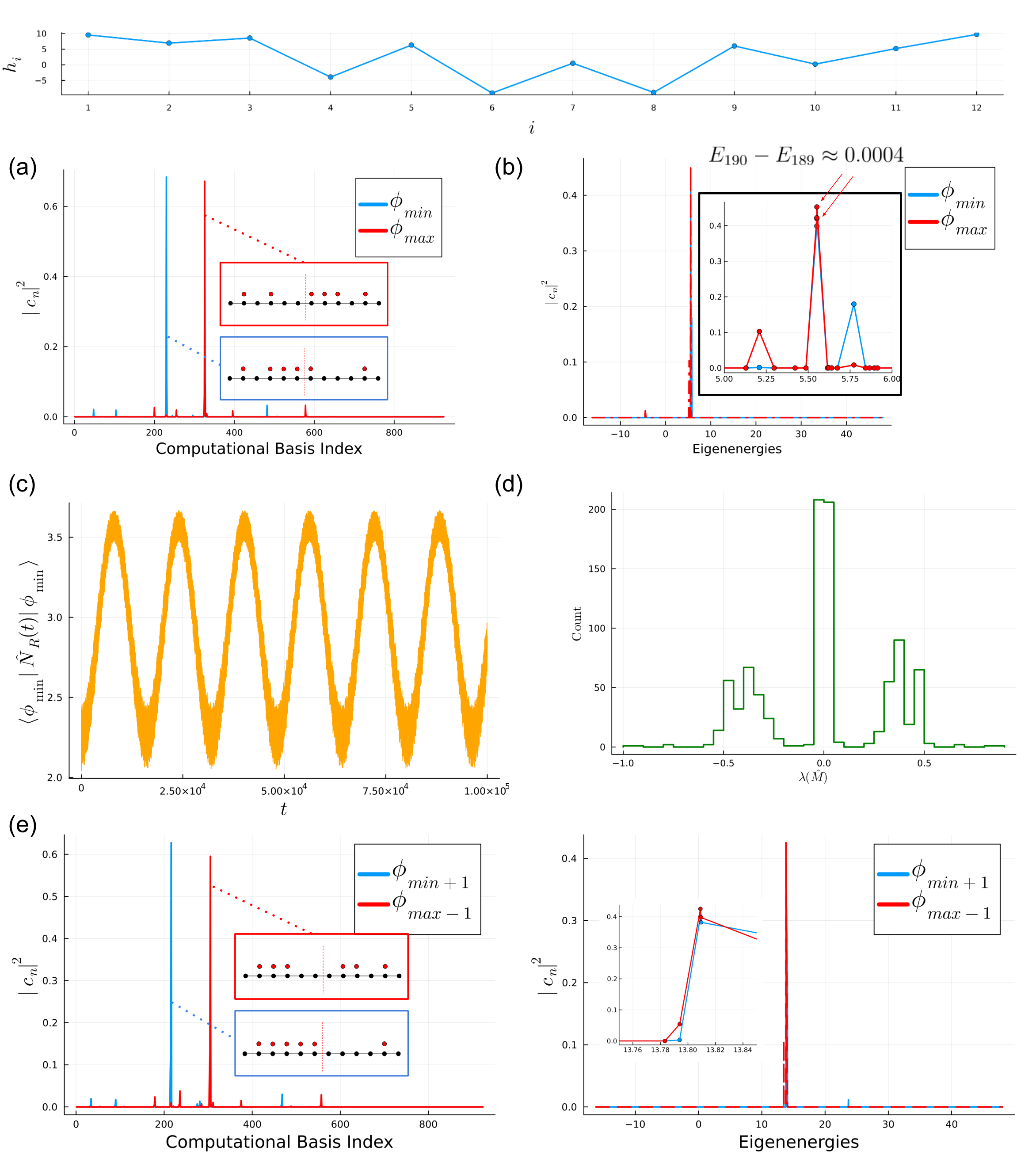}
    \caption{Illustration of an additional example of a CTR which has $\ctc=1.836, L=12, W=10$.
    The absolute-value-overlap in this case is $|\ket{\maxstate}|\cdot|\ket{\minstate}|\approx 0.90$.
    The subplots in this figure should be interpreted in the same way as in Fig.~\ref{fig:example_ctr}, and will not be repeated here for concision.
    The splitting between the two eigenstates that have most of the weight on $\ket{\maxstate}$ and $\ket{\minstate}$ is $\Delta E_{\text{res}} \approx 0.0004$. In the inset, we show a wider scale showing nearby eigenstates also having some weight on these $\ket{\maxstate}$ and $\ket{\minstate}$.
    We make a few additional notes: 
    First, in the CTC operator DOS panel (d), there are $4$ eigenstates at the edges of the DOS plot, with eigenvalues $-0.87$, $-0.83$, $0.91$, and $0.97$.
    These two extra states are additional resonances involving the same active charge configuration as in (a) but with a different LIOM background.
    The relevant product states are shown in panel (e). 
    The relevant states are detectable by the following eigenstates of $\hat M$: $\phi_{\text{max}-1}$, representing the eigenstate of $\hat M$ with the second largest eigenvalue, and $\phi_{\text{min}+1}$, representing the eigenstate of $\hat M$ with the second smallest eigenvalue.
    We plot the expansion of these states in the computational basis (left) and the energy eigenbasis (right) in panel (e), with an inset zooming into the energy eigenbasis plot at the peak..
    }
    \label{fig:example_2_ctr}
\end{figure*}

\begin{figure*}[t]
    \centering
    \includegraphics[width=0.85\linewidth]{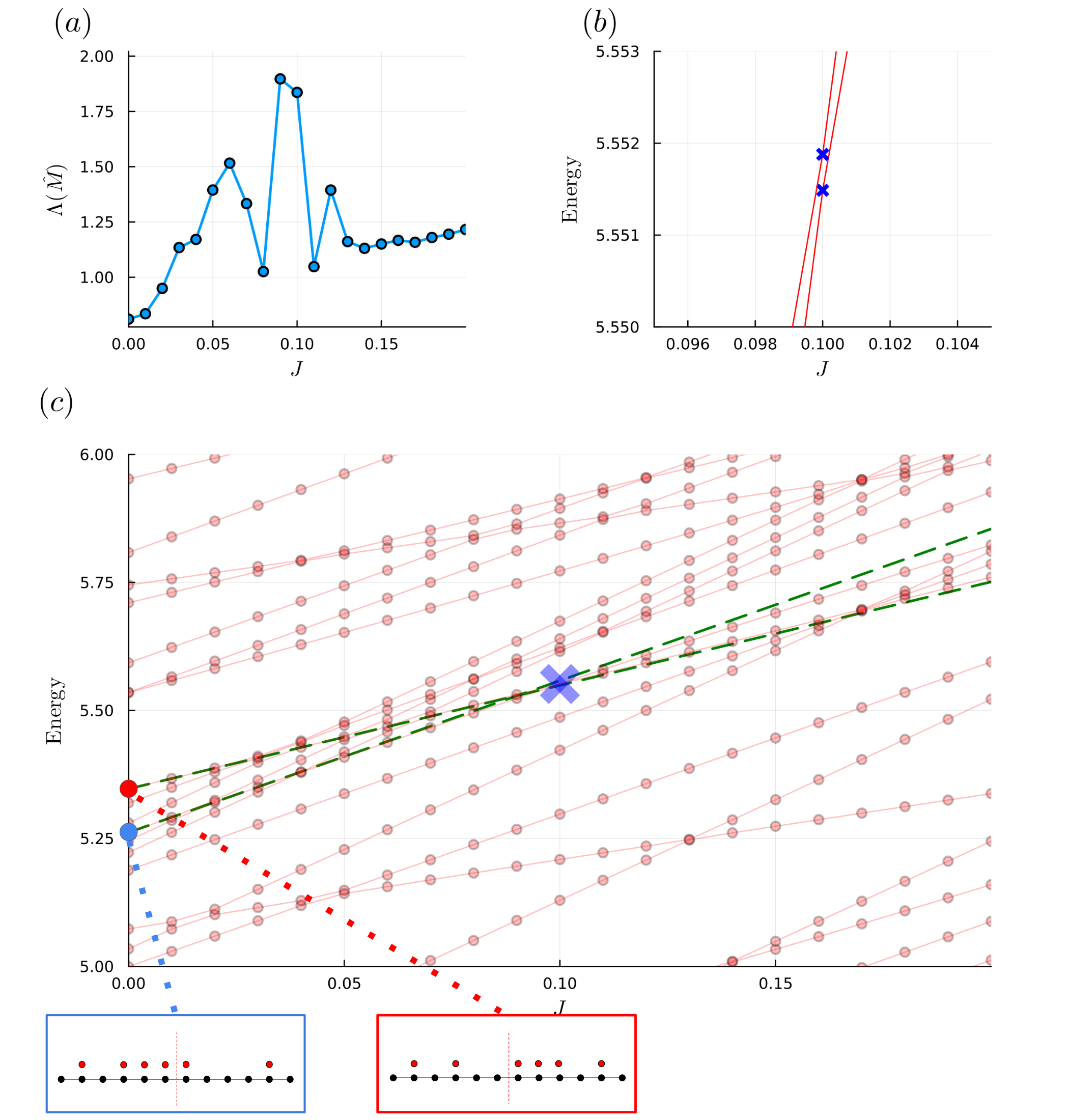}
    \caption{Illustration of the parametric level dynamics in the additional CTR example summarized in Fig.~\ref{fig:example_2_ctr}, panels (a)-(c). 
    The interpretation of this plot should be understood in the same way as in Fig.~\ref{fig:example_forbidden_crossing}, and will not be repeated here for the sake of concision. 
    }
    \label{fig:forbidden_level_example_2}
\end{figure*}

\begin{figure*}[t]
    \centering
    \includegraphics[width=0.65\linewidth]{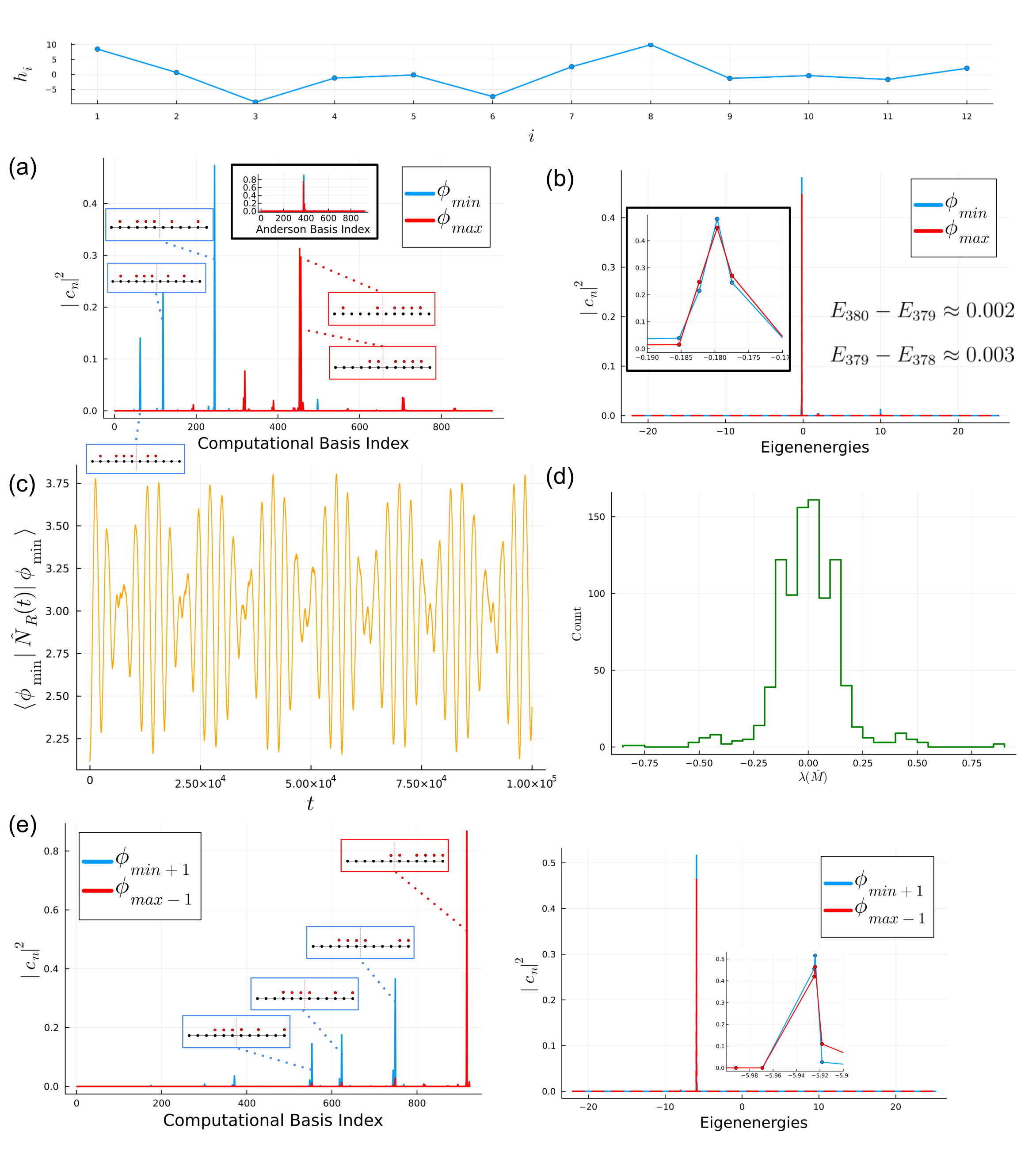}
    \caption{Illustration of a third example of a CTR which has $\ctc=1.736, L=12, W=10$.
    The absolute-value-overlap in this case is $|\ket{\phi_\text{max}}| \cdot |\ket{\phi_\text{min}}|\approx 0.98$.
    The interpretation of this figure is similar to Fig.~\ref{fig:example_2_ctr}.
    We note that in panel (a), we plot, in addition, in the inset $\ket{\maxstate}$ and $\ket{\minstate}$ expanded in the non-interacting Anderson model basis for this disorder realization, noting that these special states are localized in this basis, even though $\ket{\minstate}$ and $\ket{\maxstate}$ are less localized in the computational basis.
    This is because they include Anderson orbitals spread over 3 sites (sites $9$, $10$, $11$ for $\ket{\minstate}$ and $2$, $4$, $5$ for $\ket{\maxstate}$, i.e., single-particle resonances due to areas of flat potential energy we see in the profile $h_i$ on the top of the figure).
    We also note from panel (b) that this is a more complicated resonance apparently involving three energy eigenstates, with energy differences (illustrated in panel b) that are small compared to the average level spacing $\Delta E\approx 0.05$ but rather large compared to other examples of CTRs.
    This indicates that the off-diagonal matrix element was rather large for this instance.
    Finally, in panel (d), there are 4 eigenstates at the edges of the DOS plot, with eigenvalues -0.90, -0.88, 0.79, and 0.84.
    Similarly to Fig.~\ref{fig:example_2_ctr}(e), we show an analysis of this extra pair of resonating eigenstates in panel (e).
    We observe that in this case, the two CTRs in this spectrum have very different active regions.}
    \label{fig:example_3_ctr}
\end{figure*}

\begin{figure*}[t]
    \centering
    \includegraphics[width=0.85\linewidth]{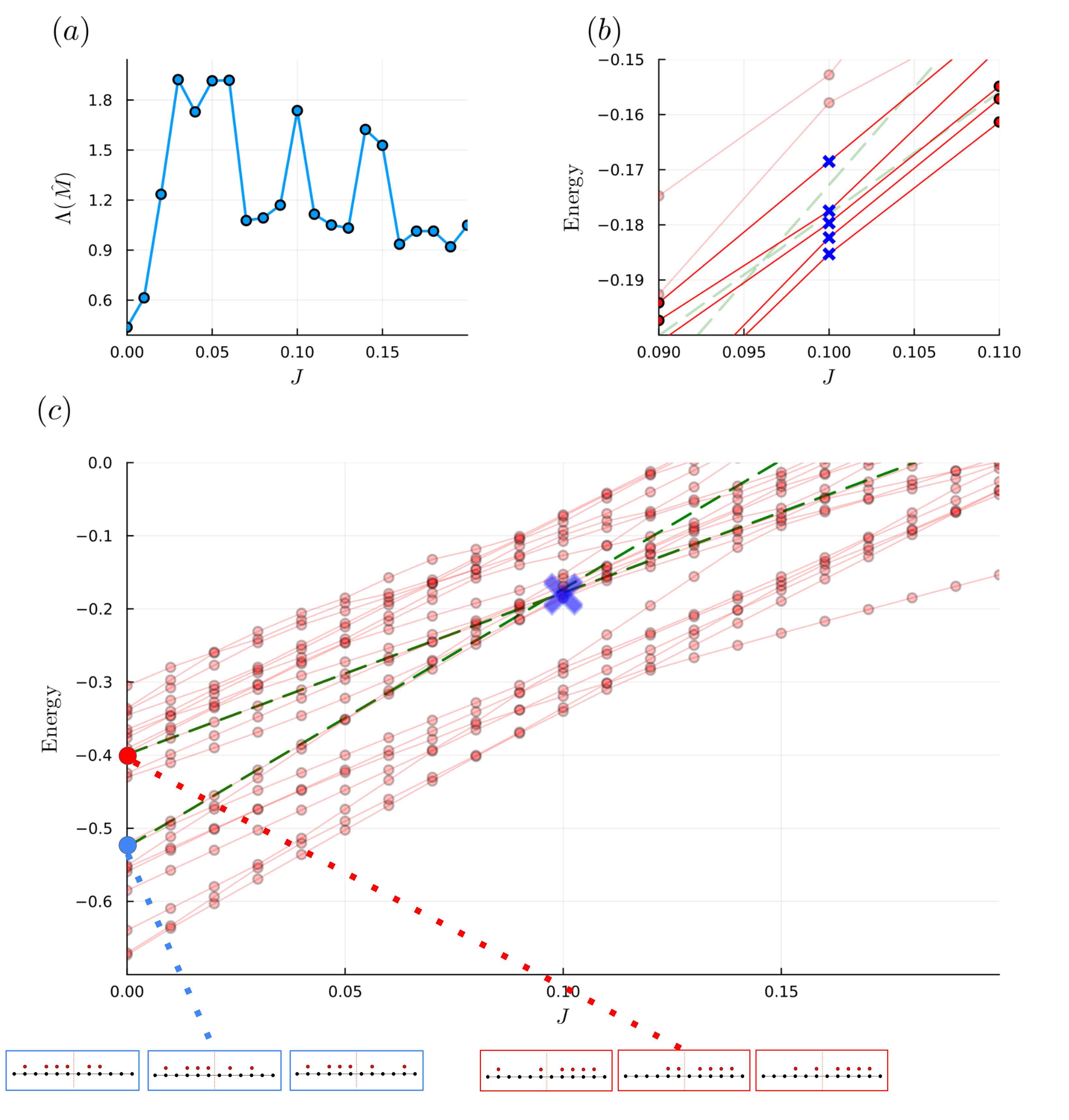}
    \caption{Illustration of the parametric level dynamics in the second example of a disorder realization with a CTR of $1.74$ particles, which was summarized in Fig.~\ref{fig:example_3_ctr}. 
    The interpretation of this plot should be understood in the same way as in Fig.~\ref{fig:example_forbidden_crossing}, and will not be repeated here for the sake of concision. 
    One extra note here is that the Anderson orbital configurations labeled by the red and blue circles at $J=0$ have significant weight on multiple product states. These product states are highlighted in the corresponding colors for each configuration.
    }
    \label{fig:forbidden_level_example_3}
\end{figure*}
In this section, we present some analysis for two additional examples of CTRs involving 2 particles. In particular, we focus on examples where the CTRs are more complex in structure than the one presented in the main text.

The first additional example of a disorder realization hosting a CTR is summarized in Fig.~\ref{fig:example_2_ctr}.
This example has $\ctc \approx 1.8$, $L=12$, $W=10$, and absolute-value-overlap $|\ket{\maxstate}| \cdot |\ket{\minstate}|\approx 0.9$.
The corresponding parametric level dynamics are illustrated in Fig.~\ref{fig:forbidden_level_example_2}.
We note that this CTR is an example of a ``messier'' CTR where the corresponding eigenstates do not have an equal superposition of two product states.
Here, $\ket{\minstate}$ and $\ket{\maxstate}$ rather closely correspond to the product states $\ket{010111100010},\ket{010100111010}$ respectively.
However, interestingly, we verified that the energy eigenstates which have significant weight in $\ket{\minstate}$ and $\ket{\maxstate}$ appear to have 3 product states significantly involved, specifically $\ket{010111100010}$, $\ket{010110110010}$, and $\ket{010100111010}$, which is hinted at in the messier expansion of $\ket{\maxstate}$ and $\ket{\minstate}$ in the energy eigenbasis.
Furthermore, we note that in this disorder realization, the spectrum of the Hamiltonian contains 2 pairs of energy eigenstates involved in a resonance, which happens to involve the same set of active sites (the parametric energy level dynamics for the second pair is not shown).
This additional CTR is detected by the pair of $\hat M$ eigenstates with the next smallest/largest eigenvalue, and we illustrate their properties in Fig.~\ref{fig:example_2_ctr}(e).
Compared with the other CTR, the background in this CTR differs by taking the charge at position 7 and moving it to position 3.
It must be noted that out of the $O(2^{L-4})$ possible background configurations (the exact number is $70$ for $L=12$), these two background configurations were the only ones that happened to quantitatively tune the energies into resonance.
Thus, this emphasizes that the resonance collar for this set of active sites is indeed larger than the system size $L=12$, aligning with our analysis in Sec.~\ref{sec:counting_resonances}.

The second additional example is summarized in Fig.~\ref{fig:example_3_ctr}, with corresponding parametric level dynamics analysis presented in Fig.~\ref{fig:forbidden_level_example_3}.
A few comments are in order.
In the analysis of $\ket{\minstate}$ and $\ket{\maxstate}$ in Fig.~\ref{fig:example_3_ctr}(a), one sees that $\ket{\minstate}$ and $\ket{\maxstate}$ have significant weight on multiple product states which are close to each other.
This arises because the Anderson orbitals near sites $2$, $4$, $5$ are delocalized, as the potential landscape at those positions appears to be relatively flat, and similarly for sites $9$, $10$, $11$.
Furthermore, if one plots weights of $\ket{\minstate}$ and $\ket{\maxstate}$ in the Anderson orbital occupation basis at $J=0$ [see the inset in Fig.~\ref{fig:example_3_ctr}(a)], one immediately notices that $\ket{\minstate}$ and $\ket{\maxstate}$ are highly localized in this basis.
One can even trace back to find two Anderson orbital occupation configurations which correspond almost one-to-one with $\ket{\minstate}$ and $\ket{\maxstate}$, illustrated by the product states in Fig.~\ref{fig:forbidden_level_example_3}(c) at the $J=0$ line.
Thus, this is an example where it is more helpful to think of a CTR as a resonance of two Anderson orbital occupation configurations rather than product states.

Just like in the previous example, this example also has two CTRs in the spectrum which transfer 2 charges, where the second CTR is detected by states $\ket{\phi_{\text{min}+1}}$ and $\ket{\phi_{\text{max}-1}}$ in the spectrum of the CTC operator with eigenvalues close to $\lambda_{\text{min}}$ and $\lambda_{\text{max}}$ respectively, as seen in the DOS of Fig.~\ref{fig:example_3_ctr}(d).
We illustrate the structure of those resonating states in Fig.~\ref{fig:example_3_ctr}(e).
Interestingly, one sees that the range of this additional resonance is also very large, involving active sites that are more than $L/2$ apart, and would be classified as a system-wide resonance that is also a CTR of $2$ particles.
We also note that the active sites in the CTR in Fig.~\ref{fig:example_3_ctr}(a) and (e) are rather different, unlike the CTRs in the previous example Fig.~\ref{fig:example_2_ctr}(a) and (e), showing a variety of possibilities that can happen while at the same time further supporting that the collar for these resonances is larger or at the limit of the accessible system size.

Finally, the level dynamics in Fig.~\ref{fig:forbidden_level_example_3} for the primary CTR in this example are more complex than previously seen.
The CTR here could be attributed to crossings of at most 5 orbitals, as illustrated in Fig.~\ref{fig:forbidden_level_example_3}(b). 
The prediction of the level crossings using the diagonally-improved first-order perturbation theory (green curve), starting from the previously mentioned Anderson orbital occupation configurations, does not match the ED data that well (but in fact does predict that the resonance occurs).
This suggests that this resonance is a higher-order process involving the crossing of multiple levels, which was clearly facilitated by an anomalously large off-diagonal matrix element.

\section{Transport distances and dipole moments of charge transport resonances}\label{app:dipole}
\begin{figure*}[t]
    \centering
    \includegraphics[width=\linewidth]{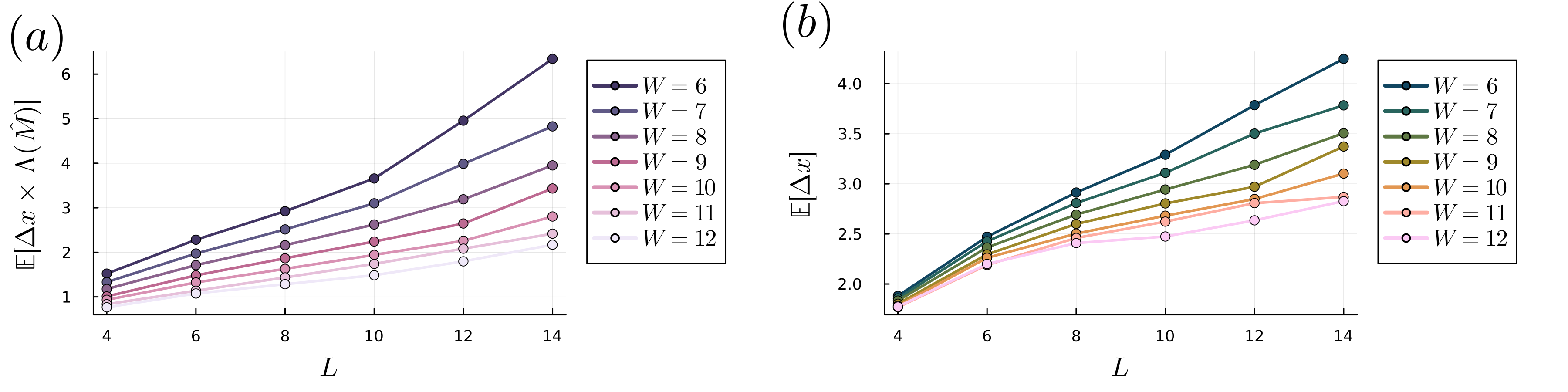}
    \caption{The analysis of the distance at which charge is transported in the CTRs that transfer the most charge. 
    Panel (a): the disorder-averaged dipole moment $p$, defined in Eq.~(\ref{eq:dipole_moment}), as a function of $L$.
    Panel (b): the disorder-averaged transport distance $\Delta x$, Eq.~(\ref{eq:ctr_distance}), as a function of $L$.
    The colored curves represent disorder strengths for $W=6$ to $W=12$ in steps of 1.
    Here, we only consider the strongly disordered regime $W\geq 6$.
    Out of 4000 disorder realizations, we select those where $|\ket{\maxstate}| \cdot |\ket{\minstate}|>0.8$ (to filter out the cases where $\ket{\maxstate}$ and $\ket{\minstate}$ do not provide information about resonances). 
    Each data point contains at least $1900$ such instances.
    Both insets show the same data, but with the data on the $y$-axis divided by $L$.
     }
\label{fig:dipole_moment_j_01}
\end{figure*}
In this Appendix, we analyze the transport distance of the charge transport resonances in the strongly disordered regime.
From the CTC data in the main text, we understand clearly that the amount of charge the CTRs can transfer increases with $L$.
However, knowing the amount of charge a resonance transfers across $i_0$ neglects how far that resonance transfers those charges.
For instance, looking at the example CTR given in Fig.~\ref{fig:example_ctr}, it appears that the 2-particle transfer process is fairly localized around $i_0$.
However, the example CTR in Fig.~\ref{fig:example_3_ctr}, on the other hand, is a 2-particle transfer process which nearly spans the entire lattice.
The existence of many resonances of the first type feels compatible with localization, while the second type does not.
This then raises the question: how can one quantify the distance of charge transfers in these CTRs, and are these CTRs becoming longer-ranged as the system size increases?
Although we cannot answer this question definitively, we present some additional data that suggests that the CTRs transporting the most charge are also transporting them across longer and longer distances as the system size is increased.

We calculate the transport distance of a resonance as follows.
First, we first define the ``center of mass'' of charge displacement to the left of the link $i_0$, using $\Delta n(i)\equiv \bra{\maxstate}\hat n_i\ket{\maxstate} - \bra{\minstate}\hat n_i\ket{\minstate}$:
\begin{equation}
    r_{\text{left}} = \frac{1}{M_{\text{left}}}\sum_{i=1}^{i_0} \Delta n(i)\times \lp \left[i_0+\frac{1}{2}\right]-i\rp~.
\end{equation}
Here, $M_{\text{left}}=\sum_{i=1}^{i_0}\Delta n(i)$.
Similarly, with $M_{\text{right}}=\sum_{i=i_0+1}^{L}\Delta n(i)$:
\begin{equation}
    r_{\text{right}} = \frac{1}{M_{\text{right}}}\sum_{i=i_0 + 1}^{L} \Delta n(i)\times \lp i-\left[i_0+\frac{1}{2}\right]\rp~.
\end{equation}
The \emph{transport distance} $\Delta x$ between $\ket{\minstate}$ and $\ket{\maxstate}$,
\begin{equation}\label{eq:ctr_distance}
    \Delta x\equiv r_{\text{right}}+r_{\text{left}}~,
\end{equation}
is a measure of roughly how far individual active charges move in the dynamical process connecting $\ket{\minstate}$ to $\ket{\maxstate}$.
This measure is not the \emph{range} of the resonance in the sense described in Sec.~\ref{sec:mbr_relation_to_ctr}.
Instead, $\Delta x$ is typically smaller than the range.
For example, if the range of A CTR of $2$ particles is $4$, then this CTR would have $\Delta x=2$.
To provide concrete examples, the CTR example in the main text, Fig.~\ref{fig:example_ctr}, has $\Delta x\approx 2.51$. 
The CTR example in Fig.~\ref{fig:example_2_ctr} has $\Delta x\approx 3.05$, while the CTR example in Fig.~\ref{fig:example_3_ctr} has $\Delta x\approx 5.90$.

Another relevant quantity is the dipole moment of $\ket{\maxstate}$ and $\ket{\minstate}$:
\begin{equation}\label{eq:dipole_moment}
p \equiv \ctc\Delta x~.
\end{equation}
Both of these quantities are intended as measures when $\ket{\minstate}$ and $\ket{\maxstate}$ are connected by time evolution and have a net change in charge across the cut. 
It breaks down as a measure of the distance of transport in cases where there is a rearrangement of charges on either side of the cut, but no net charge is moved over the cut.
For instance, tiny fluctuations in charge far from the cut can artificially suggest a large transport distance, even though no charge has moved across the cut.
Thus, when calculating these quantities, we only consider cases where $|\ket{\maxstate}|\cdot|\ket{\minstate}|\approx 1$.
When one additionally multiplies the transport distance by $\ctc$, we also filter out cases with little transported charge across the cut.

We plot the analysis of both the transport distance $\Delta x$ and the dipole moment $p$ in Fig.~\ref{fig:dipole_moment_j_01} for the strongly disordered regime.
From the figure, one can see that both the disorder-averaged dipole moment $p$ and the transport distance $\Delta x$ show a growth trend for the $L$ values considered, suggesting that the CTRs of the largest charges are also transporting these charges over a longer range.
Of course, the data must be interpreted with the understanding that the number of charges in the worst-case CTR grows with $L$. 
As mentioned previously, the transport distance is directly correlated with the number of charges transferred across the cut.
Thus, if the charge transfer amount is increasing, then one would expect the distance to increase, just as is observed here.

To disentangle whether the growth in the dipole moment and $\Delta x$ is driven by the CTRs becoming longer-ranged, or simply by the CTRs transferring more charges, we show in the insets of Fig.~\ref{fig:dipole_moment_j_01} the same data normalized by $L$. 
One sees in these insets that the dipole moment continues to increase even when divided by $L$. 
Furthermore, the inset of Fig.~\ref{fig:dipole_moment_j_01}(b) shows that $\Delta x/L$ appears to approach a flat value, suggesting that the range of the CTRs is extensively large at these system sizes.
We emphasize that the system sizes here are small, so the conclusions that can be drawn here are limited---transport processes involving 2 particles, which appear to occur localized near the cut, still have a range that is large (i.e., a good fraction of) relative to the total small system size.
This limitation can be appreciated from the quoted values of $\Delta x$ for the CTR examples in Figs.~\ref{fig:example_ctr},~\ref{fig:example_2_ctr},~\ref{fig:example_3_ctr}, which are comparable to $\doubleE[\Delta x]$ in Fig.~\ref{fig:dipole_moment_j_01}.
Still, these data provide some additional insight into the range of the observed CTRs.

\section{Charge transport capacity in the XXX model}\label{app:large_interaction_strength}
\begin{figure*}[t]
    \centering
    \includegraphics[width=\linewidth]{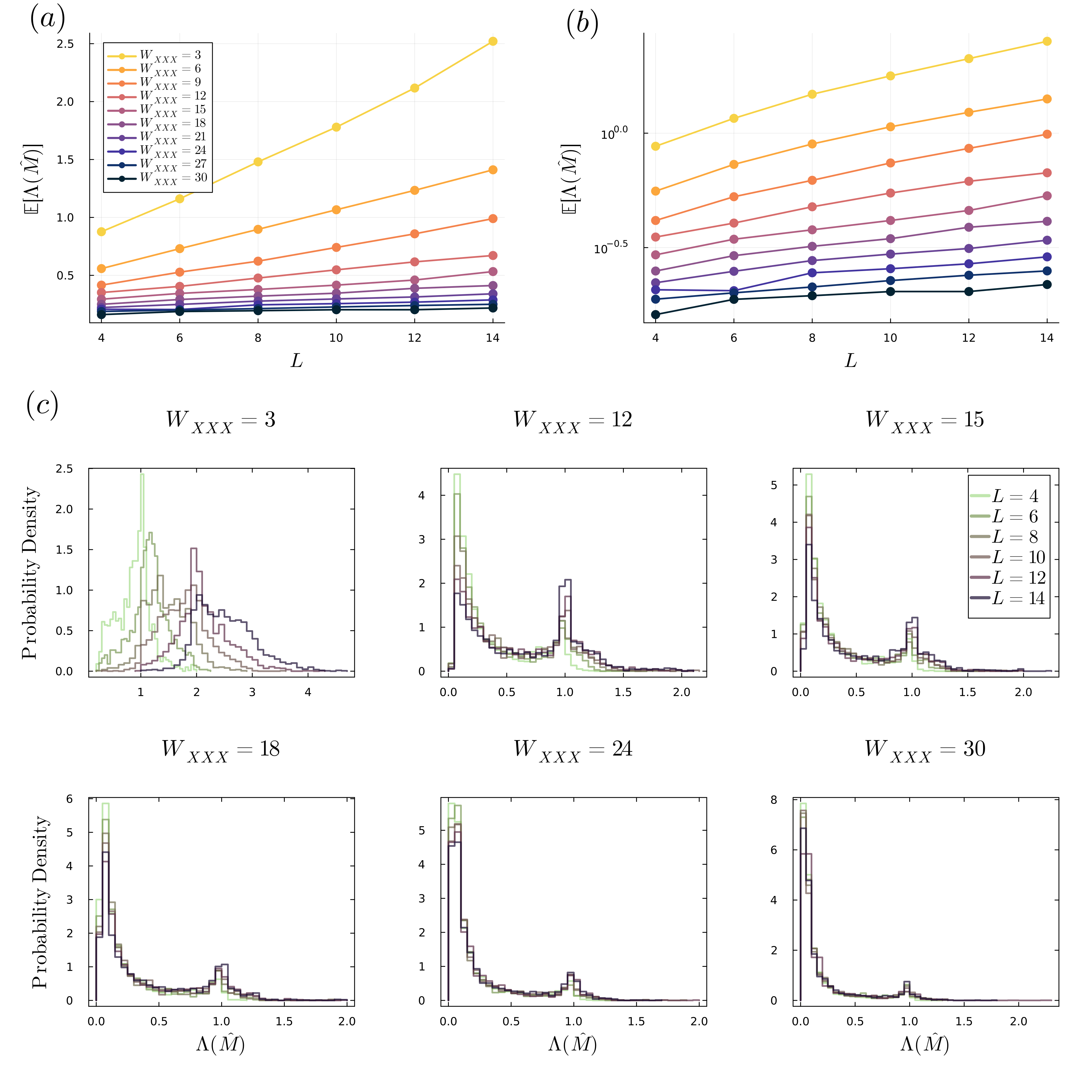}
    \caption{Analysis of the CTC in the interacting Anderson model for the parameters $t=0.5, J=1$. These are the same parameters as for the commonly studied disordered Heisenberg model (XXX chain).
    The interpretation of this figure is the same as in Fig.~\ref{fig:ctcb_main_results_J_01}, but for the specified parameter regime.
    Note that the range of $W$ studied in this parameter regime is much larger than that of the data in the main text. For each disorder strength $W$ and system size $L$, 2000 disorder realizations were used.
     }
\label{fig:ctcb_main_results_J_1}
\end{figure*}

\begin{figure*}[t]
    \centering
    \includegraphics[width=\linewidth]{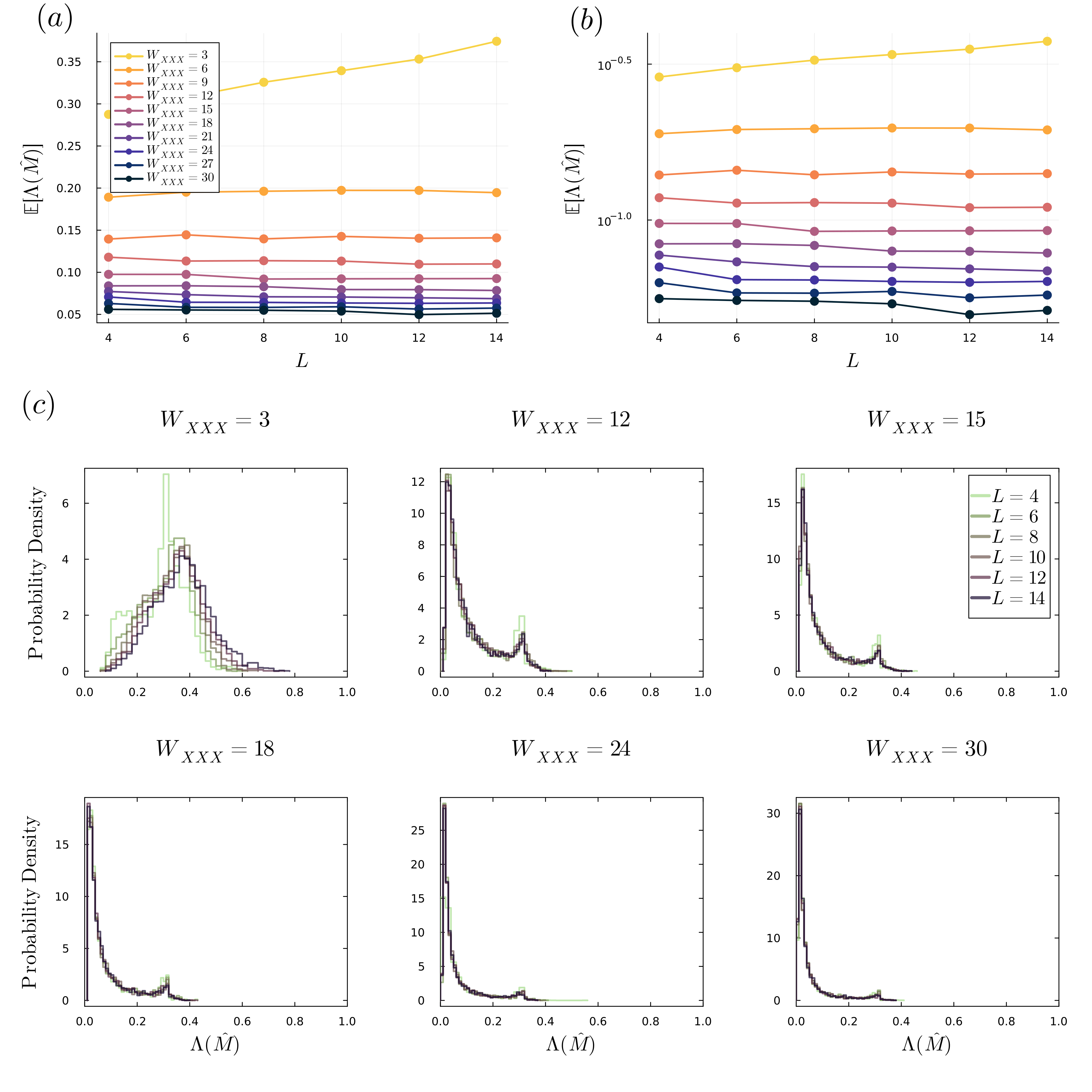}
    \caption{
    Analysis of the half-filling normalized Frobenius norm of $\hat M$, $\|\hat M\|_F^{\text{half-filling}}$, in the interacting Anderson model for the parameters $t=0.5, J=1$. These are the same parameters as for the commonly studied Heisenberg chain with a disordered $Z$ field.
    The interpretation of this figure is the same as in Fig.~\ref{fig:frobenius_norm_j_01}, but for the specified regime. 
     }
\label{fig:frobenius_main_results_J_1}
\end{figure*}

In this appendix, we present data on the charge transport capacity $\ctc$ and the normalized half-filling Frobenius norm $\|\hat M\|_F$ for the model Eq.~(\ref{eq:interacting_hamiltonian}) with the new parameters $J=1, t=0.5$. 
These are the equivalent parameters in our model to those of the commonly studied disordered Heisenberg chain or XXX model (for instance, see Eq.~(7) in Ref.~\cite{sierant_many-body_2025}).
In the disordered XXX model, the accepted lower-bound on the critical disorder strength is $W_{XXX} > 18$ \cite{sels_bath-induced_2022,morningstar_avalanches_2022}.
We will show that the overall behavior of the CTC and related quantities appears to be qualitatively similar in the XXX model to that in the $J=0.1$, $t=1$ regime.

In Fig.~\ref{fig:ctcb_main_results_J_1}, we plot an analysis of the charge transport capacity for a range of $W_{XXX}$ around the literature-estimated lower bound on $W^*_{XXX}$.
We note that the CTC behaves qualitatively similarly in the $J=1$, $t=0.5$ and $J=0.1$, $t=1$ regimes for disorder strengths stronger than the literature-estimated transitions.
This suggests that the physics of charge transport resonances discussed in the main text may apply in this regime as well.
An interesting observation in this data is that we do not observe any CTRs of 2 particles beyond $W^*_{XXX}=18$, starkly different from what is observed in the $J=0.1$ case.
A possibility is that, at these system sizes, the disorder strengths simulated here are too large for the interaction to sufficiently mix the noninteracting energy levels and produce true many-body physics.

In Fig.~\ref{fig:frobenius_main_results_J_1}, we show our analysis of $\|\hat M\|_F^{\text{half-filling}}$ for the XXX model, covering $W_{XXX}$ smaller and larger than $W_{XXX}=18$.
We do not attempt to provide an estimate of $W_{XXX}^*$ from our data due to the large amount of noise, but we note that $\|\hat M\|_F^{\text{half-filling}}$ appears to saturate with $L$ even starting from $W_{XXX}=6$.
This cautions against using the saturation of $\|\hat M\|_F^{\text{half-filling}}$ to estimate $W^*$.

\bibliography{references, references-2}

\end{document}